\documentclass[review,nopreprintline]{elsarticle}

\usepackage{float}
\usepackage{xr}
\usepackage{subcaption}
\usepackage{longtable}
\usepackage[table]{xcolor}
\usepackage{mathtools}
\usepackage{lineno}
\usepackage{booktabs}
\usepackage[graphicx]{realboxes}
\usepackage{multirow}
\usepackage{csquotes}
\modulolinenumbers[5]

\usepackage[
            ]{hyperref}

\usepackage[superscript,biblabel]{cite}
\newcommand{\beginsupplement}{%
	\setcounter{table}{0}
	\renewcommand{\thetable}{S\arabic{table}}%
	\setcounter{figure}{0}
	\renewcommand{\thefigure}{S\arabic{figure}}%
	\setcounter{section}{0}
	\renewcommand{\thesection}{Supplemental \arabic{section}}%
	\setcounter{equation}{0}
	\renewcommand{\theequation}{S\arabic{equation}}
	\setcounter{page}{1}
}

\usepackage{geometry}
\newlength{\mymargin}
\begin{document}

\begin{frontmatter}

\title{Exploring carbon dioxide removal strategies to help decarbonise Europe using high-resolution modelling}

\author[mpe,novo]{Ricardo Fernandes \corref{cor1}}
\cortext[cor1]{Lead contact and corresponding author, Email: ricardo.fernandes@mpe.au.dk}
\author[mpe]{Alberto Alamia}
\author[mpe,novo]{Sina Kalweit}
\author[DTU,novo]{Marta Victoria}

\affiliation[mpe]
{
	organisation = Department of Mechanical and Production Engineering, Aarhus University,
	city = Aarhus,
	postcode = 8200,
	country = Denmark
}

\affiliation[novo]
{
	organisation = The Novo Nordisk Foundation CO$_2$ Research Center,
	city = Aarhus,
	postcode = 8000,
	country = Denmark
}

\affiliation[DTU]
{
    organisation = Department of Wind and Energy Systems, Technical University of Denmark,
    city = Kongens Lyngby,
    postcode = 2800,
    country = Denmark
}

\begin{abstract}
The electrification of energy demand across sectors, powered by solar and wind generation, is the best strategy for achieving carbon neutrality. Carbon dioxide removal (CDR) strategies are also expected to play a crucial role by providing net-negative emissions that can offset residual CO$_2$ emissions, including those from cement manufacturing. While previous studies have assessed the role of CDRs in Europe's decarbonisation, most either focus solely on combinations of biogenic point-source capture and direct air capture (DAC) coupled with underground sequestration, or consider multiple CDR strategies at low spatial and temporal resolution, thereby limiting the representation of linkages amongst technologies. In this study, the sector-coupled European energy system model PyPSA-Eur is extended to include afforestation, perennialisation, biochar, and enhanced rock weathering (ERW) as additional CDR strategies. Using this model with a 3-hourly resolution and a network comprising 90 nodes, results show that a climate-neutral energy system equipped with these CDR strategies is 9\% less expensive. Afforestation, perennialisation, and ERW potentials are fully utilised across regions, whereas biochar is not selected due to limited solid biomass feedstock being allocated to other higher-value processes. Furthermore, when these CDR strategies are combined with underground sequestration and a continental CO$_2$ transport network, DAC is no longer required to achieve climate neutrality in Europe.
\end{abstract}


\end{frontmatter}

\begin{linenumbers}

\nolinenumbers

\clearpage

\section{Introduction}
A massive deployment of solar and wind power, together with direct and indirect electrification of other sectors, is the most cost-effective strategy to attain carbon neutrality by mid-century in Europe \cite{VICTORIA20221066} and the rest of the world \cite{Bogdanov_2021, Luderer_2021}. Even for the most advanced mitigation scenarios, a relatively small amount of carbon dioxide (CO$_2$) emissions remains unabated, such as those associated with cement manufacturing. Achieving net-zero emissions therefore requires compensating these residual emissions through carbon dioxide removal (CDR).

CDR refers to strategies designed to capture CO$_2$ from the atmosphere and sequester it for long periods of time, at least centuries, effectively attaining net-negative emissions. The potential impact of CDR strategies remains highly uncertain due to their early deployment state and the challenges associated with monitoring, reporting, and verification (MRV) of their real long-term contribution. Several CDR strategies impacting the energy system have been proposed, with six strategies standing out. First, solid biomass can be used to produce electricity and heat while capturing the emitted CO$_2$. Second, direct air capture (DAC) can be used to capture CO$_2$ from the atmosphere \cite{Creutzig_2019}, although at high cost due to the large energy input requirement \cite{Spek_2025}. When combined with underground CO$_2$ sequestration, these two options can be considered CDR strategies and are commonly referred to as bioenergy with carbon capture and storage (BECCS) \cite{Hanssen_2020} and direct air capture with carbon storage (DACCS) \cite{Spek_2025}, respectively. Third, afforestation—the establishment of forests on previously non-forested land—enhances the capacity of terrestrial ecosystems to absorb and sequester CO$_2$. Limited data on carbon sequestration rates, available land for afforestation, and impacts of forest management practices make the CDR potential of afforestation highly uncertain \cite{Wang_2025, Avitabile2024, Pilli_2022}. Fourth, biochar-based CDR converts biogenic carbon into a solid form of carbon via pyrolysis, which can be added to soils and remain stable for hundreds of years. Besides uncertainties in cost and potential \cite{Deng_2024}, large-scale biochar deployment raises the question of whether biogenic carbon could be more efficiently used elsewhere, for example, as a carbon source for synthetic fuels \cite{Millinger2025}. Fifth, the conversion of annual crops into perennial systems enhances carbon removal through continuous photosynthesis, deeper root systems, and reduced soil disturbance \cite{Chen_2022}. Limited data on soil sequestration rates and indirect emissions, such as those arising from the displacement of annual crops, are the main uncertainties of this CDR strategy. Sixth, enhanced rock weathering (ERW) involves spreading finely crushed silicate rocks over land, where they react with CO$_2$ in the atmosphere and rainwater to form carbonate minerals that remain stable for thousands of years. The main uncertainties about the potential of this CDR strategy relate to weathering rates and the rock mass that can be integrated into the soil \cite{Strefler2018}.

The initial representation of CDR strategies in Integrated Assessment Models (IAMs) included only BECCS, which was extensively used in net-zero scenarios \cite{IPCC}. However, the unsustainable land requirements and uncertain cost estimation raised doubts about this strategy \cite{Gambhir_2019, Mclaren_2020, Creutzig_2019, Van_vuuren_2018}. Recently, a more diverse representation of CDR strategies has been included in some global IAMs \cite{Strefler_2021, Fuhrman2023} and in an energy system for Europe \cite{Markkanen_2024}, unveiling the different regional potentials of each strategy. Nonetheless, existing studies typically rely on coarse spatial and temporal resolutions, limiting the representation of the linkages between CDR strategies and the operation of the energy system.

To address this limitation, we employ PyPSA-Eur, an open-source model of the European energy system including a 90-node network, 3-hourly resolution, and detailed representations of the power, heating, transport, agriculture, and industrial sectors. Previous analyses using PyPSA-Eur considered only underground CO$_2$ sequestration with a potential of 200 MtCO$_2$/a, which, combined with DAC or biogenic carbon capture, enabled net-negative emissions. This potential was consistently maxed out across a wide range of experiments \cite{Neumann2023, VANGREEVENBROEK2025101974, FERNANDES2026101593, VICTORIA20221066}. Furthermore, it has been found that increasing the sequestration potential substantially reduces system costs and allows offsetting emissions from fossil oil \cite{Millinger2025, Hofmann2025, VICTORIA20221066}. As a main novelty in this study, we extended PyPSA-Eur with a stylised representation of CDR potentials from afforestation, perennialisation, biochar, and ERW across Europe. We investigated how these additional CDR strategies collaborate and compete when modelling at high spatial resolution while representing all the relevant temporal and technological linkages within a climate-neutral sector-coupled energy system.

\section{Results}

\subsection{Additional CDR strategies enable more cost-effective decarbonisation of Europe}
\begin{figure}[!htb]
    \centering
    \includegraphics[width = 0.38\linewidth]{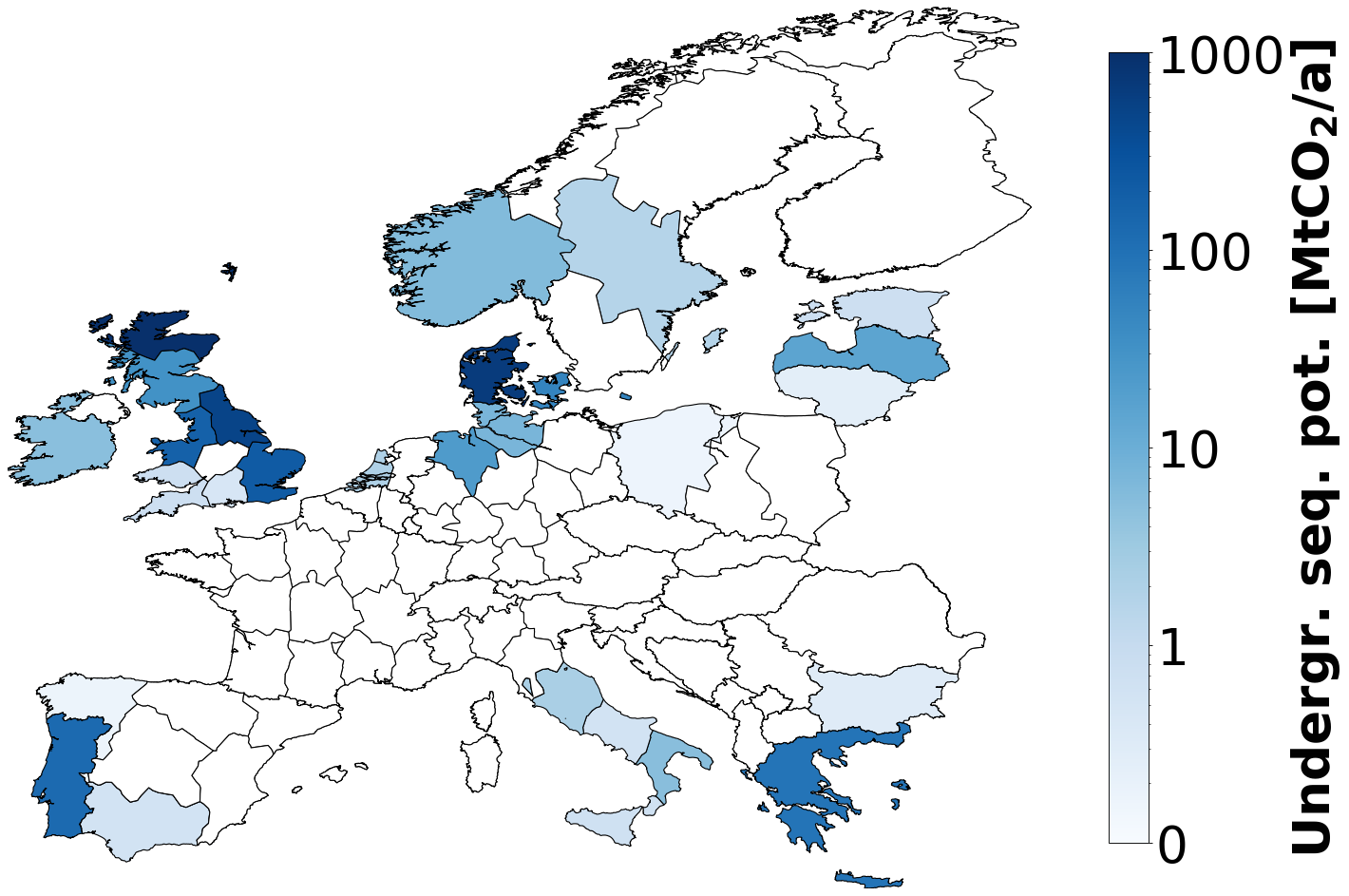}\hfill
    \includegraphics[width = 0.38\linewidth]{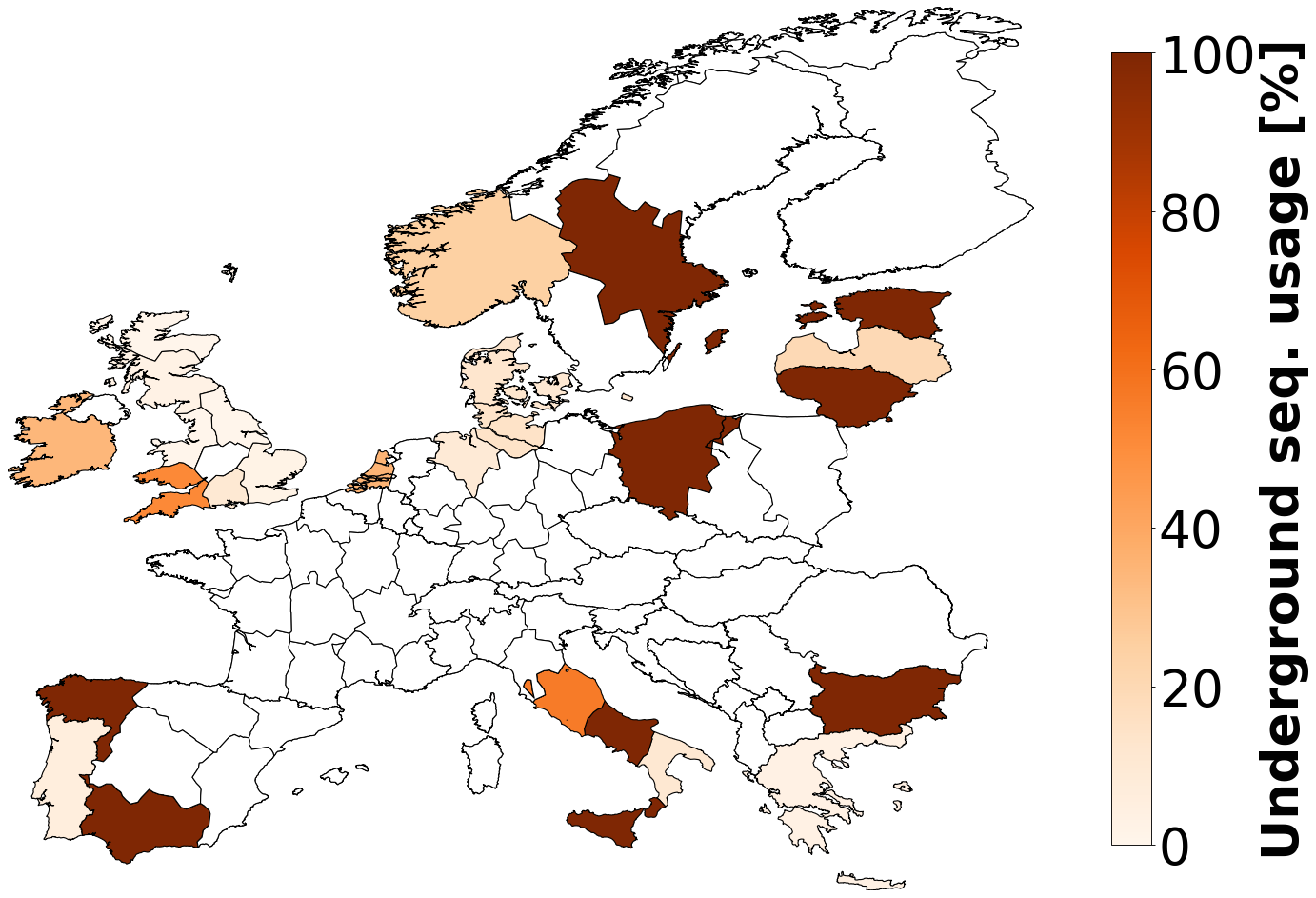}\par
    \vspace{10pt}
    \includegraphics[width = 0.38\linewidth]{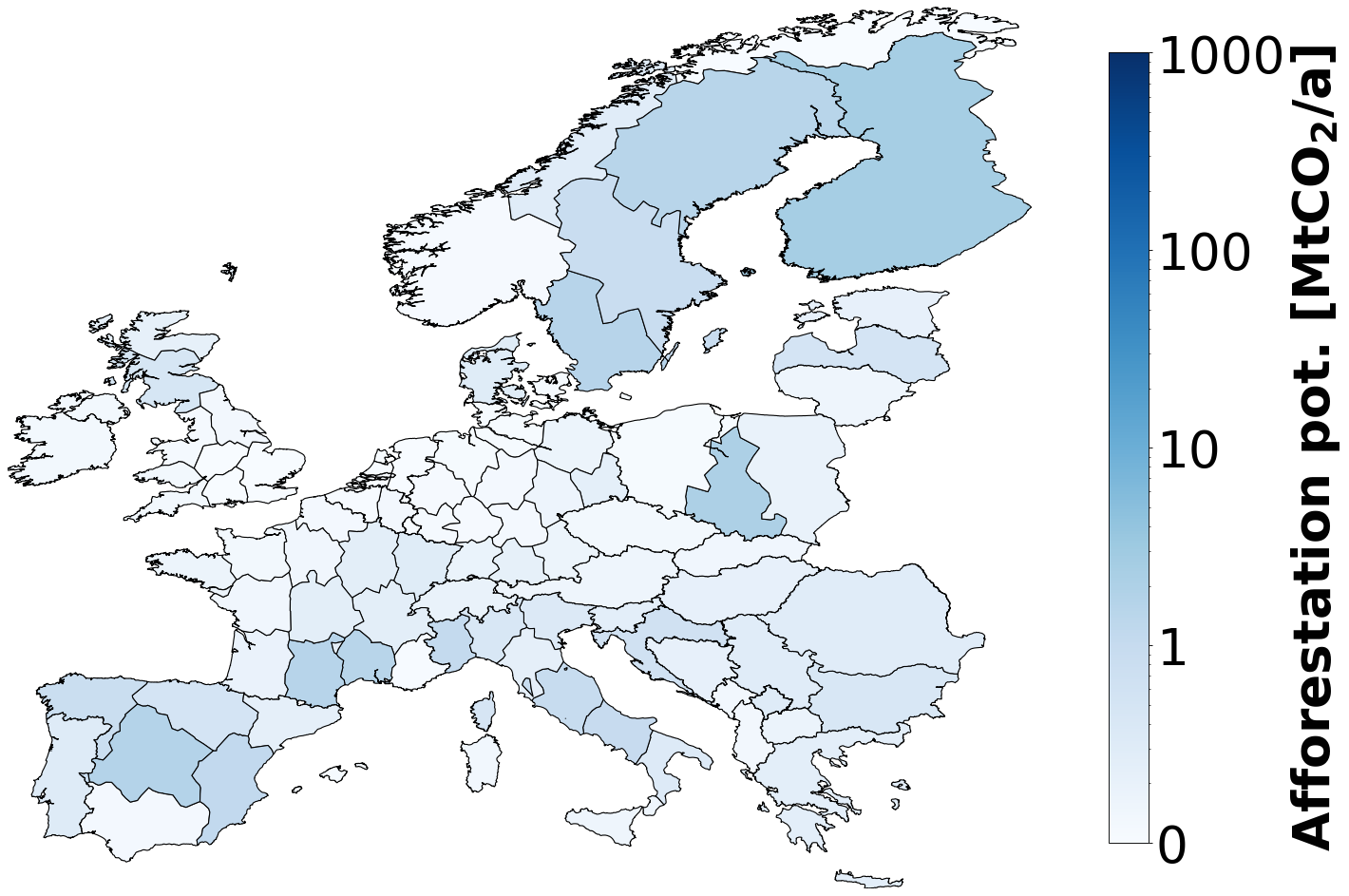}\hfill
    \includegraphics[width = 0.38\linewidth]{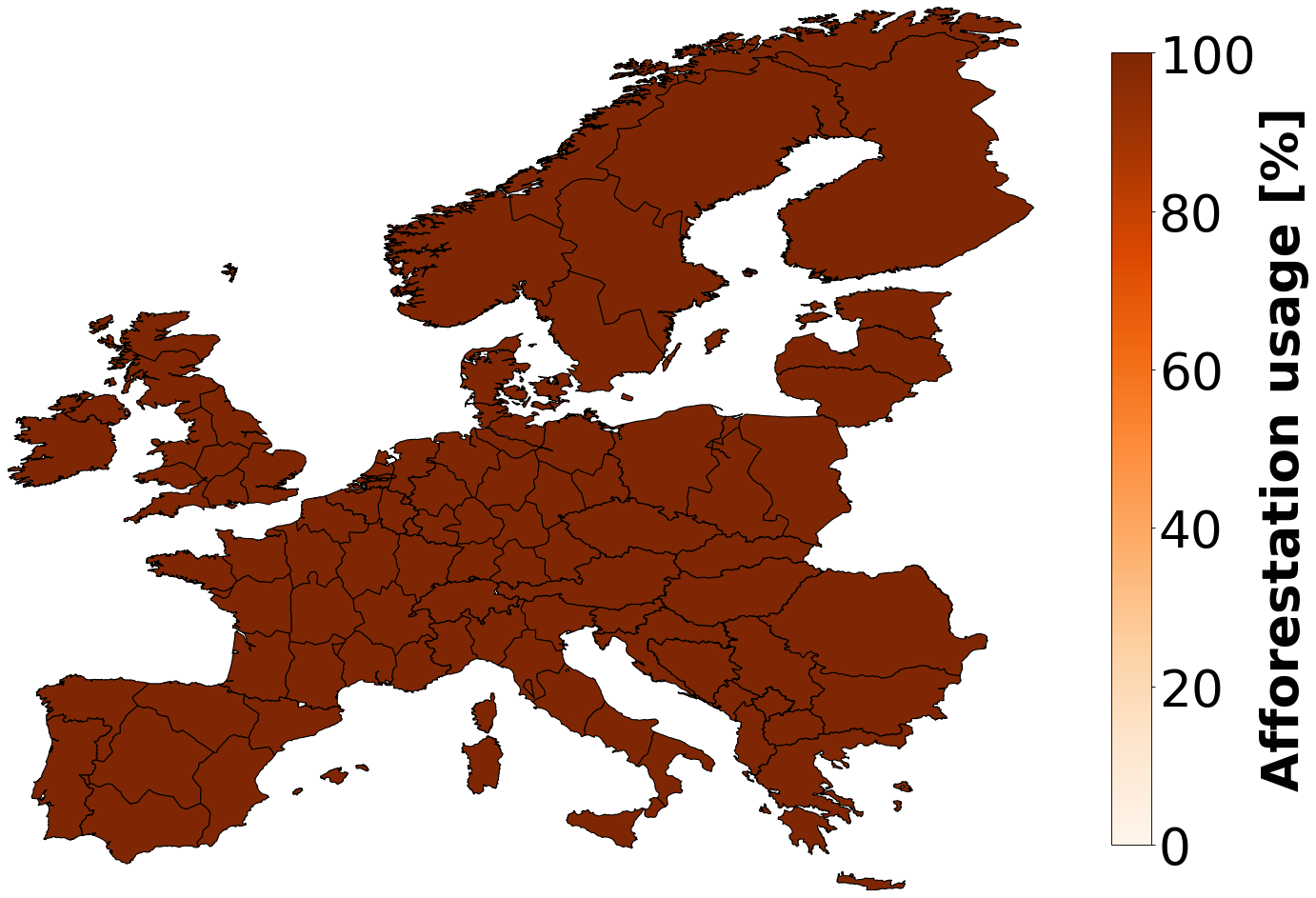}\par
    \vspace{10pt}
    \includegraphics[width = 0.38\linewidth]{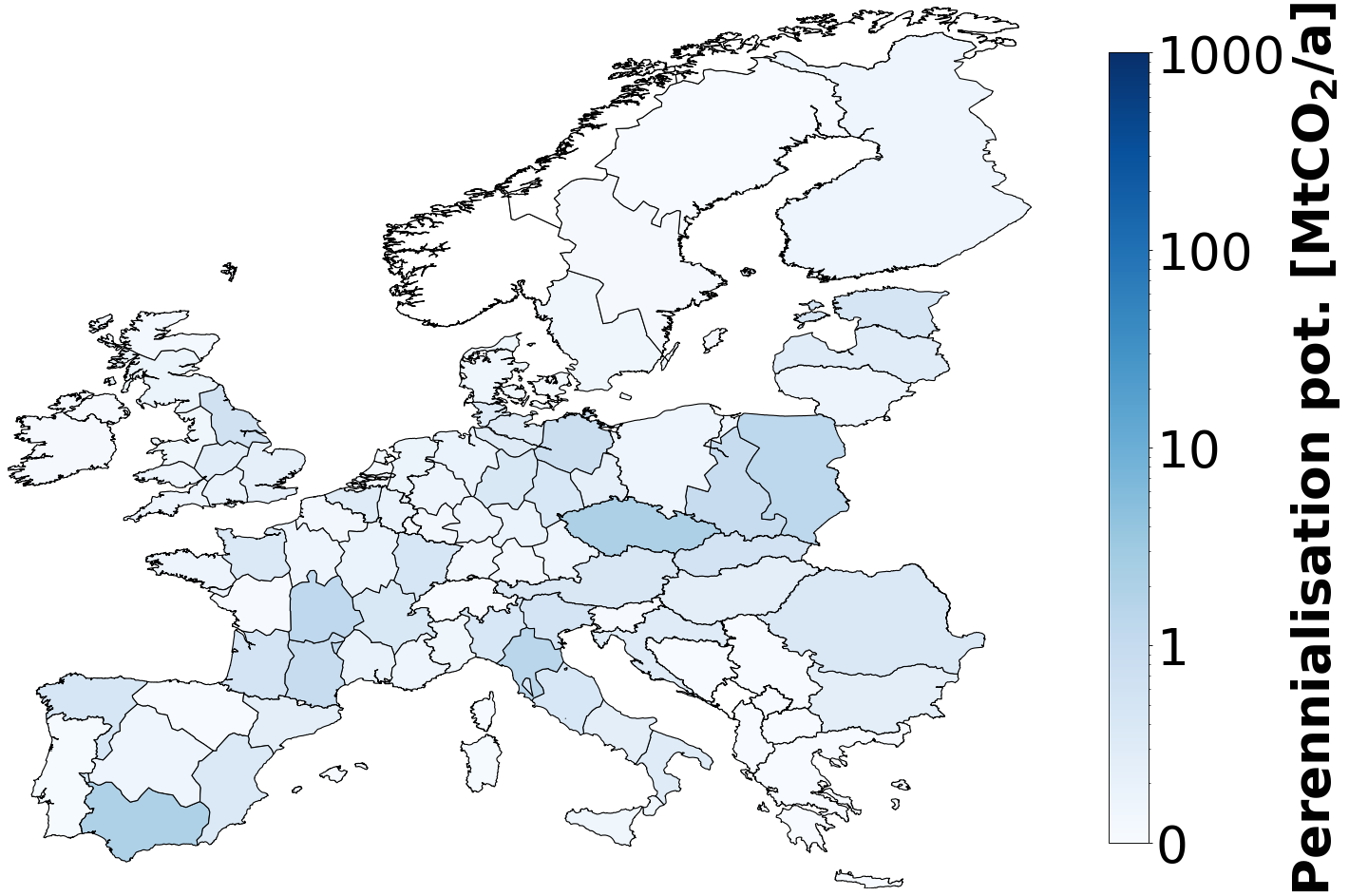}\hfill
    \includegraphics[width = 0.38\linewidth]{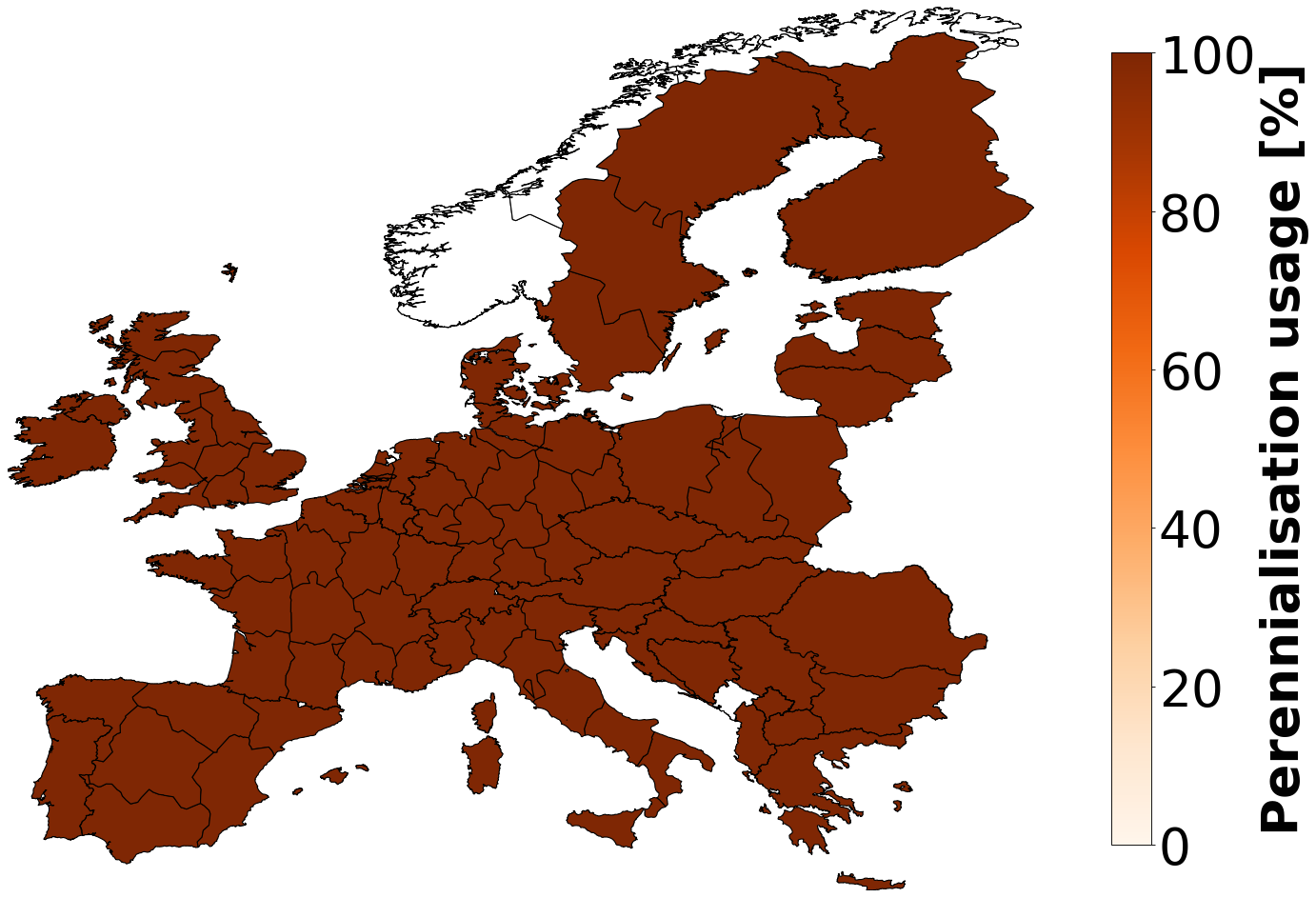}\par
    \vspace{10pt}
    \includegraphics[width = 0.38\linewidth]{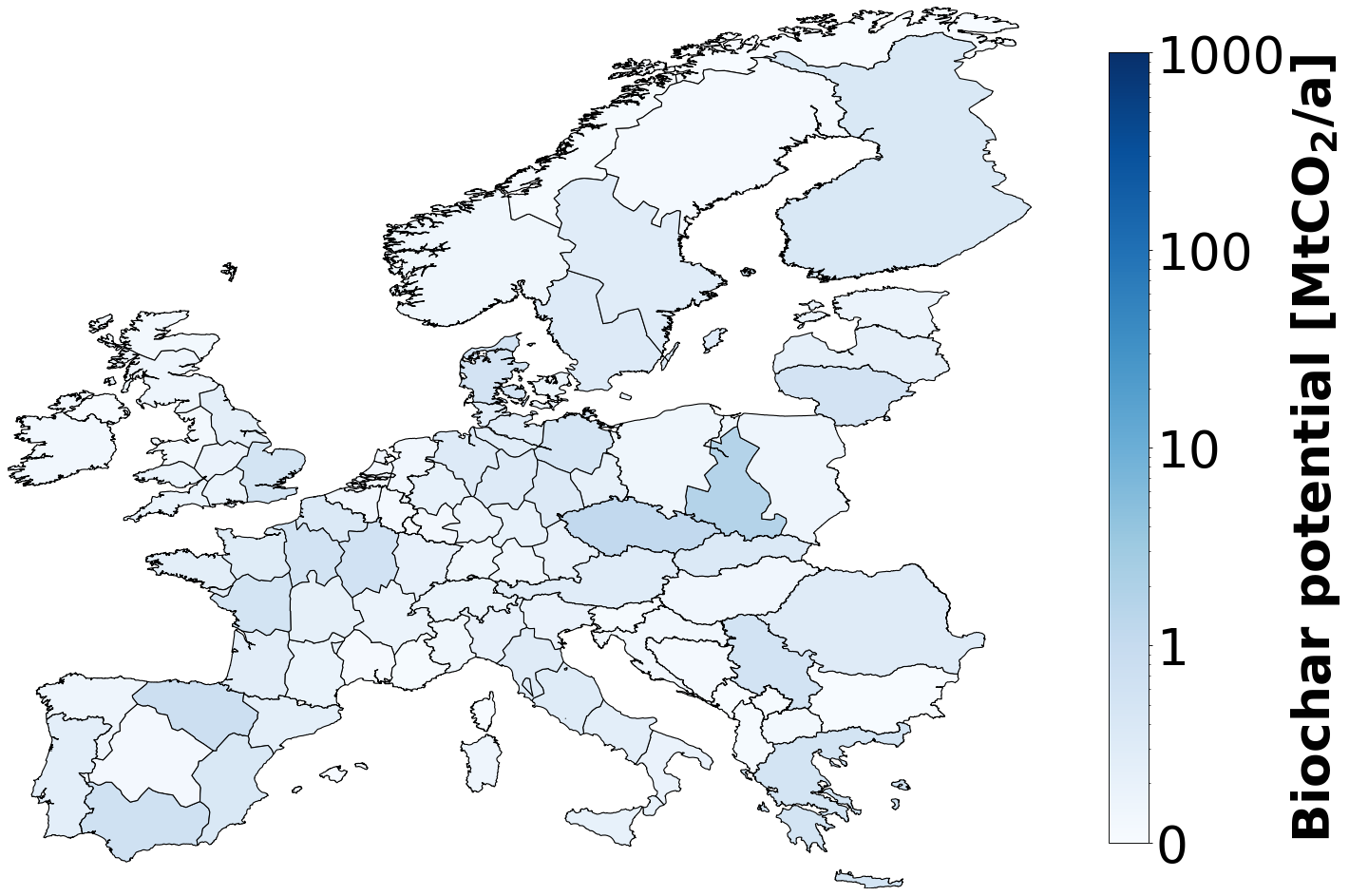}\hfill
    \includegraphics[width = 0.38\linewidth]{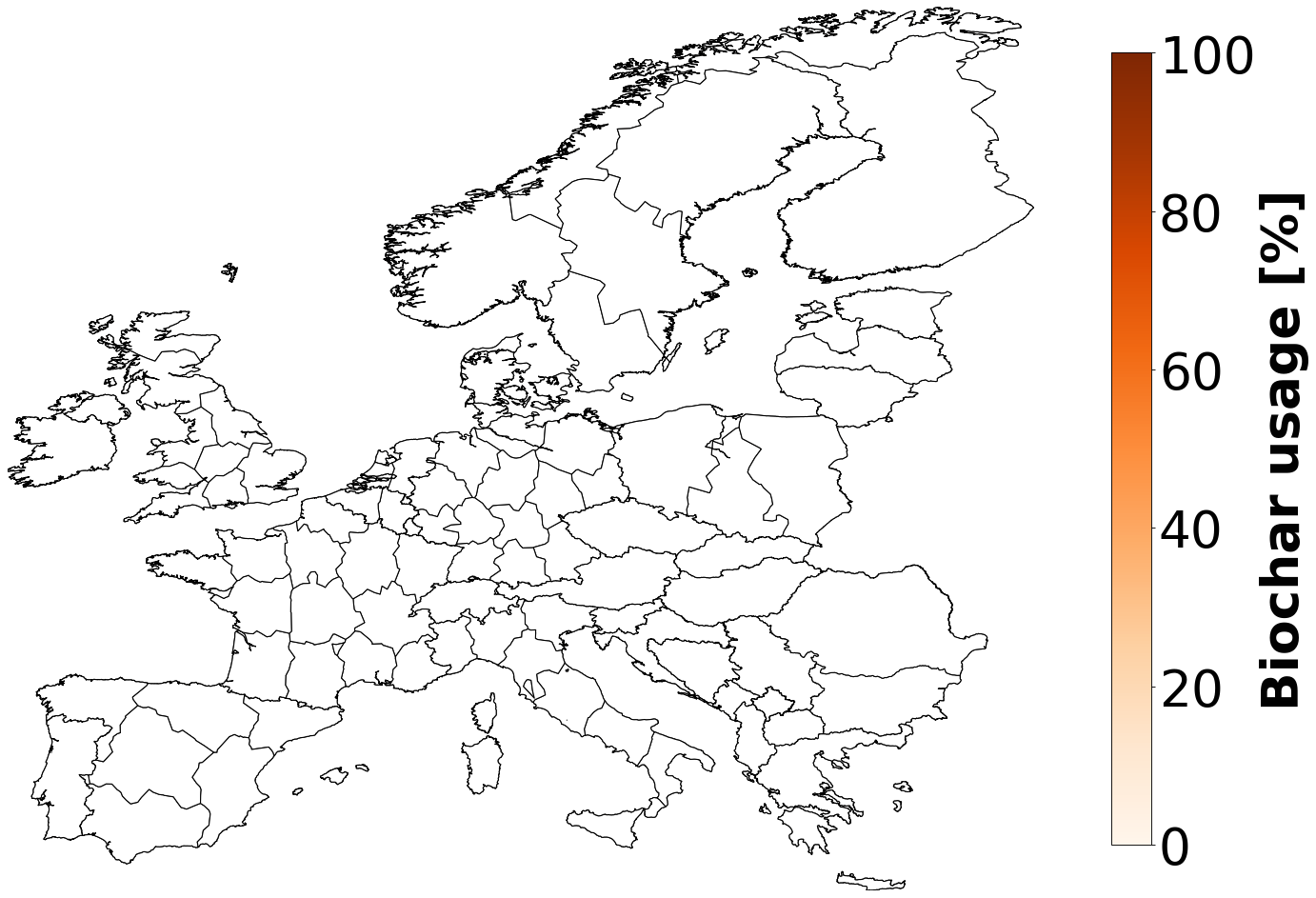}\par
    \vspace{10pt}
    \subfloat[]{\label{figure_cdr_potential}\includegraphics[width = 0.38\textwidth]{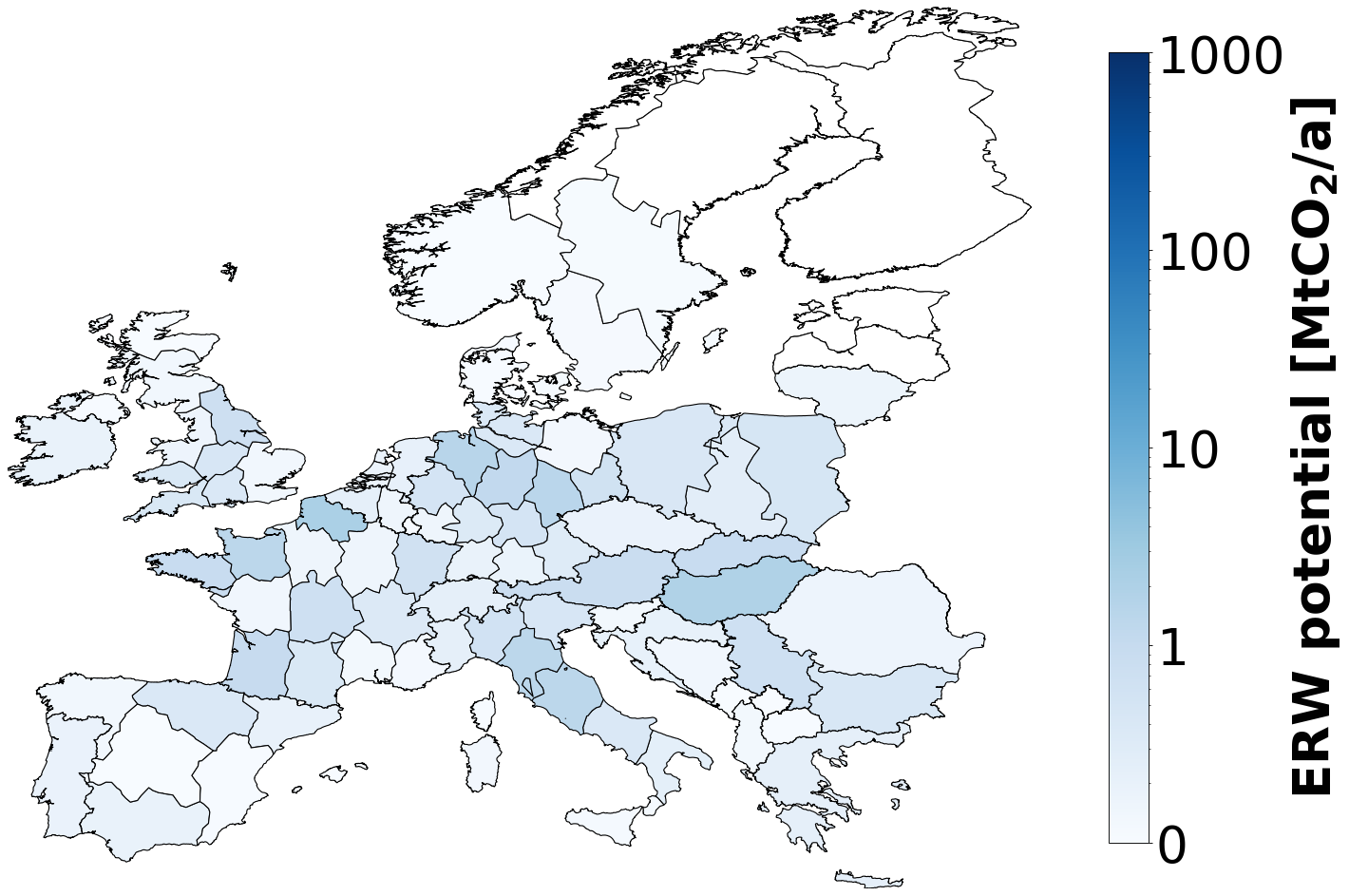}}\hfill
    \subfloat[]{\label{figure_cdr_usage}\includegraphics[width = 0.38\textwidth]{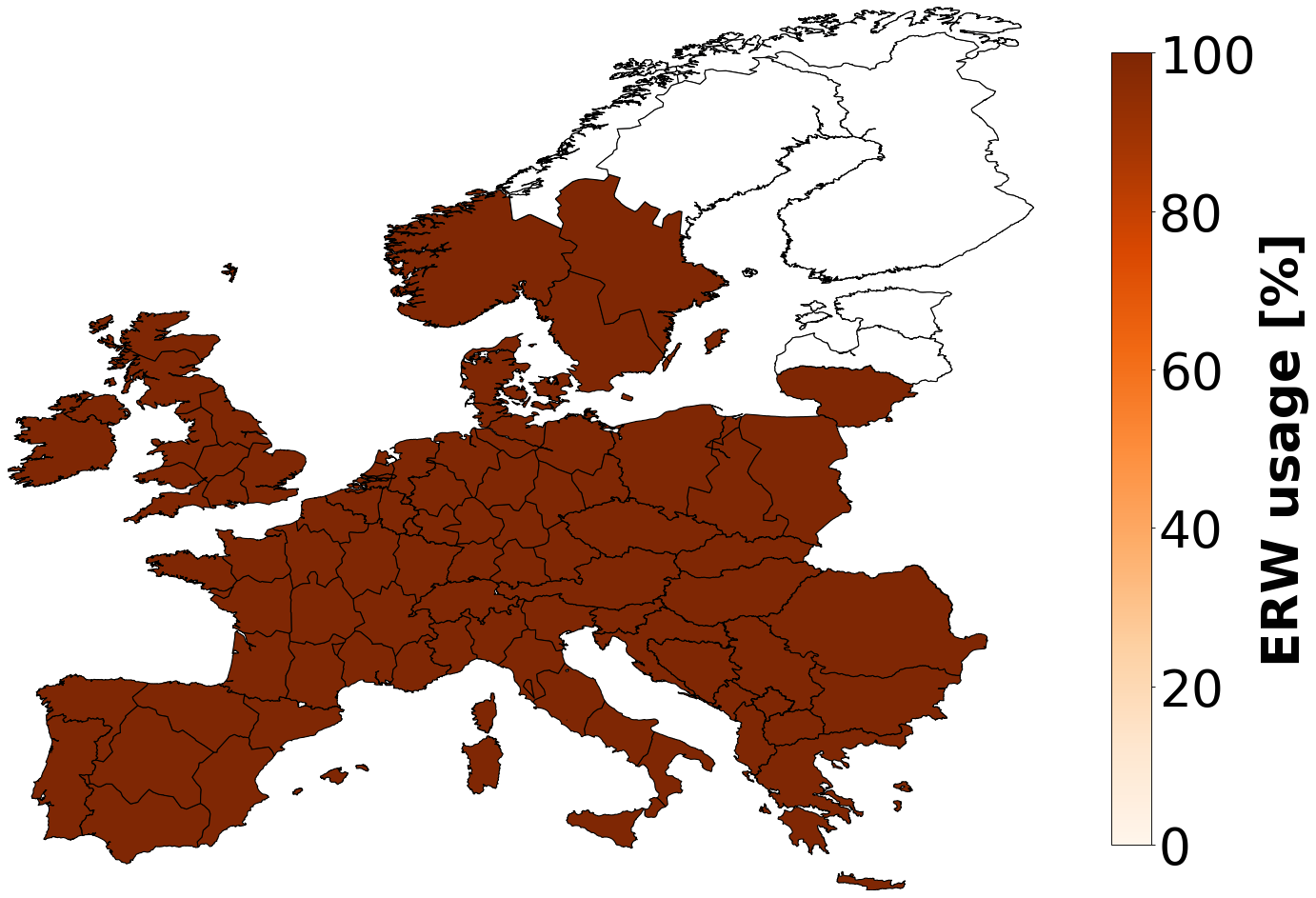}}\par
    \caption[CDRs potentials and usages across Europe]{\textbf{CDRs potentials and usages across Europe}. CDRs (A) potentials and (B) usages.}
    \label{figure_cdr_potential_usage}
\end{figure}

\begin{figure}[H]
    \centering
    \includegraphics[width = 0.84\textwidth]{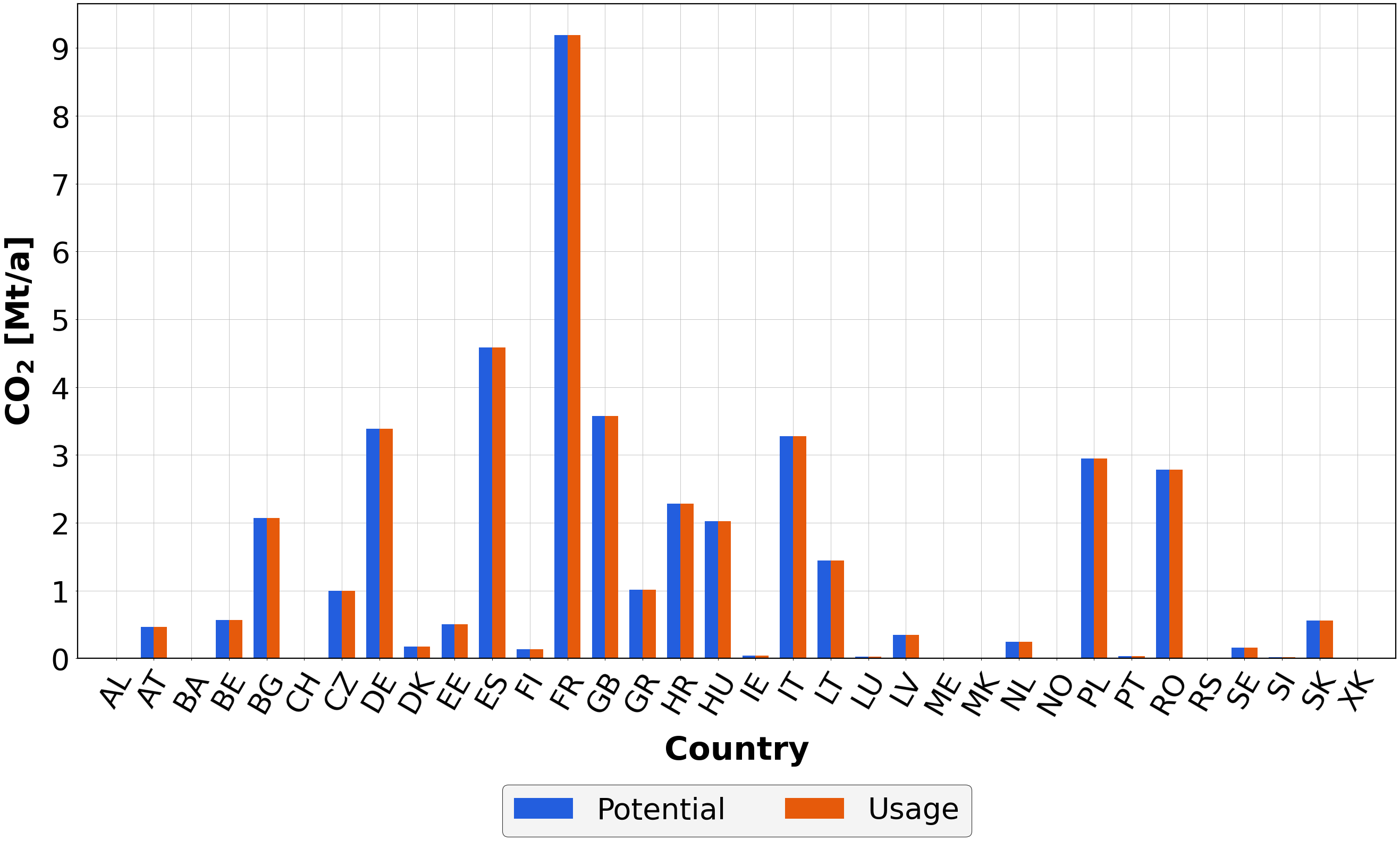}
    \caption[Perennialisation CO$_2$ removal potential and usage per country]{\textbf{Perennialisation CO$_2$ removal potential and usage per country}. Perennialisation potential is fully used in every country in a climate-neutral energy system equipped with all CDR strategies, amounting to a total of 43 MtCO$_2$/a across Europe.}
    \label{supplemental:figure_perennials_co2_store_potential_and_usage}
\end{figure}

\begin{figure}[H]
    \centering
    \includegraphics[width = 0.84\textwidth]{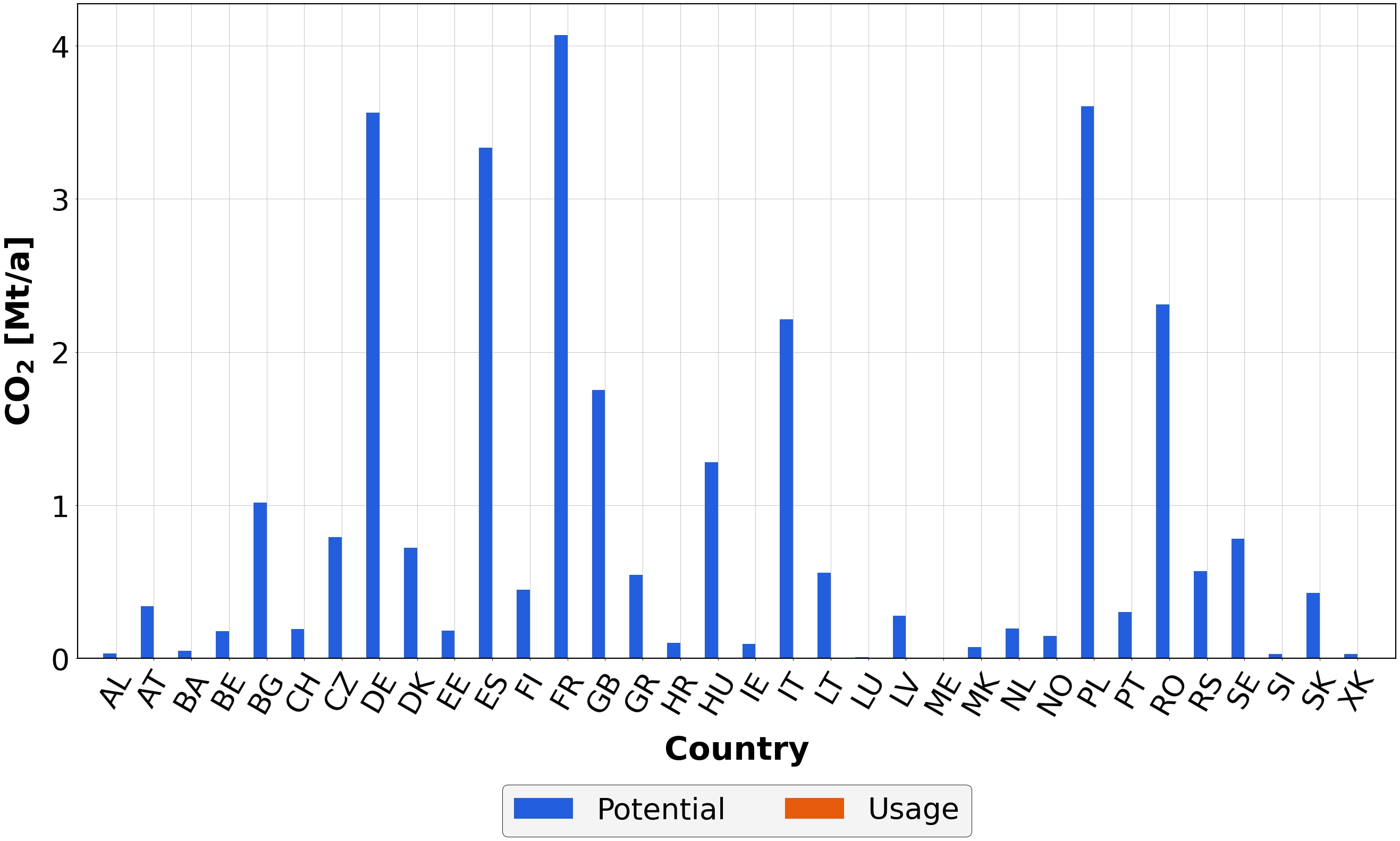}
    \caption[Biochar CO$_2$ removal potential and usage per country]{\textbf{Biochar CO$_2$ removal potential and usage per country}. Due to the solid biomass required for biochar production being prioritised for other processes, biochar potential is not used in any country as a viable CDR strategy in a climate-neutral energy system equipped with all CDR strategies.}
    \label{supplemental:figure_biochar_co2_store_potential_and_usage}
\end{figure}

\begin{figure}[H]
    \centering
    \includegraphics[width = 0.84\textwidth]{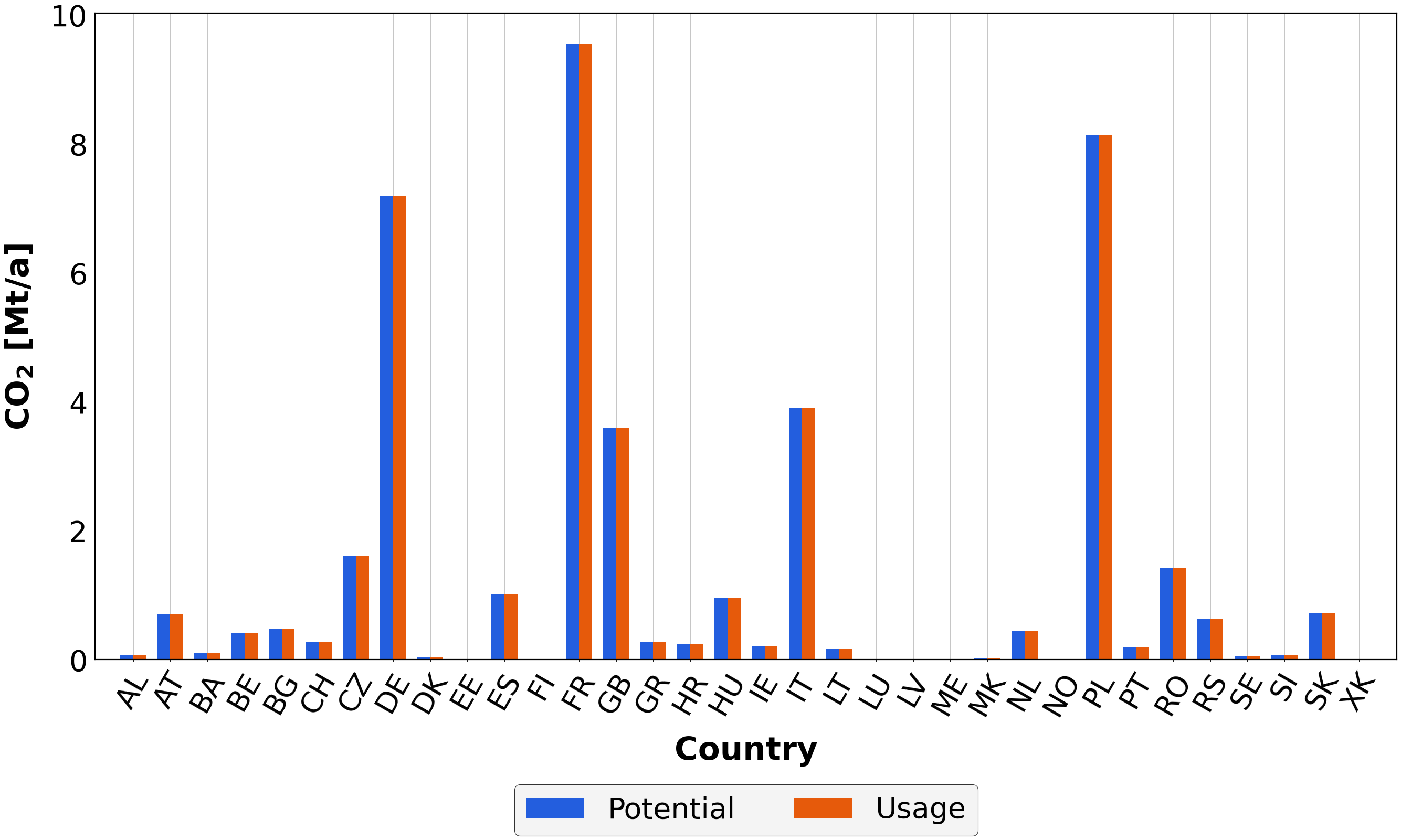}
    \caption[ERW CO$_2$ removal potential and usage per country]{\textbf{ERW CO$_2$ removal potential and usage per country}. ERW potential is fully used in every country in a climate-neutral energy system equipped with all CDR strategies, amounting to a total of 43 MtCO$_2$/a across Europe.}
    \label{supplemental:figure_erw_co2_store_potential_and_usage}
\end{figure}

Using the open-source model PyPSA-Eur \cite{Neumann2023, HORSCH2018207}, we co-optimised the capacities and dispatch of energy generation, storage, transmission, and conversion in the different sectors, along with technologies for CO$_2$ capture, conversion, and sequestration in Europe under a net-zero CO$_2$ emissions constraint. Amongst the four additional CDR strategies examined, afforestation, perennialisation, and ERW are utilised to their full potential. In contrast, biochar is not selected by the model (Figures~\ref{figure_cdr_potential_usage}, \ref{supplemental:figure_afforestation_co2_store_potential_and_usage}, \ref{supplemental:figure_perennials_co2_store_potential_and_usage}, \ref{supplemental:figure_biochar_co2_store_potential_and_usage}, and \ref{supplemental:figure_erw_co2_store_potential_and_usage}).

\begin{figure}[!htb]
    \centering
    \includegraphics[width = 0.84\textwidth]{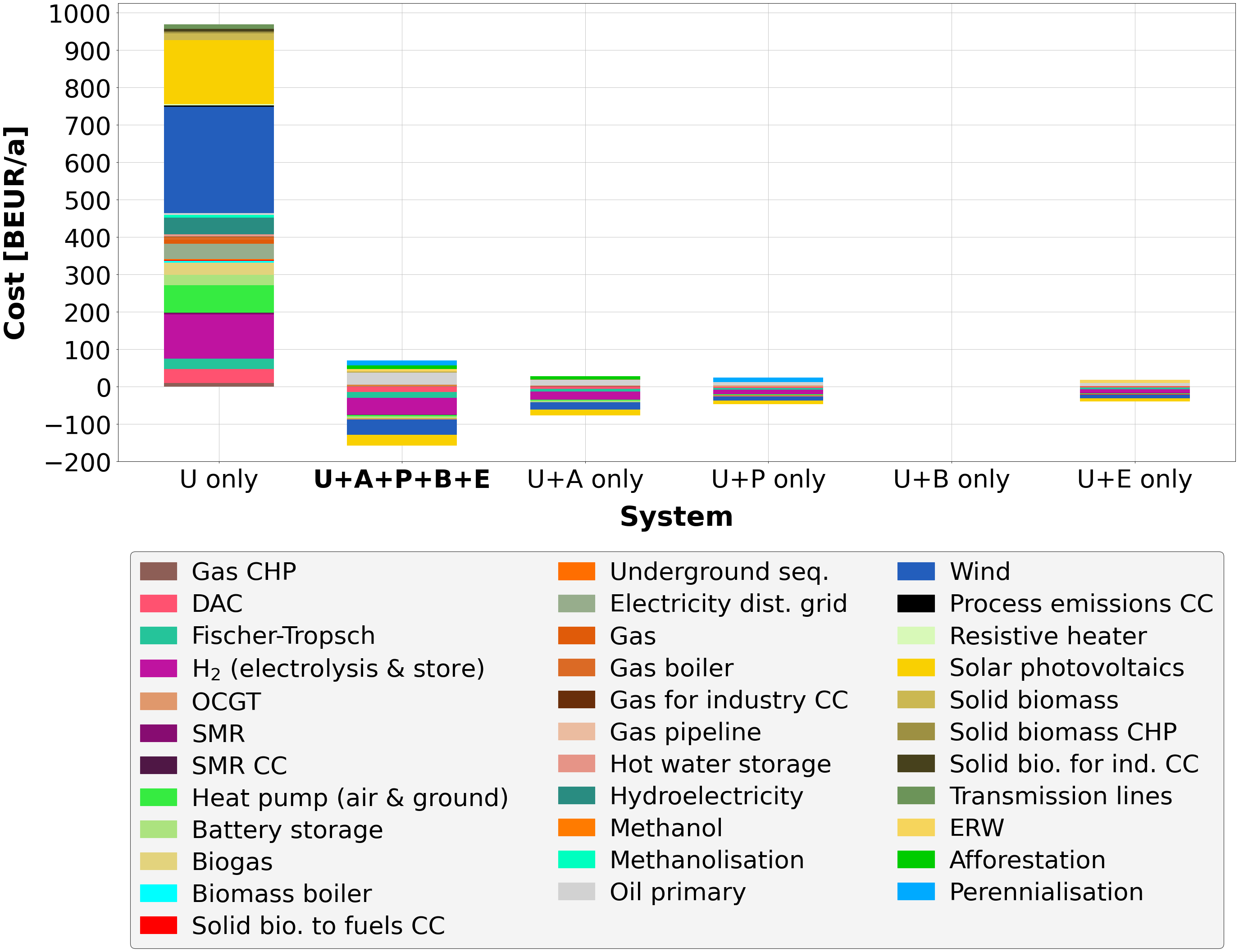}
    \caption[Total cost and technology configuration of a climate-neutral European energy system]{\textbf{Total cost and technology configuration of a climate-neutral European energy system}. Compared with an energy system relying exclusively on underground sequestration (U only), total system costs are reduced by 9\% when the additional CDR strategies are included (U+A+P+B+E). The availability of CDRs reduces the production of electrolytic H$_2$ by lowering the demand for synthetic oil through Fischer-Tropsch, which in turn decreases the required deployment of wind and solar generation capacities. Technologies with annual costs below 0.1 BEUR are omitted for readability.}
    \label{figure_cdr_sensitivity_analysis}
\end{figure}

\begin{figure}[H]
    \centering
    \includegraphics[width = 0.84\textwidth]{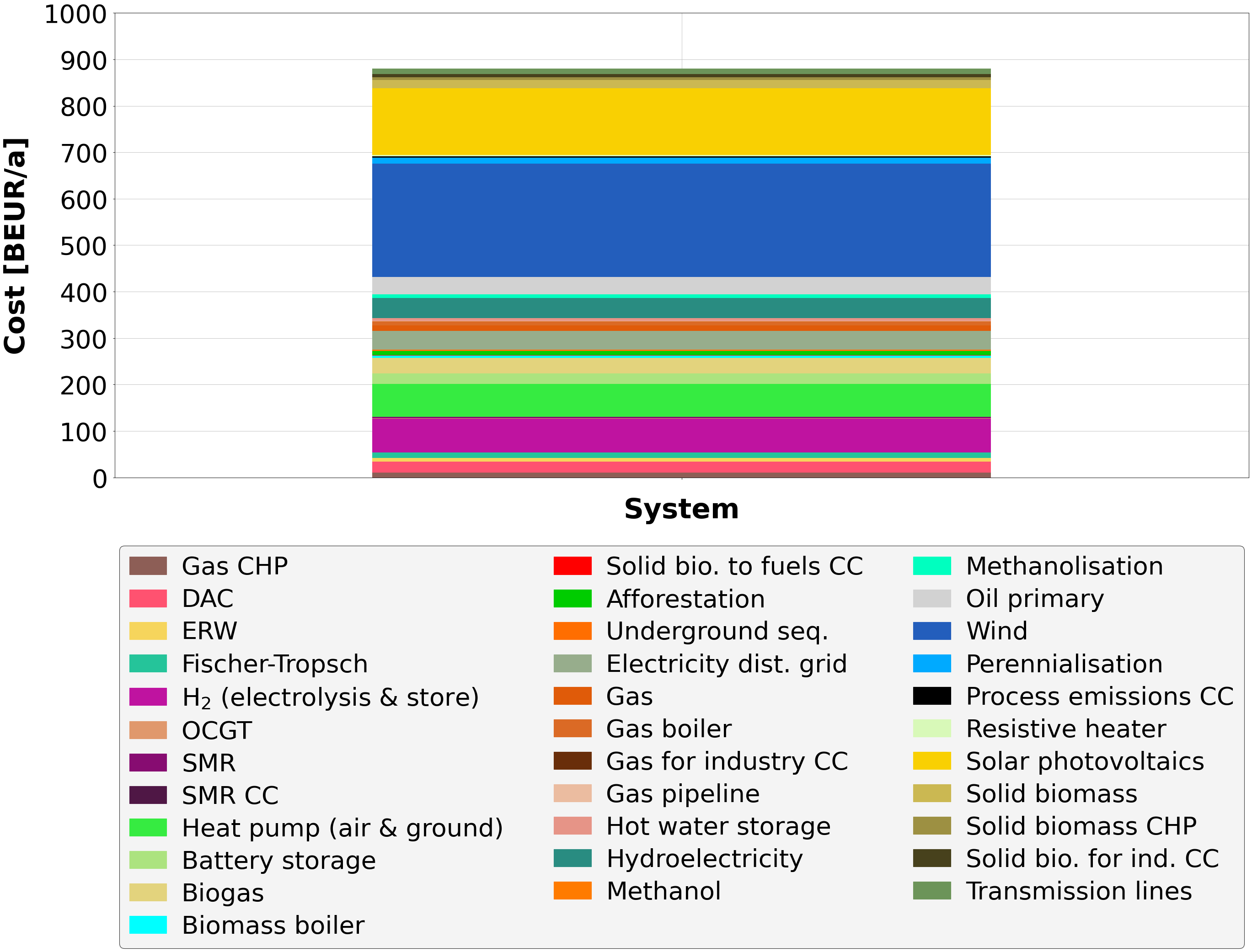}
    \caption[Total cost and technology configuration of a climate-neutral European energy system]{\textbf{Total cost and technology configuration of a climate-neutral European energy system}. An energy system equipped with all CDR strategies costs 880 BEUR per year. Wind and solar photovoltaics, electrolytic H$_2$ production and storage, heating electrification, and hydroelectricity together account for 65\% of this cost. Technologies with annual costs below 0.1 BEUR are omitted for readability.}
    \label{supplemental:figure_total_system_cost_and_technology_configuration}
\end{figure}

The climate-neutral European energy system is estimated to cost 880 BEUR per year when equipped with underground sequestration and the additional CDR strategies (Figure~\ref{supplemental:figure_total_system_cost_and_technology_configuration}). At the continental level, the selected additional CDRs collectively reduce total system costs by 9\% compared to a system without them, with individual contributions of approximately 5\%, 2\%, and 2\% for afforestation, perennialisation, and ERW, respectively (Figure~\ref{figure_cdr_sensitivity_analysis}). At the national level, CDRs reduce system costs in most countries, although a few countries experience cost increases (Table~\ref{supplemental:table_total_system_cost_per_country}).

Wind and solar photovoltaics, electrolytic hydrogen (H$_2$) production and storage, heating electrification, and hydroelectricity together account for 65\% of the total cost of a system equipped with all CDR strategies (Table~\ref{supplemental:table_cost_per_technology}). The three selected additional CDR strategies provide an aggregated removal potential of 169 MtCO$_2$/a, complementing the assumed underground sequestration potential of 200 MtCO$_2$/a. Given that unabatable industrial process emissions amount to 134 MtCO$_2$/a, the remaining net-negative emissions are utilised to offset emissions from fossil oil use in aviation and industry, thereby reducing reliance on costly Fischer–Tropsch synthetic fuels, as well as emissions from methanol use in shipping (Figure~\ref{supplemental:figure_co2_emissions_vs_co2_capture}). Furthermore, even when the underground sequestration potential is increased to 600 MtCO$_2$/a, the selected additional CDR strategies remain almost fully utilised (Figure~\ref{supplemental:figure_underground_seq_potential_co2_capture_vs_co2_sequestration_conversion}).

Although CDR strategies are selected for achieving climate neutrality in Europe and for reducing total system costs, their deployment becomes necessary only under stringent CO$_2$ emissions constraints. Amongst the modelled options, underground sequestration is the first to be deployed, being selected at a 30\% CO$_2$ emissions target relative to 1990 levels. This is closely followed by afforestation at the same target, while ERW and perennialisation are only necessary at a 20\% CO$_2$ emissions target (Figure~\ref{supplemental:figure_cdr_usage_vs_co2_emissions_limit}).

Moreover, the CO$_2$ shadow price decreases from 723 EUR/tCO$_2$ in an energy system relying solely on underground sequestration to 649 EUR/tCO$_2$ when complemented by the full portfolio of additional CDR strategies.

\subsection{Biomass is preferentially used in higher-value processes instead of biochar production}
Although the cost per tonne of CO$_2$ removed via biochar is lower than that of afforestation, perennialisation, and ERW (Table~\ref{table_cdrs_parameters}), the model does not select this option. This is because the available solid biomass, on which biochar production depends, is instead allocated to other uses, primarily for the provision of medium-temperature industrial heat, but also for combustion in combined heat and power (CHP) units and boilers, as well as for conversion to biofuels. Increasing the potential availability of solid biomass does not result in biochar being selected, as the additional solid biomass continues to be preferentially allocated to these processes. Specifically, it is first used to further increase biofuels production, while the remaining solid biomass is subsequently combusted in boilers and CHP units to generate heat and power (Figures~\ref{supplemental:figure_jrc_enspreso_solid_biomass_availability_scenario_co2_capture_vs_co2_sequestration_conversion} and \ref{supplemental:figure_solid_biomass_jrc_enspreso_potential_and_usage}).

In addition, neither transporting solid biomass to regions suitable for biochar pyrolysis, nor relocating biochar to regions with available land for application, nor integrating the heat generated during pyrolysis into district heating systems increases the contribution of biochar to CO$_2$ removal (Figures~\ref{supplemental:figure_solid_biomass_transport_co2_capture_vs_co2_sequestration_conversion}, \ref{supplemental:figure_biochar_transport_co2_capture_vs_co2_sequestration_conversion}, and \ref{supplemental:figure_biochar_pyrolysis_heat_output_co2_capture_vs_co2_sequestration_conversion}).

\begin{table}[!htb]
    \small
    \centering
    \caption[Main parameters used to model the CDR strategies]{\textbf{Main parameters used to model the CDR strategies}. The parameters are derived from the Energy System Technology Data, which are primarily based on data published by the Danish Energy Agency.}
    \label{table_cdrs_parameters}
    \begin{tabular}{*6c}
        \toprule
        \textbf{Parameter}                      & \textbf{Underground seq.} & \textbf{Afforestation} & \textbf{Perennialisation} & \textbf{Biochar} & \textbf{ERW} \\
        \midrule
        Electricity input [MWh$_{el}$/tCO$_2$]  & -                         & -                      & 0.32                      & 0.32             & 0.19         \\
        Biomass input [MWh$_{biomass}$/tCO$_2$] & -                         & -                      & -                         & 7.67             & -            \\
        Basalt input [t$_{basalt}$/tCO$_2$]     & -                         & -                      & -                         & -                & 3.33         \\
        Biogas output [MWh$_{biogas}$/tCO$_2$]  & -                         & -                      & 0.86                      & -                & -            \\
        Heat output [MWh$_{heat}$/tCO$_2$]      & -                         & -                      & -                         & 3.79             & -            \\
        Capital cost [MEUR/(tCO$_2$/h)]         & -                         & -                      & 6.05                      & 8.94             & -            \\
        Fixed O\&M cost [\% capital cost]       & -                         & -                      & -                         & 3.42             & -            \\
        Variable O\&M cost [EUR/tCO$_2$]        & -                         & -                      & 190                       & 48               & 190          \\
        Average seq. cost [EUR/tCO$_2$]         & 10                        & 113                    & -                         & -                & -            \\
        \bottomrule
    \end{tabular}
\end{table}

\subsection{DAC is not needed in a system equipped with CDRs and a CO$_2$ transport network}
Equipping the European energy system with a CO$_2$ transport network substantially reduces reliance on costly DAC by enabling greater CO$_2$ capture from solid biomass used in industry and from process emissions. This is because such a network allows captured CO$_2$ from these point sources to be transported to neighbouring regions for underground sequestration or synthetic fuels production, which would otherwise not be possible, as regions where this CO$_2$ is captured may lack sufficient sequestration potential or demand for synthetic fuels to convert (utilise) it. Furthermore, when the energy system is also equipped with the additional CDR strategies alongside underground sequestration, DAC becomes entirely redundant, as the remaining atmospheric CO$_2$ is removed by these strategies instead (Figure~\ref{supplemental:figure_co2_transport_and_cdrs_co2_capture_vs_co2_sequestration_conversion}).

Similarly, although to a lesser extent, equipping the energy system with an H$_2$ transport network reduces the need for DAC by enabling increased CO$_2$ capture also from solid biomass used in industry and from process emission sources in regions where electrolytic H$_2$ production is cost-effective. The produced H$_2$ can then be transported to regions where synthetic fuels production can be carried out more economically. When the energy system is also equipped with the additional CDR strategies, DAC is further reduced, as part of the remaining atmospheric CO$_2$ is removed by these strategies instead (Figure~\ref{supplemental:figure_h2_transport_and_cdrs_co2_capture_vs_co2_sequestration_conversion}).

\subsection{Selected additional CDR strategies are fully utilised across space}
\begin{figure}[!htb]
    \centering
    \includegraphics[width = 0.84\textwidth]{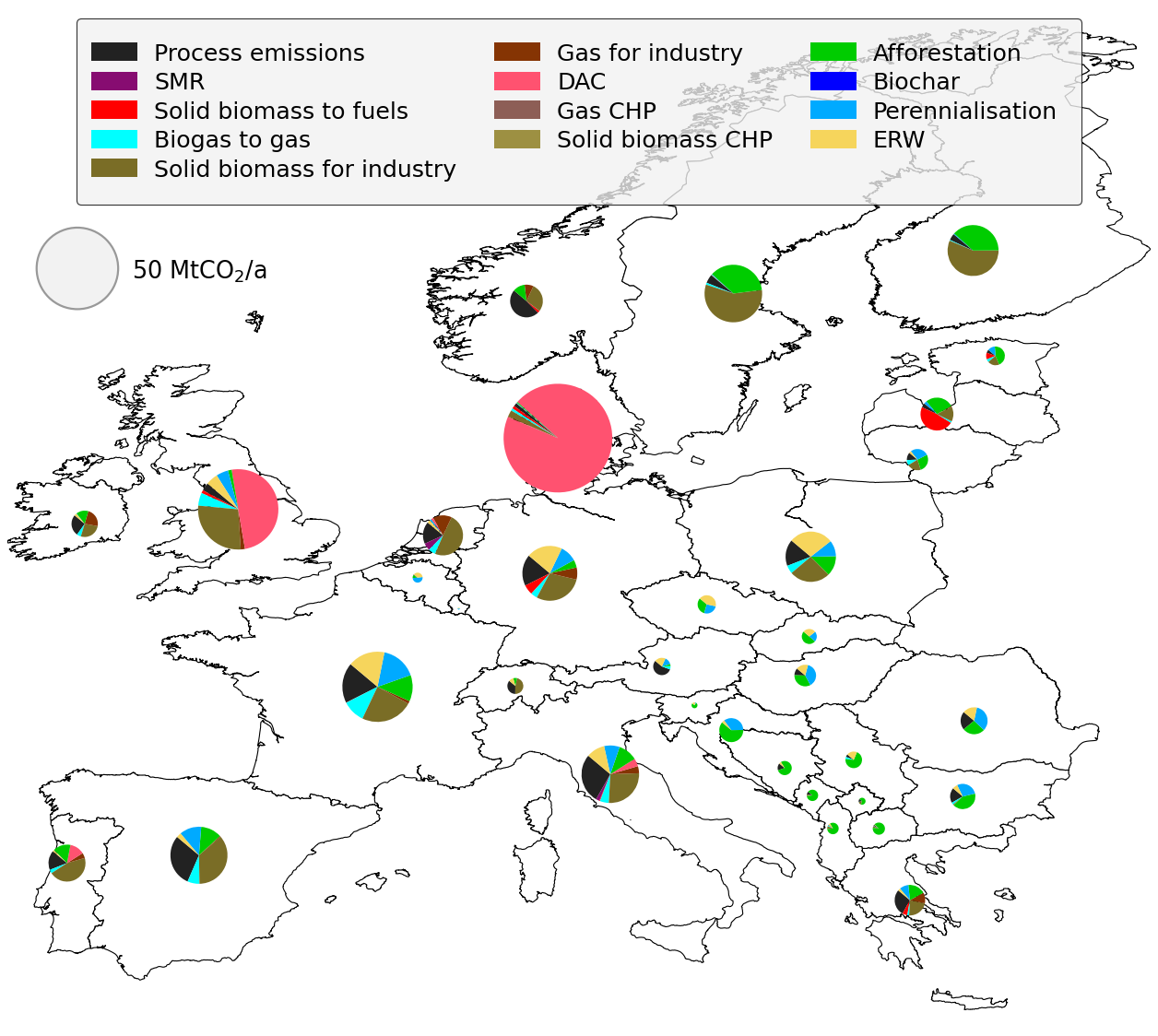}
    \caption[CO$_2$ capture across Europe]{\textbf{CO$_2$ capture across Europe}. Afforestation is widely deployed across the continent as a key strategy for capturing emitted CO$_2$, particularly in the Northeastern and Southeastern regions. Perennialisation and ERW are also widely deployed, particularly in Central Europe, France, Italy, and Great Britain. Point-source CO$_2$ capture and especially DAC are extensively deployed in countries with access to low-cost energy (enabled by favourable renewable capacity factors) and substantial underground sequestration potential, notably Great Britain, Denmark, Italy, and Portugal. Capture from process emissions and gas combustion do not entail CO$_2$ removal. Each pie chart aggregates all nodes belonging to the respective modelled country.}
    \label{figure_co2_capture_spatial_pattern}
\end{figure}

\begin{figure}[!htb]
    \centering
    \includegraphics[width = 0.84\textwidth]{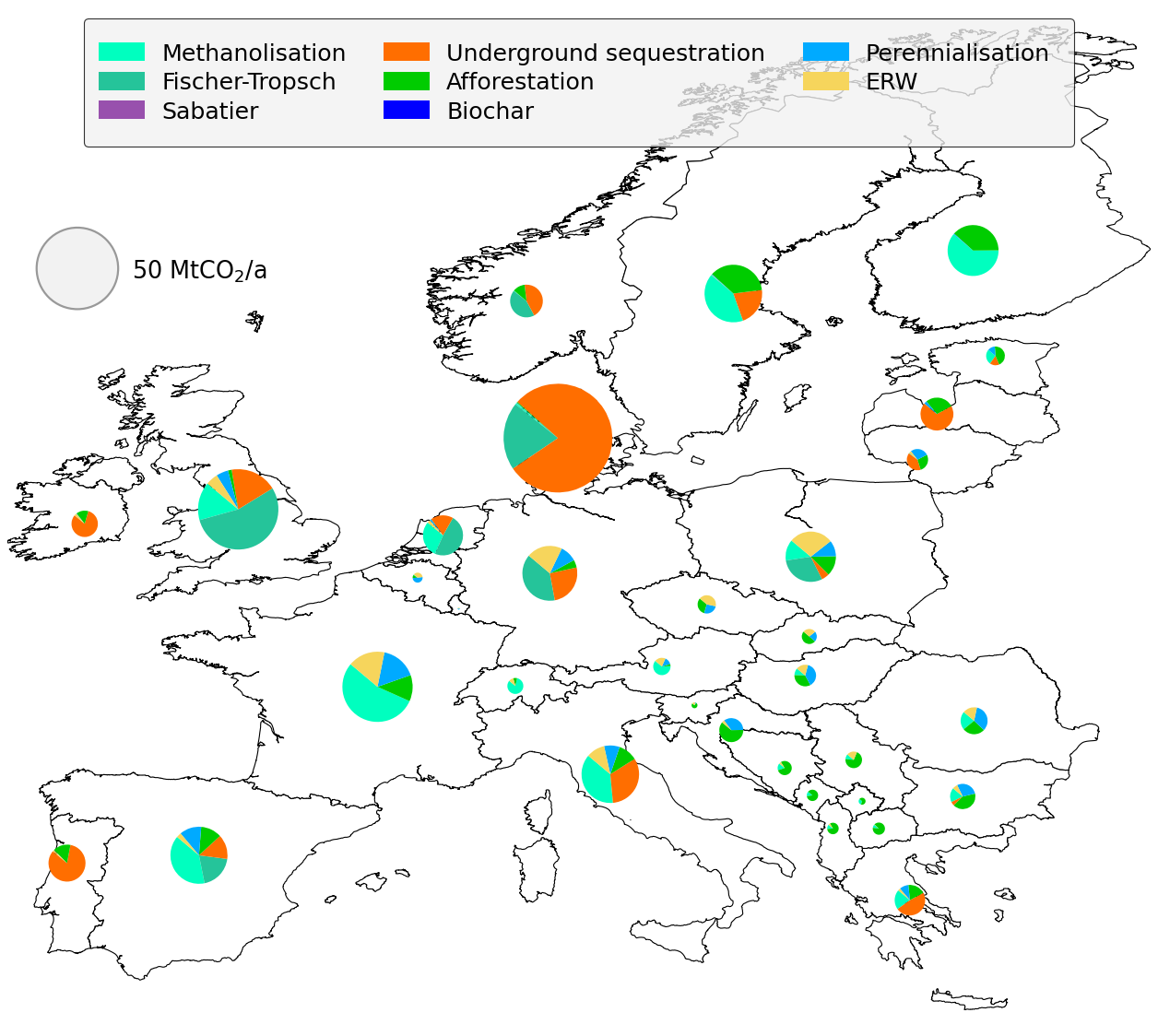}
    \caption[CO$_2$ conversion and sequestration across Europe]{\textbf{CO$_2$ conversion and sequestration across Europe}. As a direct consequence of the adopted CO$_2$ capture strategy (Figure~\ref{figure_co2_capture_spatial_pattern}), afforestation is widely deployed across the continent for sequestering captured CO$_2$, particularly in the Northeastern and Southeastern regions. Perennialisation and ERW are also widely deployed, particularly in Central Europe, France, Italy, and Great Britain. Underground sequestration is extensively deployed in countries that rely heavily on point-source capture and DAC to manage captured CO$_2$. Each pie chart aggregates all nodes belonging to the respective modelled country.}
    \label{figure_co2_sequestration_and_conv_spatial_pattern}
\end{figure}

\begin{figure}[H]
    \centering
    \includegraphics[width = 0.84\textwidth]{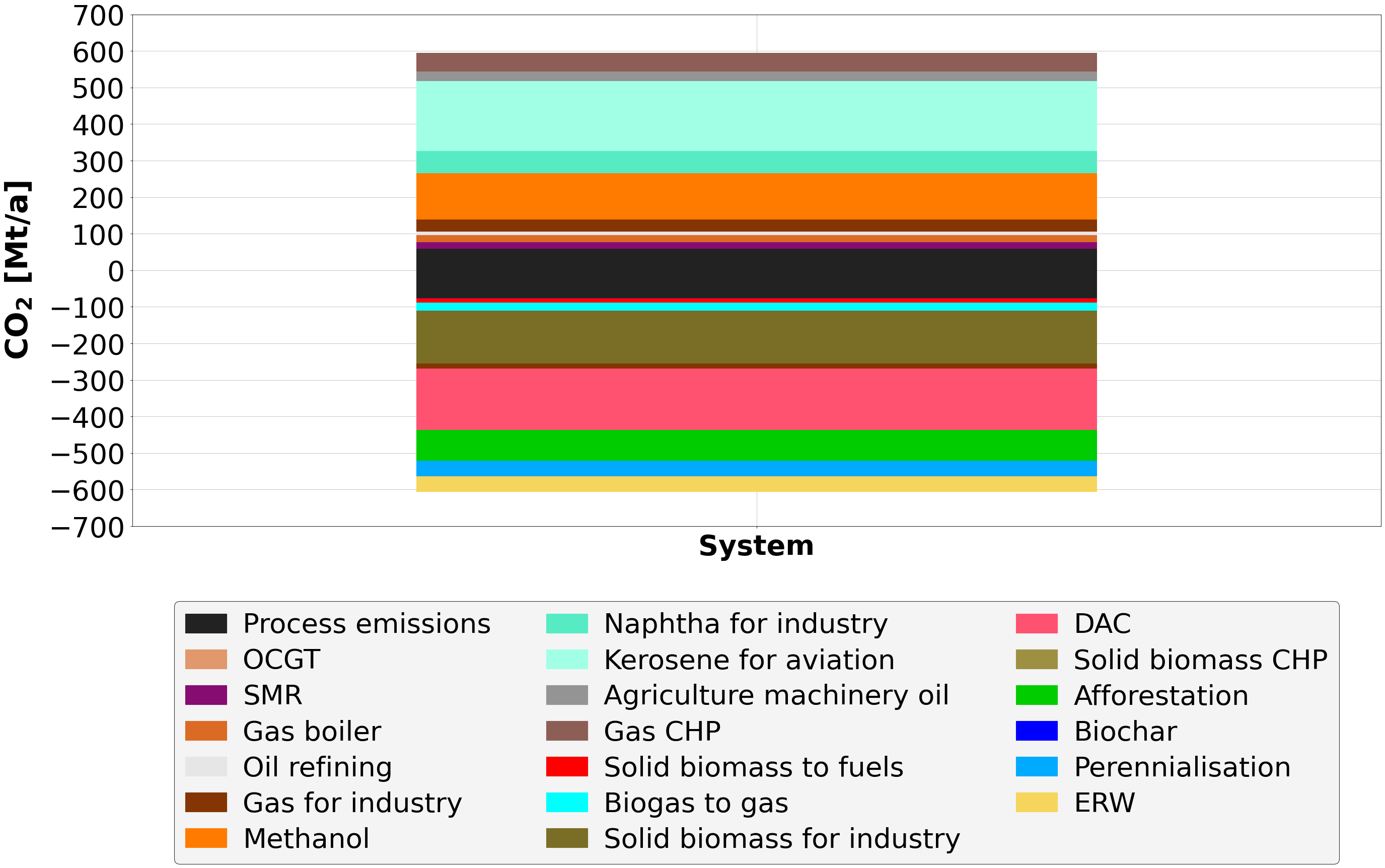}
    \caption[CO$_2$ emissions and capture in a climate-neutral European energy system]{\textbf{CO$_2$ emissions and capture in a climate-neutral European energy system}. A substantial amount of CO$_2$ emissions come from oil used in the industry and aviation sectors, as well as from methanol used in the shipping sector. These CO$_2$ emissions are captured mainly through DAC, point-source capture from solid biomass used in industry, and key additional CDR strategies, namely afforestation, perennialisation, and ERW.}
    \label{supplemental:figure_co2_emissions_vs_co2_capture}
\end{figure}

From a spatial perspective, afforestation, perennialisation, and ERW are utilised to their full potential in every modelled region (Figure~\ref{figure_cdr_potential_usage}). Afforestation is widely deployed across Europe as a key CO$_2$ removal strategy, particularly in Northeastern and Southeastern countries, removing approximately 14\% of the total 606 MtCO$_2$ emitted annually (Figures~\ref{figure_co2_capture_spatial_pattern}, \ref{figure_co2_sequestration_and_conv_spatial_pattern}, \ref{supplemental:figure_afforestation_co2_store_potential_and_usage}, and \ref{supplemental:figure_co2_emissions_vs_co2_capture}). Perennialisation and ERW are also widely deployed across the continent, particularly in Central Europe, France, Italy, and Great Britain, each removing around 7\% of the total annual CO$_2$ emissions. Regarding underground sequestration, despite many regions across the continent not fully utilising their potential, the Europe-wide annual limit of 200 MtCO$_2$ is fully reached, corresponding to one third of the total annual CO$_2$ emissions. Tables~\ref{supplemental:table_underground_co2_sequestration_potential}, \ref{supplemental:table_afforestation_land_potential_biomass_density_and_sequestration_potential}, \ref{supplemental:table_perennials_land_and_sequestration_potentials}, \ref{supplemental:table_biochar_land_and_sequestration_potentials}, and \ref{supplemental:table_ERW_land_and_sequestration_potentials} summarise the CO$_2$ sequestration and removal potentials of each CDR strategy across the modelled European countries.

\begin{figure}[!htb]
    \centering
    \includegraphics[width = 0.84\textwidth]{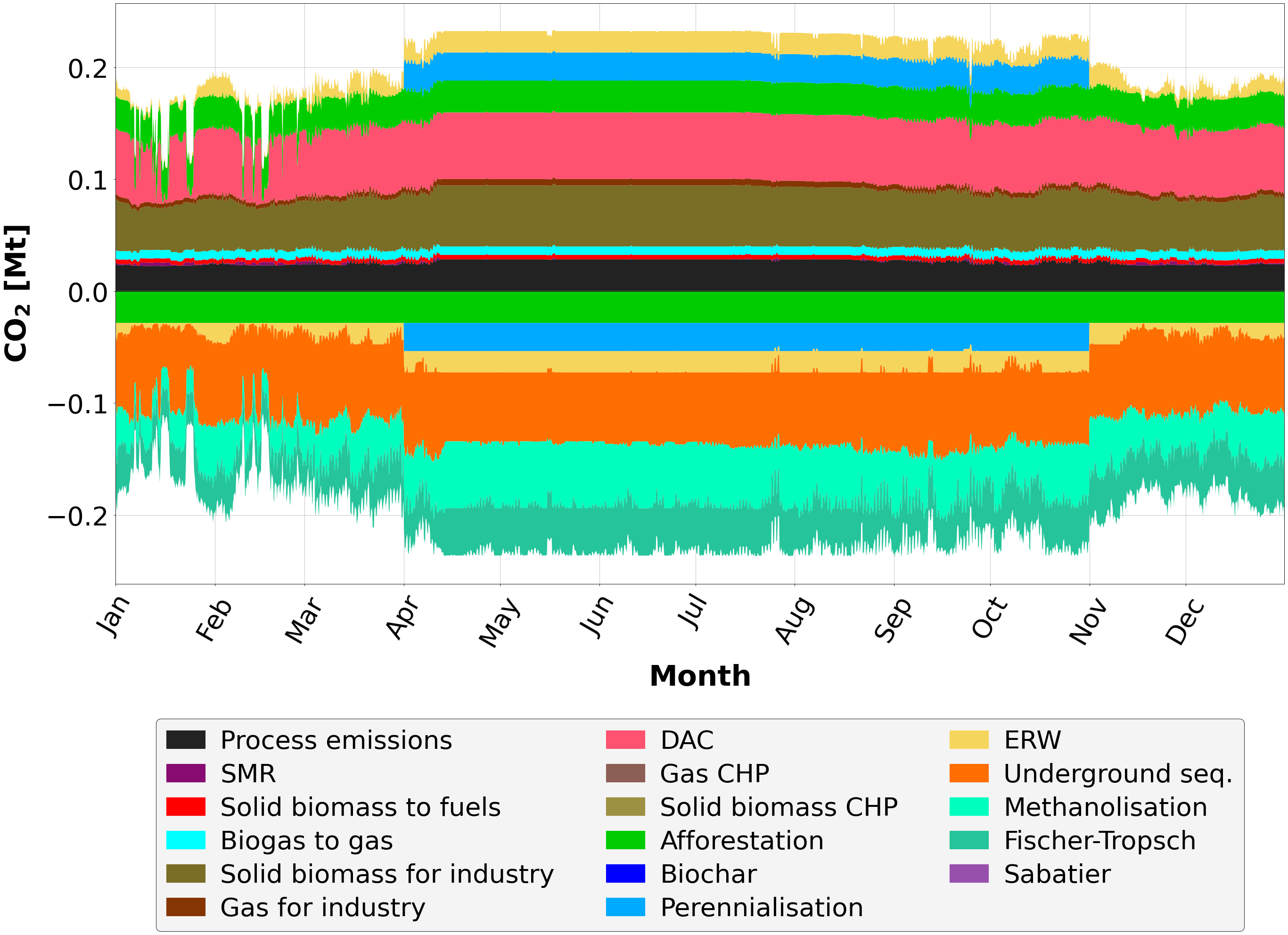}
    \caption[Temporal CO$_2$ capture, conversion, and sequestration across Europe]{\textbf{Temporal CO$_2$ capture, conversion, and sequestration across Europe}. CO$_2$ capture from point sources and DAC, as well as its conversion into synthetic fuels, exhibit a seasonal pattern, with higher activity in summer and lower activity in winter, reflecting the greater availability of low-cost electricity. Underground sequestration exhibits the opposite pattern, as less captured CO$_2$ is sequestered in summer when more is diverted to conversion processes. Due to model simplifications, afforestation exhibits no seasonal pattern and operates at a constant rate throughout the modelled year. Perennialisation also operates constantly but is restricted to the months from May to October, corresponding to the harvesting period of perennial crops. In contrast, ERW exhibits a seasonal pattern, with higher activity in summer driven by the availability of low-cost renewable electricity.}
    \label{figure_co2_capture_vs_co2_sequestration_conversion_temporal_pattern}
\end{figure}

\begin{figure}[H]
    \centering
    \subfloat[]{\label{supplemental:figure_underground_temporal_pattern}\includegraphics[width = 0.475\linewidth]{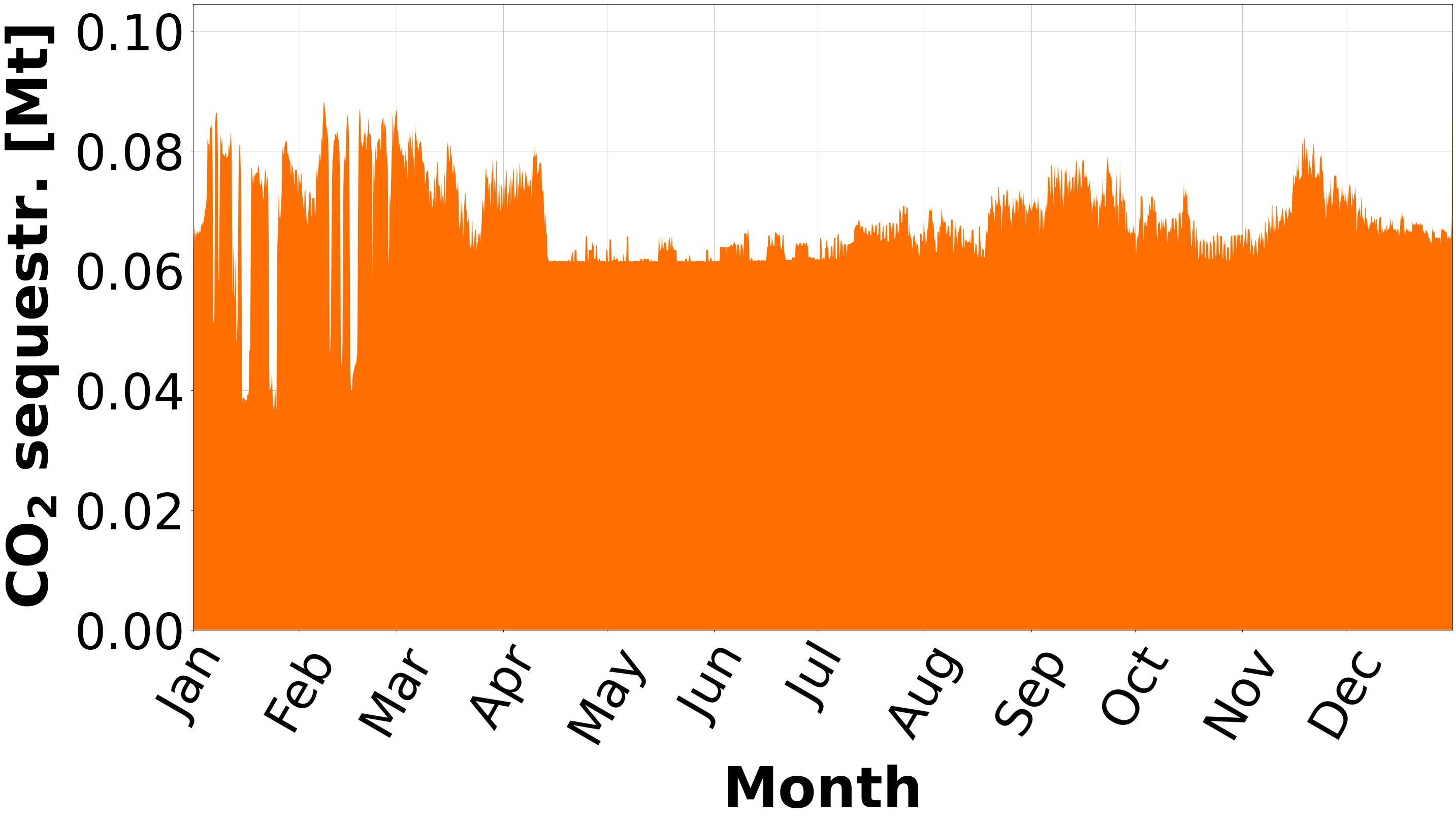}}\par
    \vspace{20pt}
    \subfloat[]{\label{supplemental:figure_afforestation_temporal_pattern}\includegraphics[width = 0.475\linewidth]{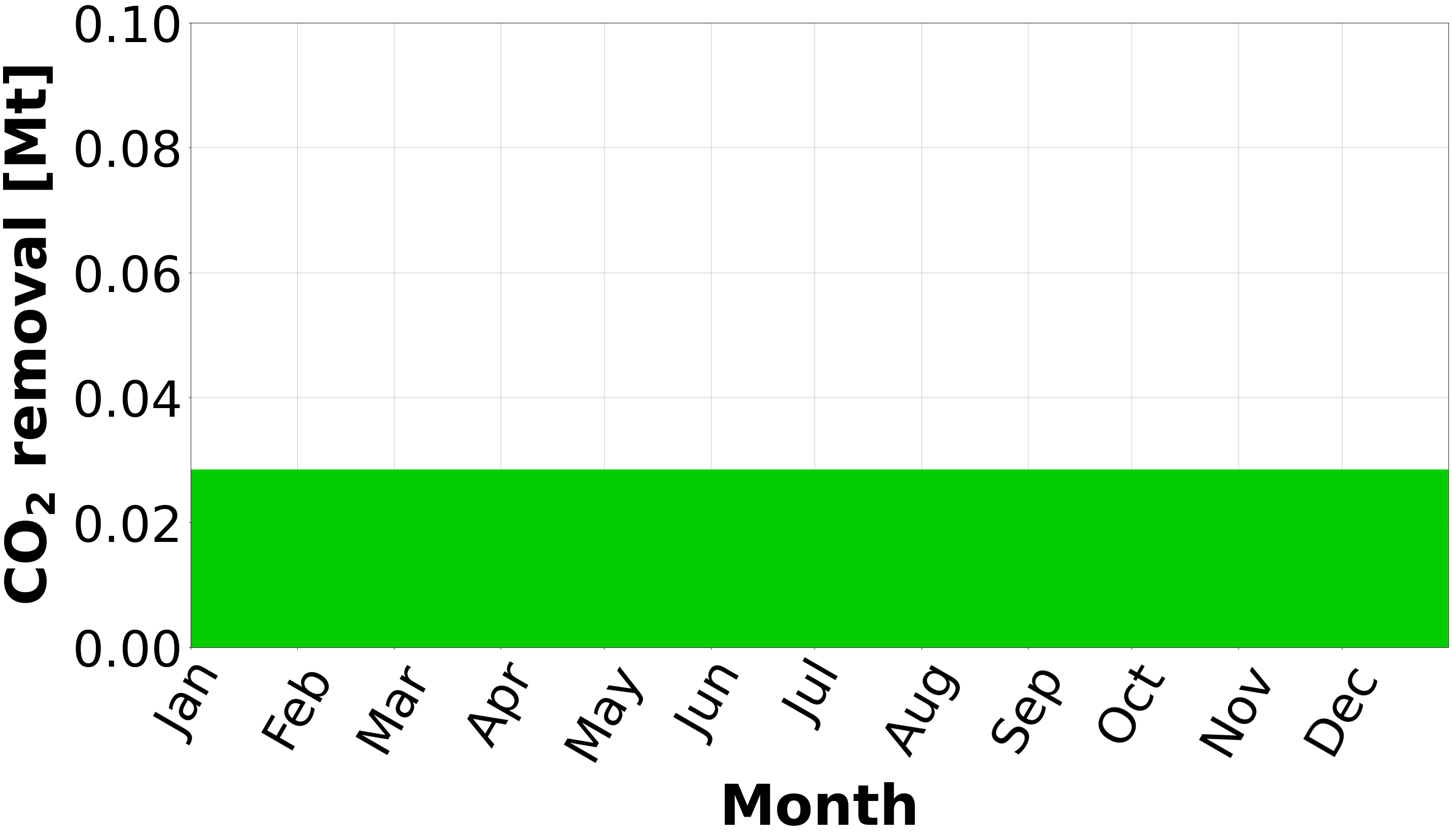}}\hfill
    \subfloat[]{\label{supplemental:figure_perennials_temporal_pattern}\includegraphics[width = 0.475\linewidth]{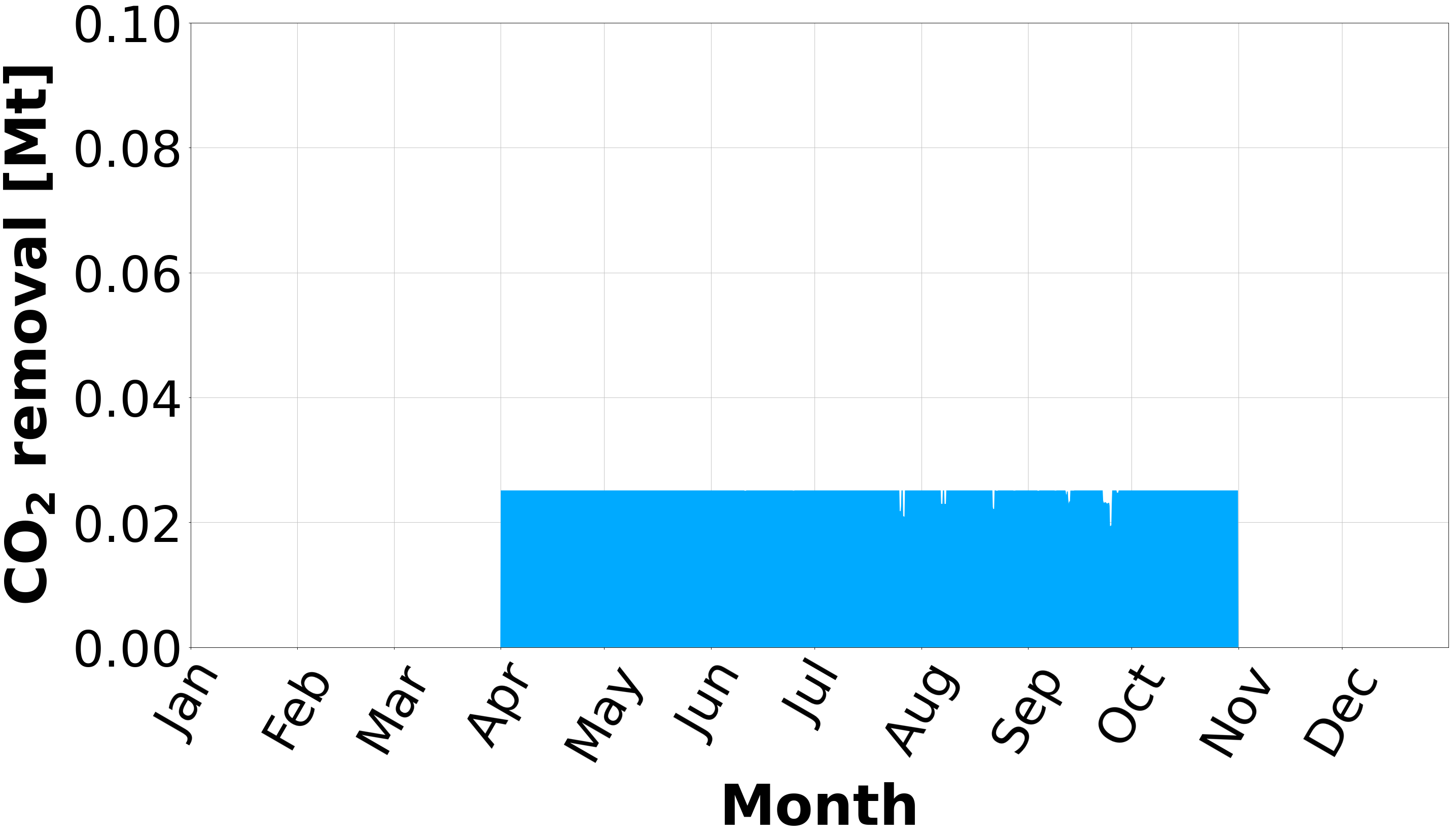}}\par
    \vspace{20pt}
    \subfloat[]{\label{supplemental:figure_biochar_temporal_pattern}\includegraphics[width = 0.475\linewidth]{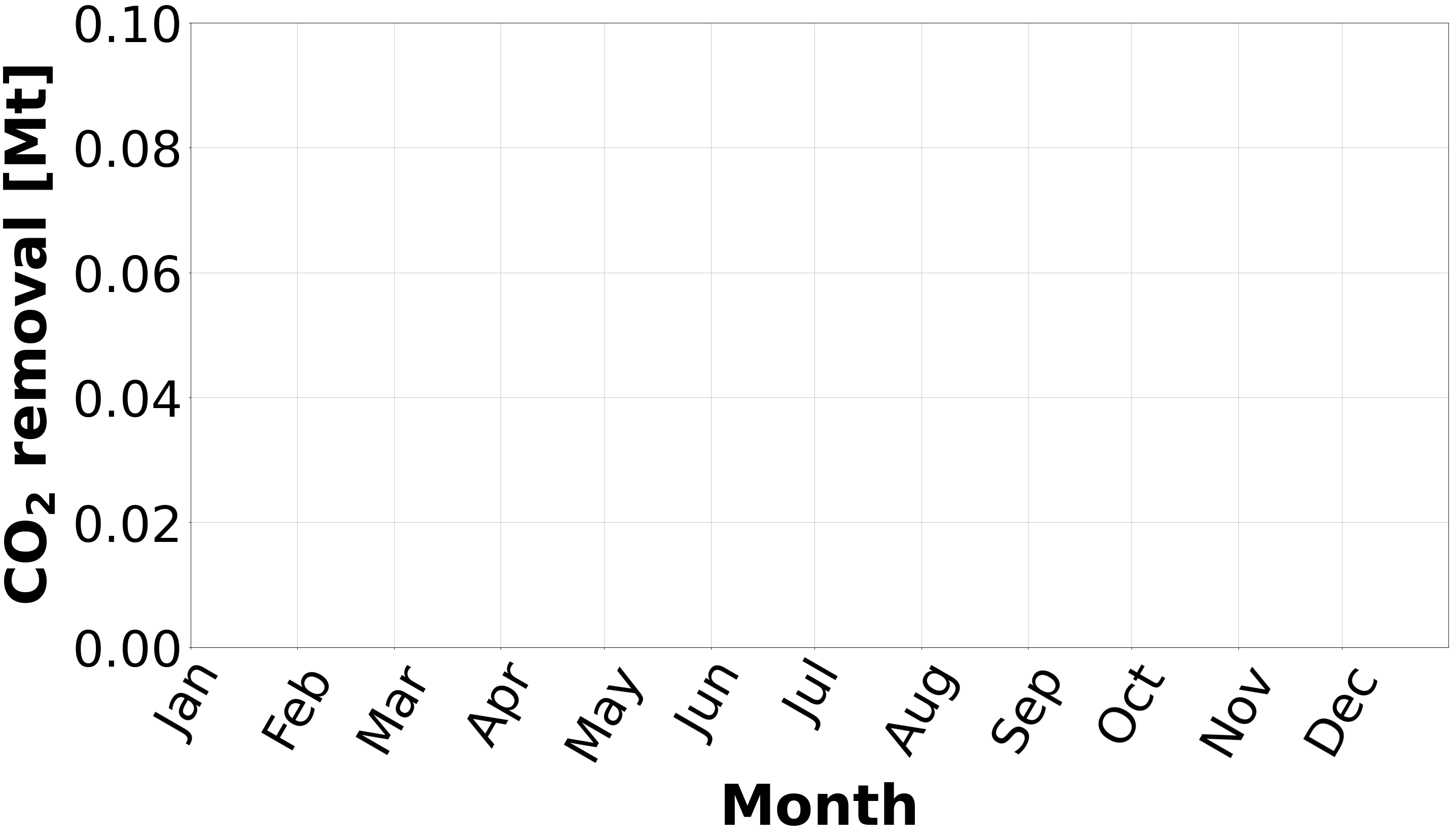}}\hfill
    \subfloat[]{\label{supplemental:figure_erw_temporal_pattern}\includegraphics[width = 0.475\textwidth]{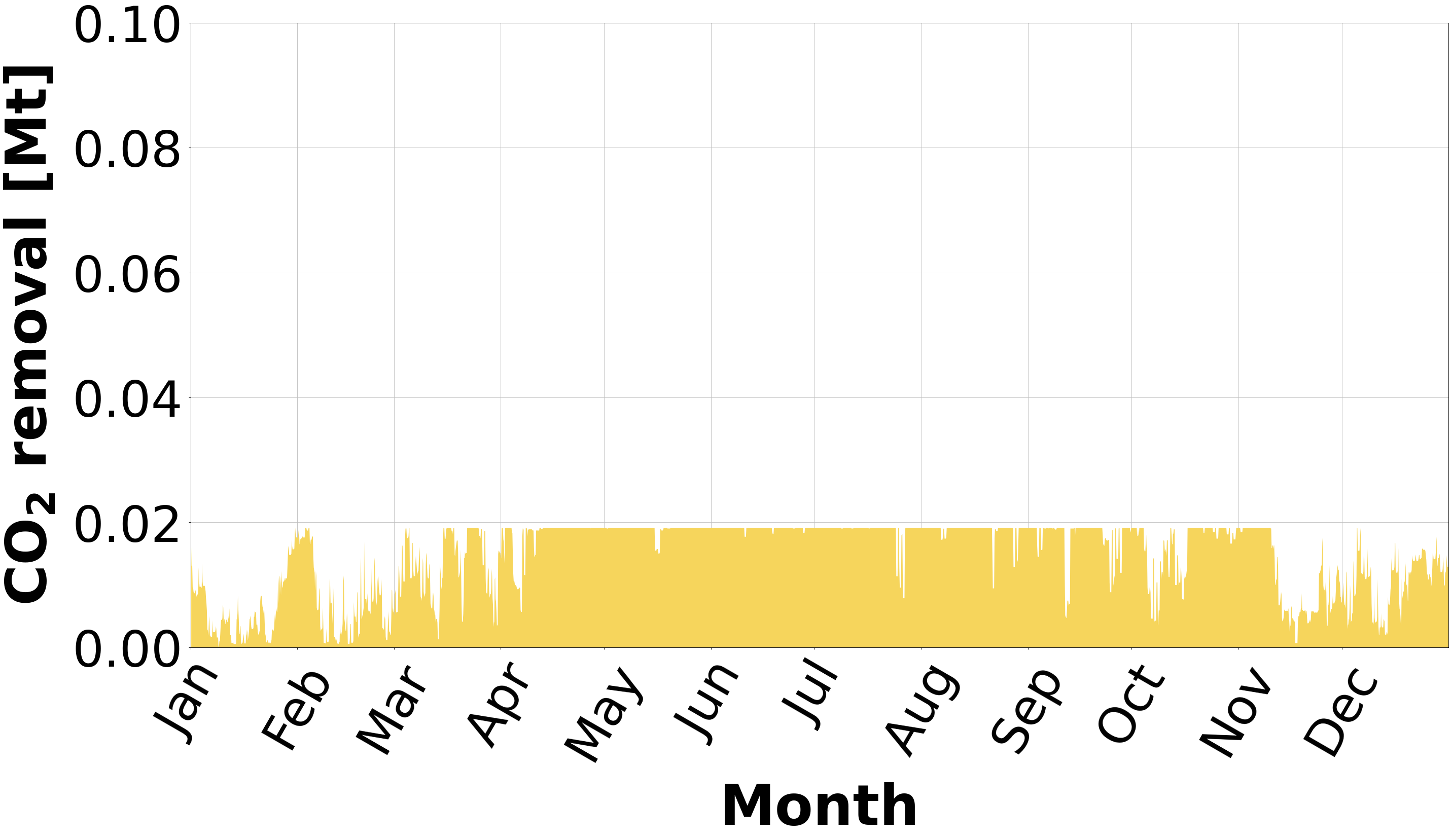}}\par
    \caption[Temporal CDR usages across Europe]{\textbf{Temporal CDR usages across Europe}. Usage patterns of (A) underground sequestration, (B) afforestation, (C) perennialisation, (D) biochar, and (E) ERW throughout the modelled year. Underground sequestration exhibits a seasonal pattern, with higher usage in winter and lower in summer, as part of the captured CO$_2$ is allocated to synthetic fuels production in the latter period (Figure~\ref{supplemental:figure_co2_conversion_temporal_pattern}). ERW shows the opposite pattern, with higher usage in summer as a result of the availability of low-cost electricity for basalt crushing. Due to model simplifications, afforestation removes CO$_2$ at a constant rate throughout the year; however, in reality, forest CO$_2$ uptake varies seasonally. Perennialisation is limited to specific months (May–October), corresponding to the harvesting period of perennial crops.}
    \label{supplemental:figure_cdr_temporal_pattern}
\end{figure}

\begin{figure}[H]
    \centering
    \subfloat[]{\label{supplemental:figure_fischer-tropsch_temporal_pattern}\includegraphics[width = 0.475\linewidth]{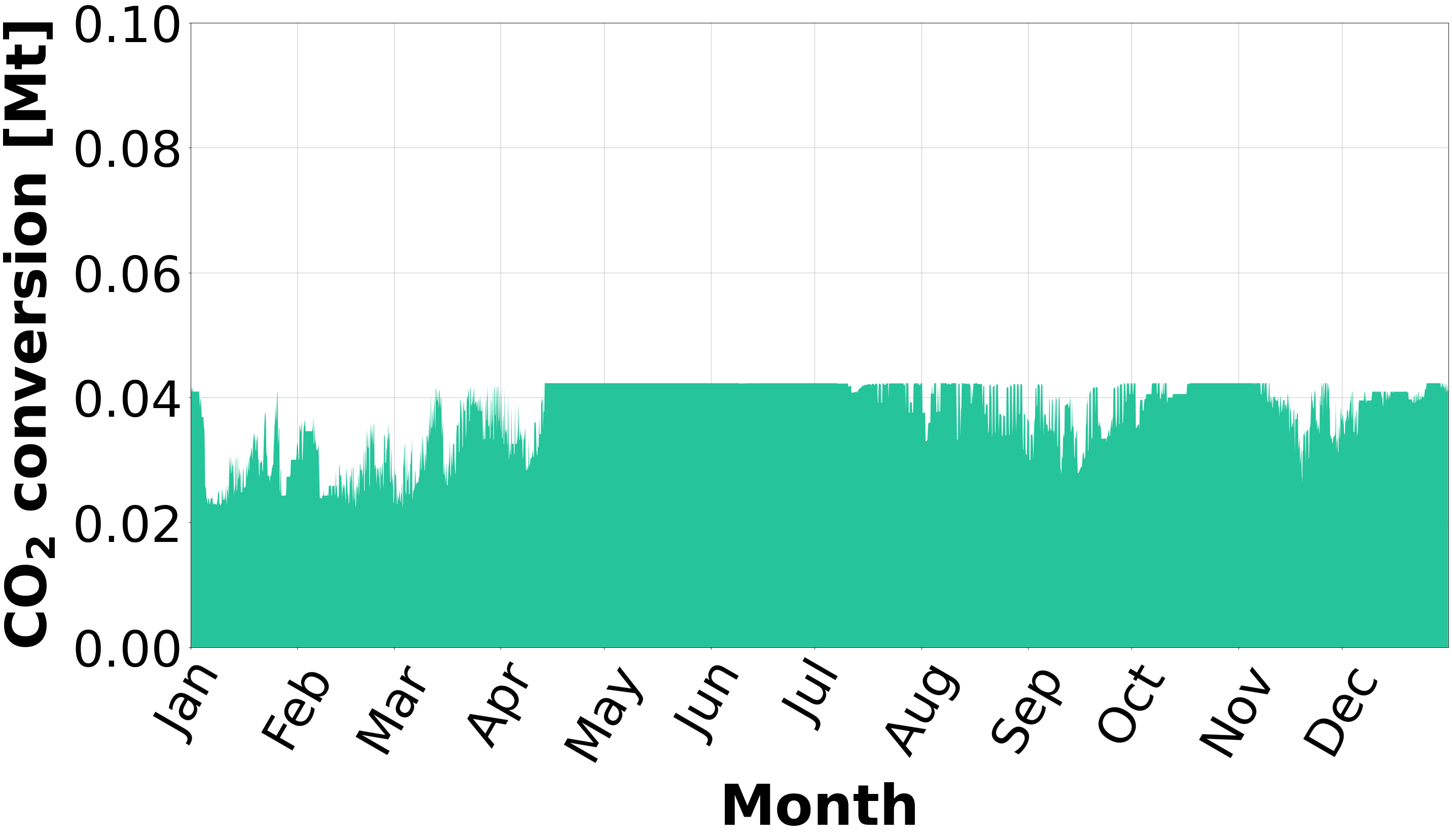}}\par
    \vspace{20pt}
    \subfloat[]{\label{supplemental:figure_methanolisation_temporal_pattern}\includegraphics[width = 0.475\linewidth]{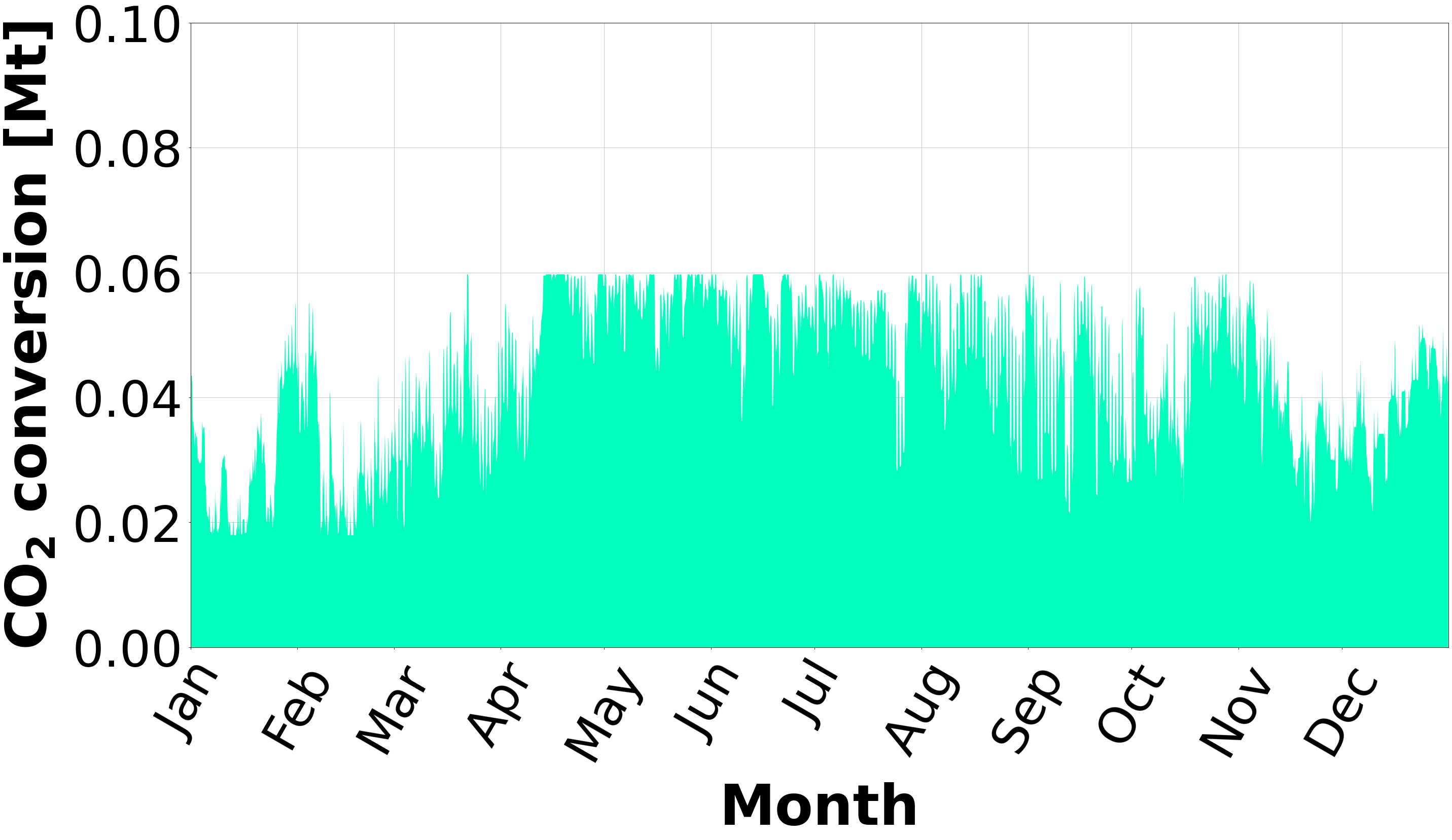}}\par
    \vspace{20pt}
    \subfloat[]{\label{supplemental:figure_sabatier_temporal_pattern}\includegraphics[width = 0.475\linewidth]{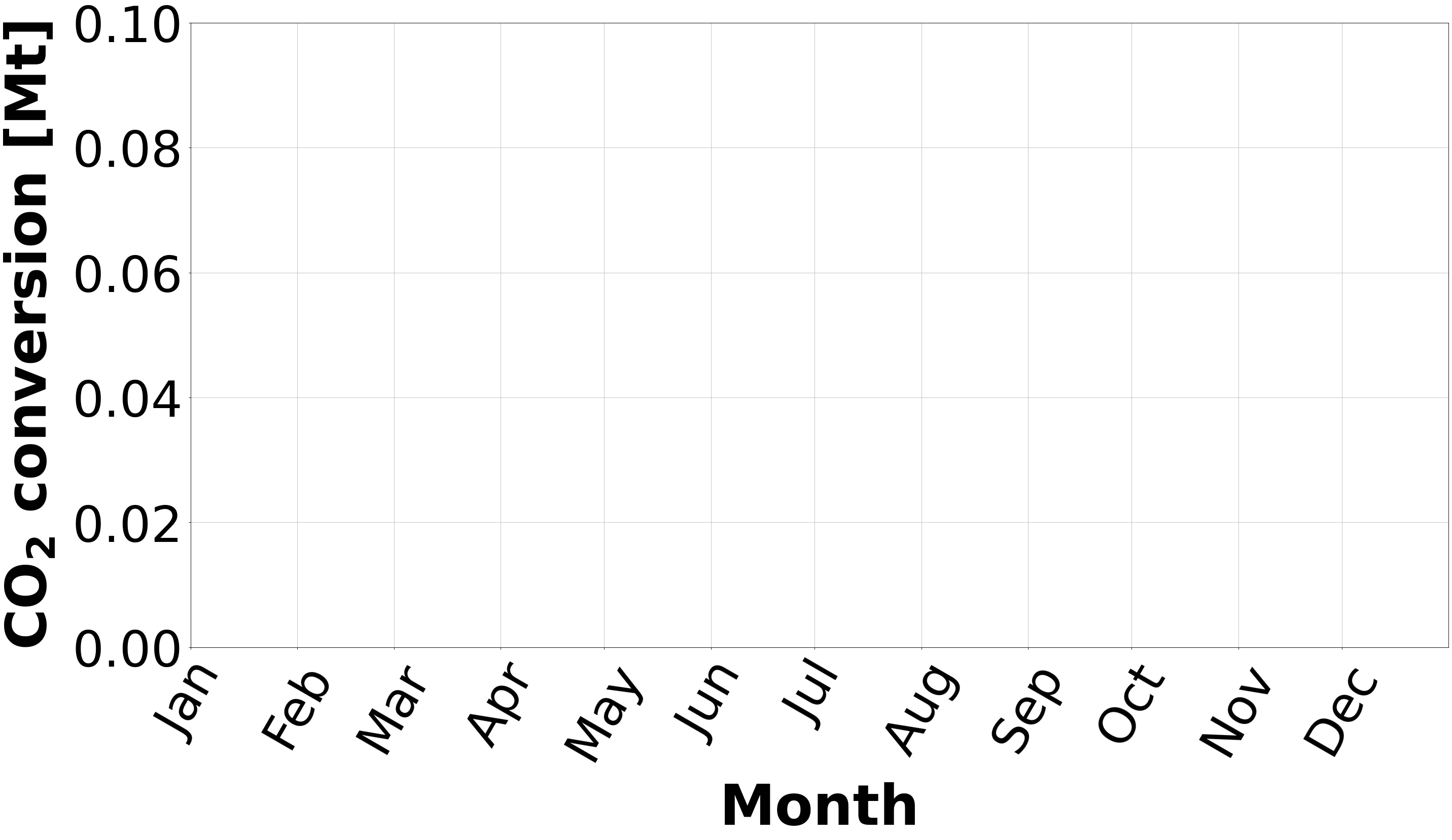}}\par
    \caption[Temporal CO$_2$ conversion across Europe]{\textbf{Temporal CO$_2$ conversion across Europe}. CO$_2$ conversion patterns of (A) Fischer-Tropsch, (B) methanolisation, and (C) Sabatier throughout the modelled year. With the exception of Sabatier, which is not cost-competitive and therefore not deployed, Fischer–Tropsch and methanolisation exhibit seasonal patterns, with higher CO$_2$ conversion in summer. This is driven by the increased availability of low-cost renewable electricity during this period, which not only powers these conversion processes but also enables higher levels of CO$_2$ capture and H$_2$ electrolysis—key inputs for synthetic fuels production.}
    \label{supplemental:figure_co2_conversion_temporal_pattern}
\end{figure}

From a temporal perspective, underground sequestration exhibits a seasonal pattern, with relatively stable levels in summer due to steady CO$_2$ capture from DAC and point sources using low-cost renewable electricity. In winter, however, greater variability occurs, as DAC and CO$_2$ capture from solid biomass used in industry are occasionally substantially reduced, leading to temporary declines in CO$_2$ sequestered underground. Afforestation exhibits no seasonal pattern because the model assumes constant CO$_2$ removal throughout the modelled year. Perennialisation is assumed to occur from May to October, corresponding to the harvesting period of perennial crops, during which biogas is also produced. ERW exhibits higher activity in summer when low-cost renewable electricity is available to power basalt crushing, although CO$_2$ removal is higher during rainy seasons (Figures~\ref{figure_co2_capture_vs_co2_sequestration_conversion_temporal_pattern} and \ref{supplemental:figure_cdr_temporal_pattern} and Section~\ref{supplemental:carbon_dioxide_removal_cdr}). The production of synthetic oil via Fischer–Tropsch and synthetic methanol via methanolisation likewise exhibits seasonality, occurring predominantly in summer (Figure~\ref{supplemental:figure_co2_conversion_temporal_pattern}). This reflects the abundance of low-cost renewable electricity during this period, which enables both CO$_2$ capture via DAC and H$_2$ production via electrolysis—key inputs for synthetic fuels production—as well as the associated conversion processes.

\subsection{High-resolution modelling better captures system dynamics}
In this section, we briefly assess the impact of the chosen spatial and temporal resolutions through a sensitivity analysis of both assumptions.

\begin{figure}[H]
    \centering
    \includegraphics[width = 0.84\textwidth]{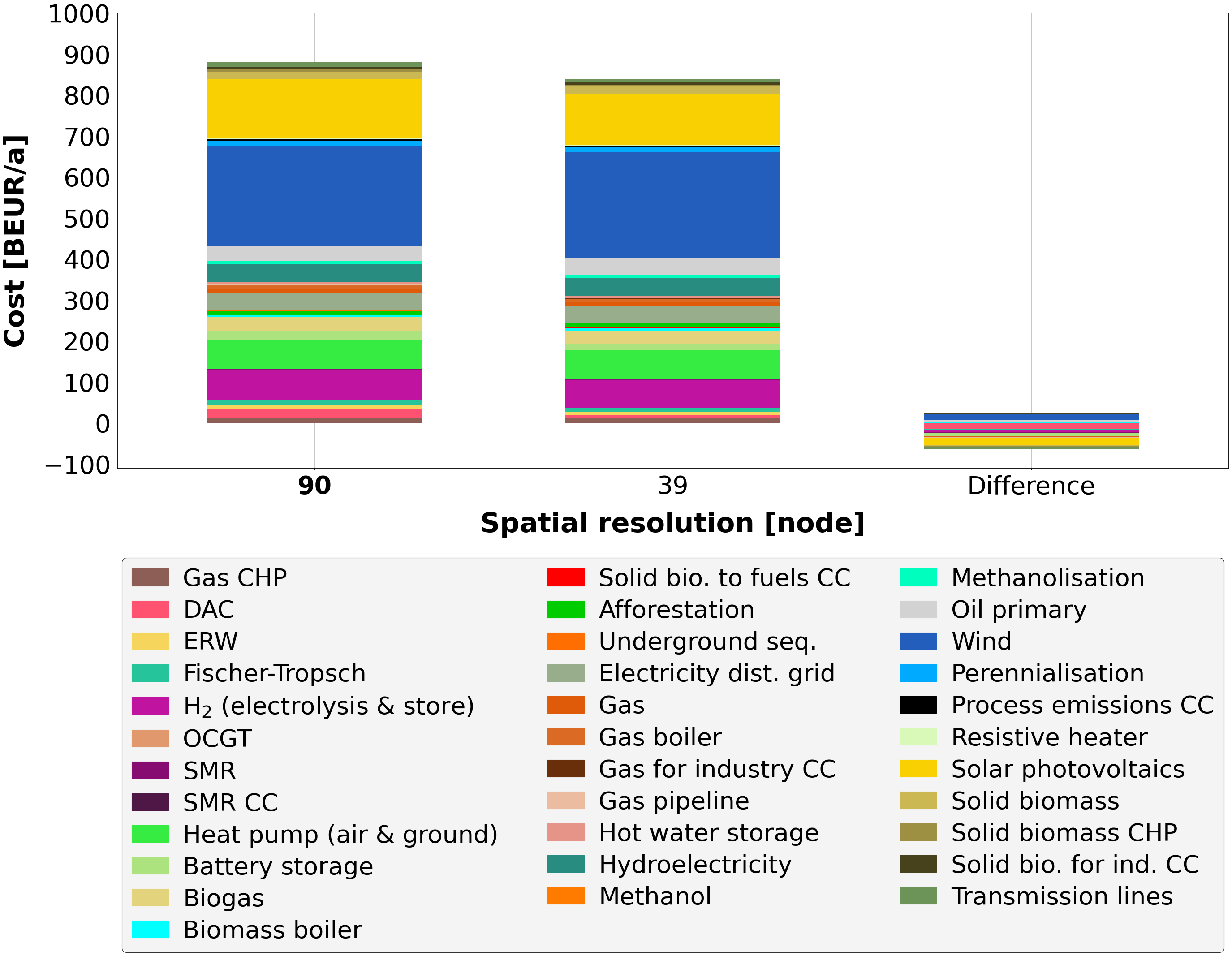}
    \caption[Cost spatial resolution sensitivity analysis]{\textbf{Cost spatial resolution sensitivity analysis}. Coarsening the model's spatial resolution from 90 to 39 nodes reduces costs by 5\%, mainly due to lower investments in solar and DAC. Spatial aggregation primarily masks regional heterogeneity while preserving temporal balancing requirements, thereby limiting its impact on total system costs. Technologies with annual costs below 0.1 BEUR are omitted for readability.}
    \label{supplemental:figure_spatial_resolution_sensitivity_analysis_cost}
\end{figure}

\begin{figure}[H]
    \centering
    \includegraphics[width = 0.84\textwidth]{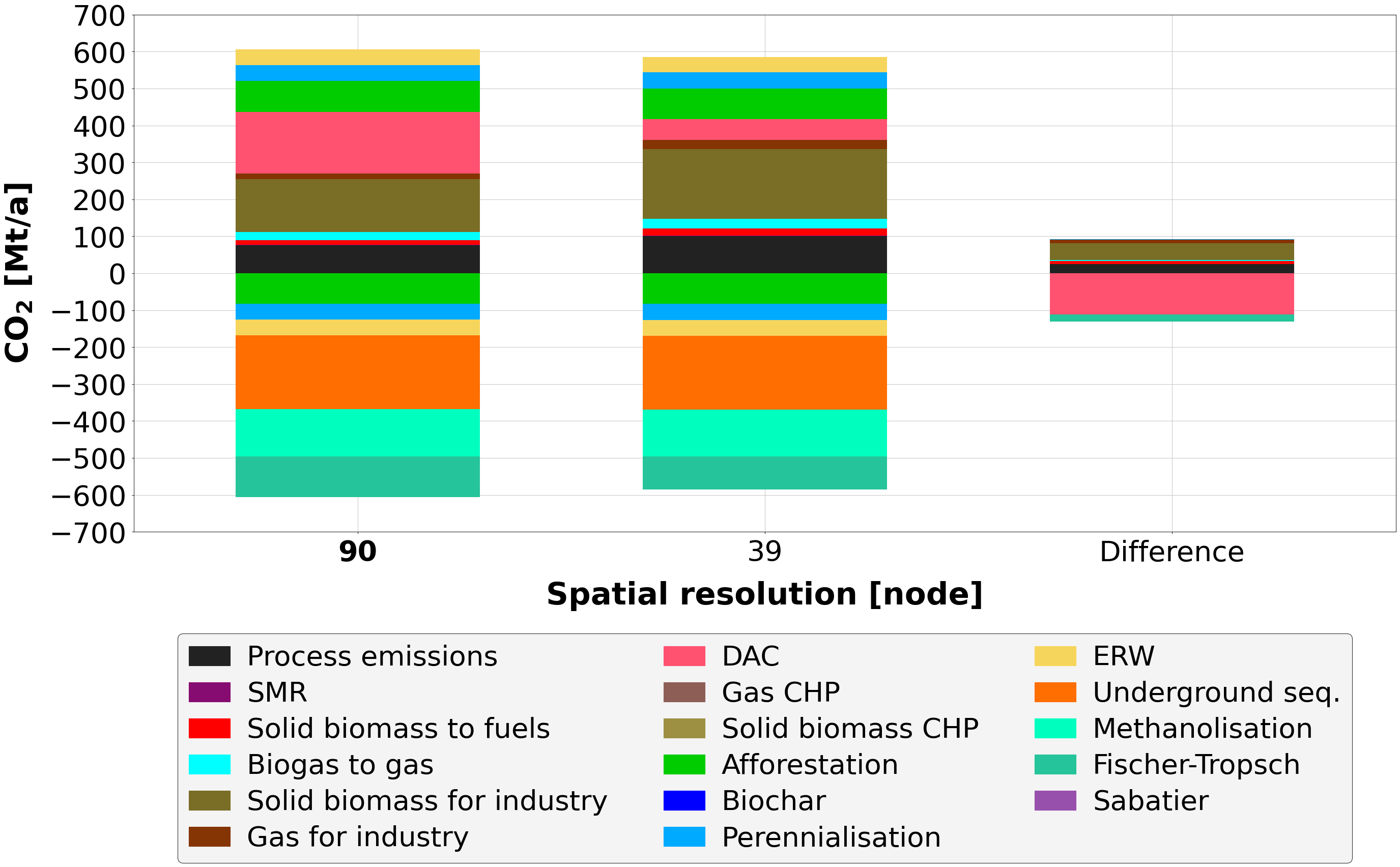}
    \caption[CO$_2$ capture, conversion, and sequestration spatial resolution sensitivity analysis]{\textbf{CO$_2$ capture, conversion, and sequestration spatial resolution sensitivity analysis}. Despite a substantial reduction in CO$_2$ capture via DAC under the coarser spatial resolution, CO$_2$ balances and the contributions of the additional CDR strategies remain largely unchanged across the two spatial resolutions. This reduction is compensated by increased CO$_2$ capture from process emissions and from solid biomass used in industry.}
    \label{supplemental:figure_spatial_resolution_sensitivity_analysis_co2_capture_vs_co2_sequestration_conversion}
\end{figure}

\begin{figure}[H]
    \centering
    \includegraphics[width = 0.84\textwidth]{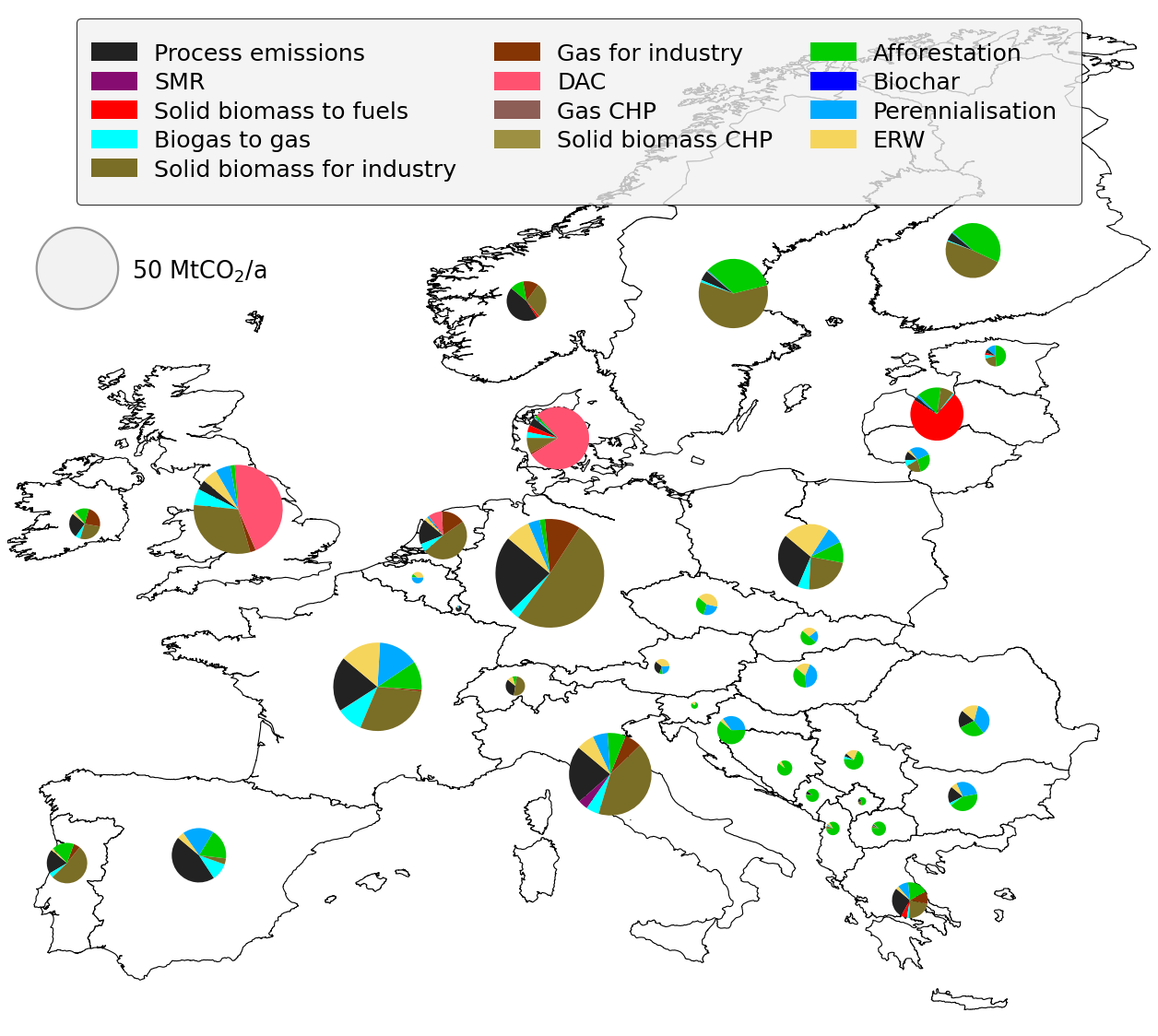}
    \caption[CO$_2$ capture across Europe under a 39-node spatial resolution]{\textbf{CO$_2$ capture across Europe under a 39-node spatial resolution}. The spatial pattern of CO$_2$ capture differs markedly from that obtained under a 90-node resolution (Figure~\ref{figure_co2_capture_spatial_pattern}). Countries such as France, Germany, Italy, Poland, and Latvia exhibit substantially higher levels of CO$_2$ capture under a 39-node resolution, while Denmark exhibits a substantial reduction.}
    \label{supplemental:figure_co2_capture_spatial_pattern_39_nodes}
\end{figure}

\begin{figure}[H]
    \centering
    \includegraphics[width = 0.84\textwidth]{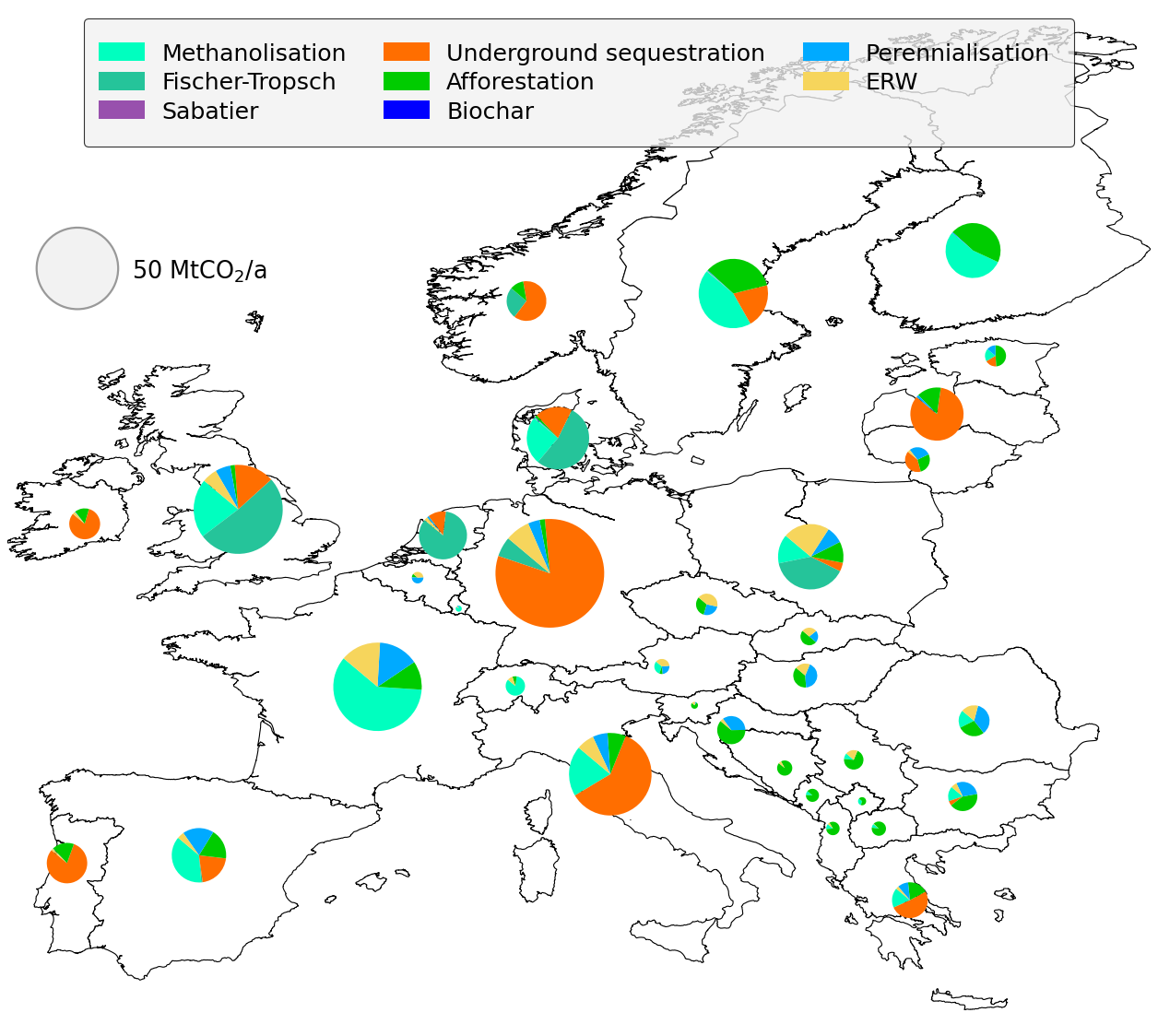}
    \caption[CO$_2$ conversion and sequestration across Europe under a 39-node spatial resolution]{\textbf{CO$_2$ conversion and sequestration across Europe under a 39-node spatial resolution}. The spatial patterns of CO$_2$ conversion and sequestration differ markedly from those obtained under a 90-node resolution (Figure~\ref{figure_co2_sequestration_and_conv_spatial_pattern}). Countries such as France, Germany, Italy, Poland, and Latvia exhibit substantially higher levels of CO$_2$ conversion and sequestration under a 39-node resolution, while Denmark exhibits a substantial reduction.}
    \label{supplemental:figure_co2_sequestration_and_conv_spatial_pattern_39_nodes}
\end{figure}

First, coarsening the model's spatial resolution from 90 to 39 nodes largely preserves the system configuration while reducing total system costs by 5\%, mainly due to lower investments in solar and DAC (Figure~\ref{supplemental:figure_spatial_resolution_sensitivity_analysis_cost}). Consistent with previous findings \cite{horsch_the_role_2017}, this reduction reflects the benefits of regional integration and the smoothing of energy transport network bottlenecks, which are apparent in the 90-node resolution but less pronounced at coarser resolution. While CO$_2$ balances and the contributions of the additional CDR strategies remain largely unchanged across the two spatial resolutions, CO$_2$ capture via DAC is significantly reduced under the coarser spatial resolution. This reduction is partially offset by increased CO$_2$ capture from process emissions and from the use of solid biomass in industry (Figure~\ref{supplemental:figure_spatial_resolution_sensitivity_analysis_co2_capture_vs_co2_sequestration_conversion}). Although CO$_2$ balances remain relatively similar, the spatial distribution and deployment levels of CO$_2$ capture, conversion, and sequestration technologies differ markedly across Europe between the two spatial resolutions (Figures~\ref{supplemental:figure_co2_capture_spatial_pattern_39_nodes} and \ref{supplemental:figure_co2_sequestration_and_conv_spatial_pattern_39_nodes}).

\begin{figure}[H]
    \centering
    \includegraphics[width = 0.84\textwidth]{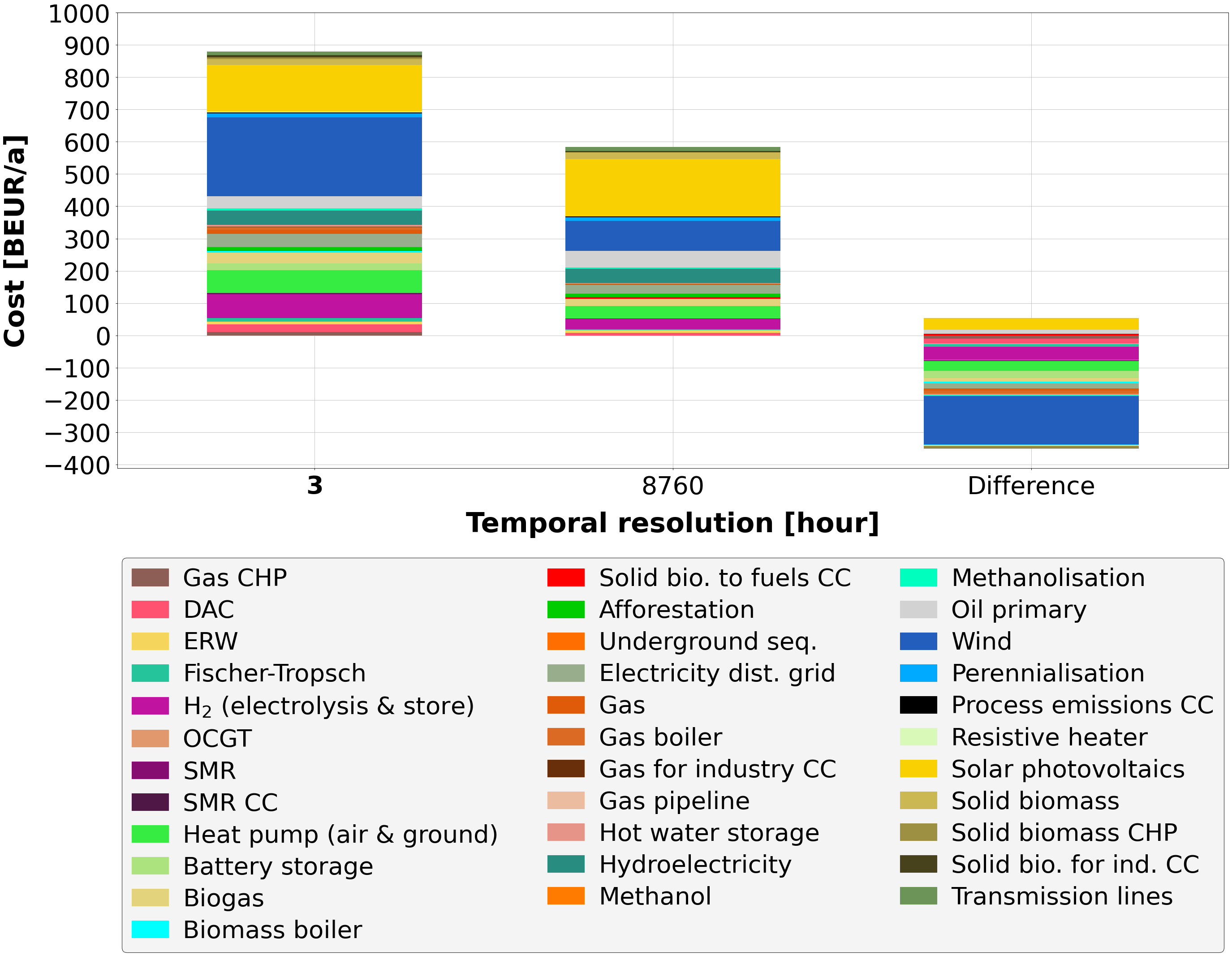}
    \caption[Cost temporal resolution sensitivity analysis]{\textbf{Cost temporal resolution sensitivity analysis}. Coarsening the model's temporal resolution from 3 to 8760 hours (one representative year) reduces costs by more than 30\%, mainly due to lower investments in wind and electrolytic H$_2$ production. Temporal aggregation smooths short-term variability in demand and renewable generation, underestimating peak capacity, storage, and flexibility requirements, thereby substantially impacting total system costs. Technologies with annual costs below 0.1 BEUR are omitted for readability.}
    \label{supplemental:figure_temporal_resolution_sensitivity_analysis_cost}
\end{figure}

\begin{figure}[H]
    \centering
    \includegraphics[width = 0.84\textwidth]{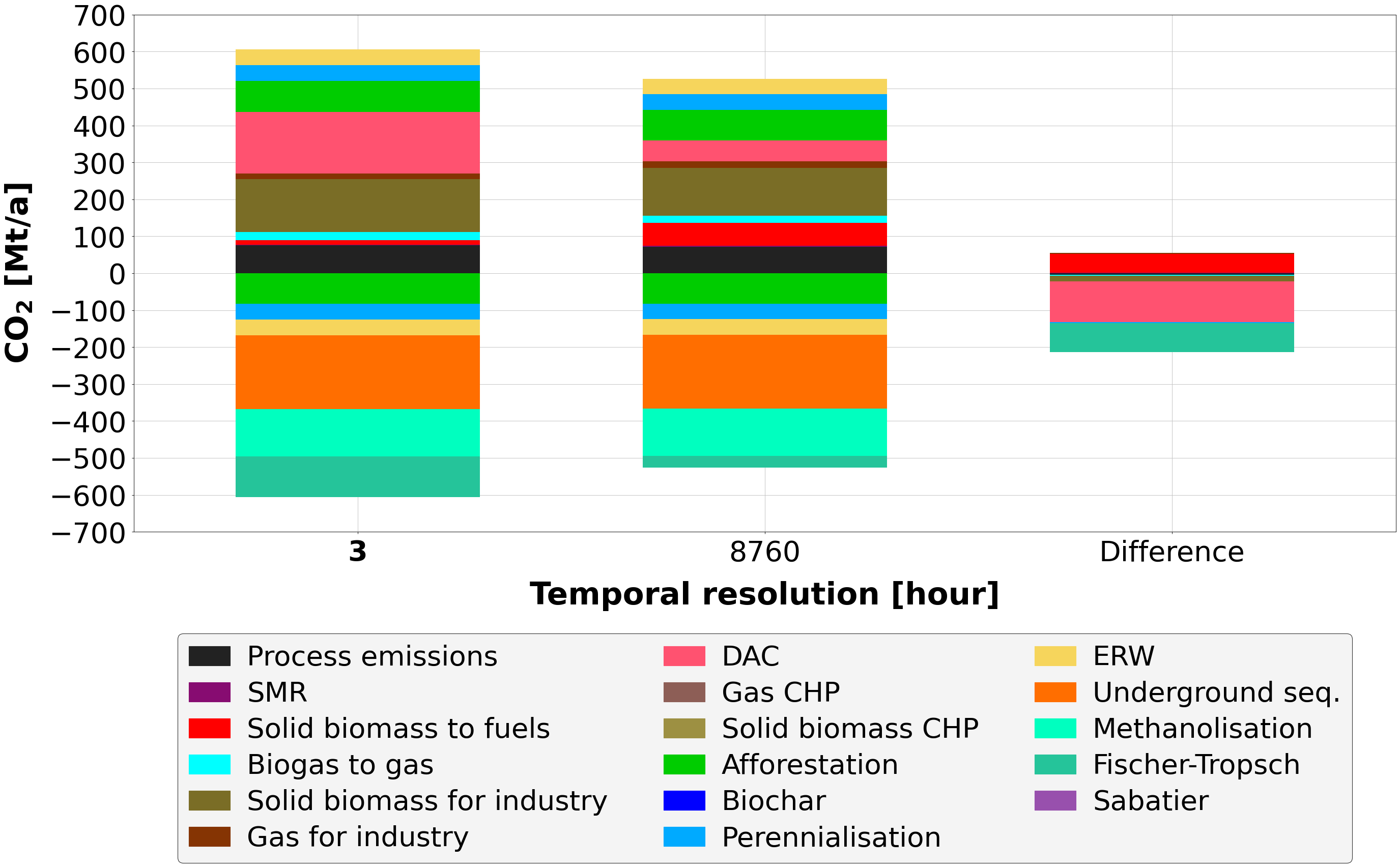}
    \caption[CO$_2$ capture, conversion, and sequestration temporal resolution sensitivity analysis]{\textbf{CO$_2$ capture, conversion, and sequestration temporal resolution sensitivity analysis}. While the contributions of the additional CDR strategies remain unchanged across the two temporal resolutions, CO$_2$ balances are substantially reduced under the coarser resolution, primarily due to a pronounced phase-out of DAC, which is partially offset by increased CO$_2$ capture from biofuels production.}
    \label{supplemental:figure_temporal_resolution_sensitivity_analysis_co2_capture_vs_co2_sequestration_conversion}
\end{figure}

Second, coarsening the model's temporal resolution to a single time step representing an entire year, as commonly done in some IAMs, reduces total system costs by more than 30\% (Figures~\ref{supplemental:figure_temporal_resolution_sensitivity_analysis_cost} and \ref{supplemental:figure_model_temporal_sensitivity_analysis}). As expected, this reduction is primarily driven by the implicit balancing of renewable generation across timescales within the year, which effectively occurs at no cost and therefore lowers the required investments in installed wind capacity and H$_2$ electrolysis. In addition, while the contributions of the additional CDR strategies remain unchanged across the two temporal resolutions, CO$_2$ balances are substantially reduced under the coarser resolution. This reduction is mainly driven by a decrease in DAC and synthetic oil production (Figure~\ref{supplemental:figure_temporal_resolution_sensitivity_analysis_co2_capture_vs_co2_sequestration_conversion}).

\section{Discussion}
The inclusion of the selected additional CDR strategies—afforestation, perennialisation, and ERW—in the European energy system substantially reduces total system costs by decreasing reliance on comparatively expensive CO$_2$ capture and conversion technologies, such as DAC, point-source capture, and Fischer–Tropsch. Specifically, by limiting the deployment of these capital- and energy-intensive options, these CDR strategies improve the overall economic feasibility of achieving a climate-neutral energy system by 89 BEUR annually (a 9\% reduction). In particular, an annual investment of 9.4 BEUR in afforestation alone results in annual system cost savings of 49 BEUR (5\%). Although afforestation is not inherently cheaper than underground sequestration alone, it does not require additional processes or infrastructure to remove CO$_2$ from the atmosphere, and therefore reduces the need for investment in DAC, renewable generation, and associated infrastructure. This underscores that afforestation is not only a critical strategy for advancing Europe's decarbonisation, but also a highly cost-effective measure that delivers substantial system-wide economic benefits.

\begin{figure}[H]
    \centering
    \includegraphics[width = 0.84\textwidth]{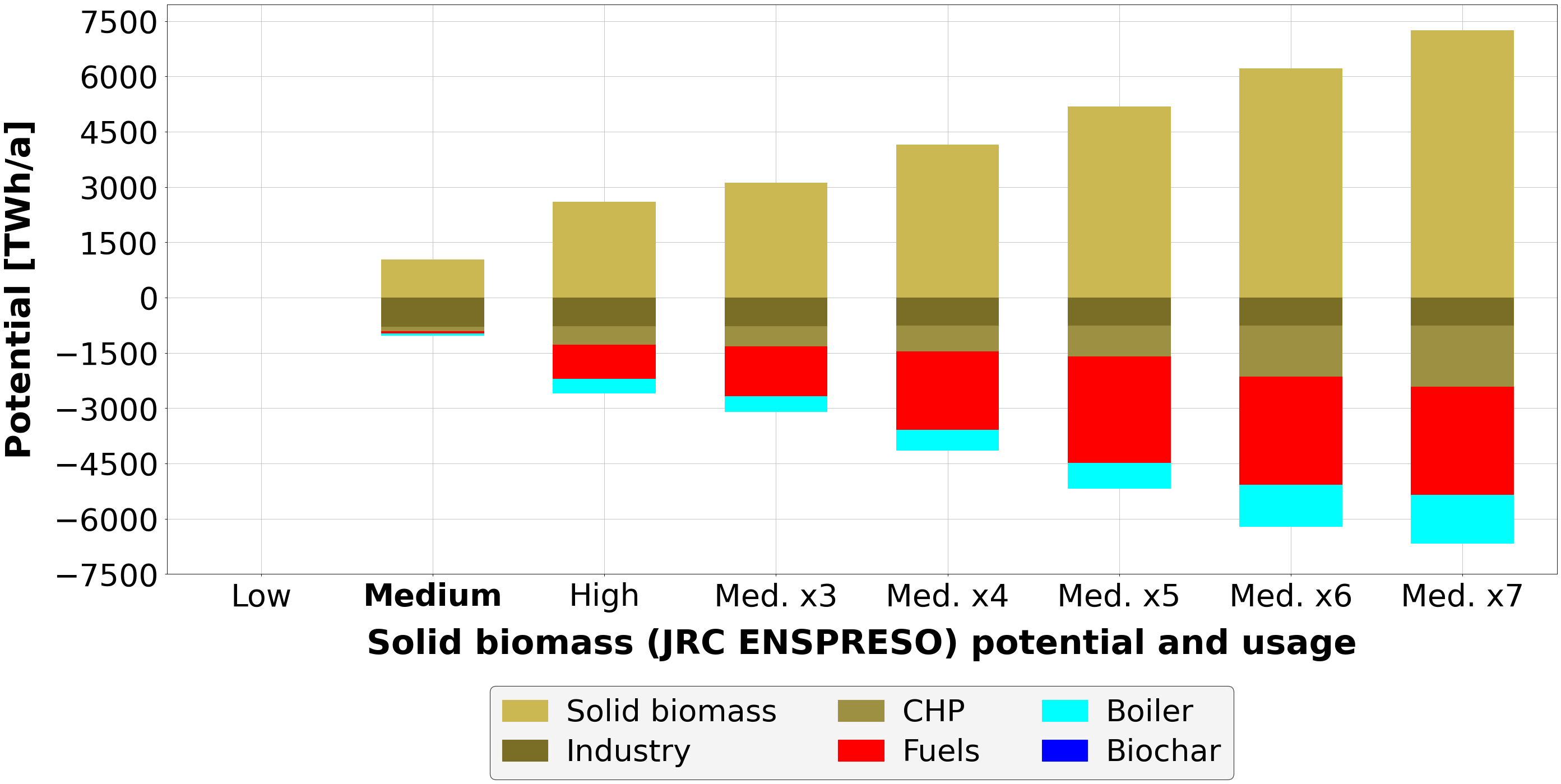}
    \caption[Solid biomass (JRC ENSPRESO) potential and usage]{\textbf{Solid biomass (JRC ENSPRESO) potential and usage}. Solid biomass in the model (based on the JRC ENSPRESO medium scenario) is primarily allocated to industrial heat supply. Even with increased solid biomass availability, biochar remains uncompetitive despite its low cost per tonne of CO$_2$ removed, as the additional solid biomass is mainly directed to CHP units, fuels production, and boiler combustion.}
    \label{supplemental:figure_solid_biomass_jrc_enspreso_potential_and_usage}
\end{figure}

With the exception of biochar, all other additional CDR strategies are fully utilised across all regions, delivering an aggregated CO$_2$ removal of 169 MtCO$_2$/a, in addition to the assumed underground sequestration potential of 200 MtCO$_2$/a. Biochar is not selected as a viable strategy due to its relatively inefficient biomass utilisation, with less than 50\% of the carbon content sequestered and without energy production. The solid biomass required for its production is preferentially allocated to fuels production and heat supply during winter demand peaks, while capturing the associated CO$_2$ emissions. This is consistent with Millinger et al. \cite{Millinger2025}, who found that large biomass imports into Europe are cost-effective for meeting heating demand and producing bio-based fuels. By contrast, perennialisation removes CO$_2$ while simultaneously expanding the available budget of fossil-neutral carbon within the energy system. In our analysis, perennialisation increases available biogas by 37 TWh/a in addition to the baseline potential of 336 TWh/a. Although the additional solid biomass potentially available from afforestation is not represented in our model, the main results are expected to remain robust, as even under substantially higher solid biomass availability, biochar does not play a meaningful role in decarbonising Europe (Figure~\ref{supplemental:figure_solid_biomass_jrc_enspreso_potential_and_usage}).

Markkanen et al. \cite{Markkanen_2024} also estimated the usage of CDR strategies in Europe under different transition scenarios. In their optimistic technology-development scenario, the CO$_2$ removal potentials are consistently higher across all strategies than in our study (66 MtCO$_2$/a for biochar, 331 MtCO$_2$/a for afforestation, and 208 MtCO$_2$/a for ERW). As in our work, solid biomass availability is based on the JRC ENSPRESO medium scenario. Consistent with our results, the removal potentials for afforestation and ERW are fully utilised. However, contrary to us, biochar is selected in most of their scenarios. The lack of detailed hourly balancing for the entire year in their model may overlook competing, time-sensitive uses of solid biomass—such as heat provision during peak demand periods and synthetic fuels production—that are prioritised in our high-resolution model.

\begin{figure}[H]
    \centering
    \includegraphics[width = 0.84\textwidth]{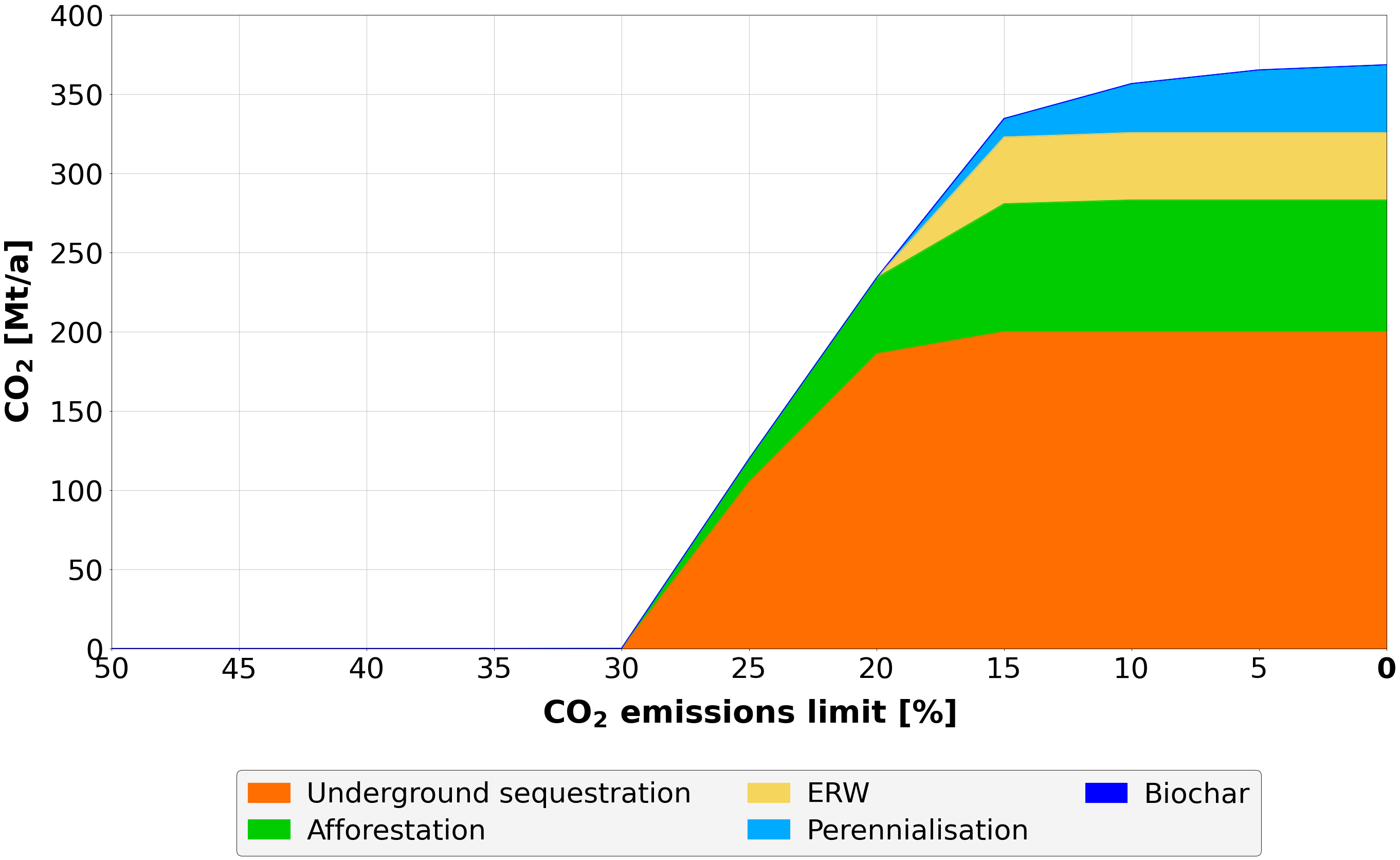}
    \caption[CO$_2$ emissions limit sensitivity analysis]{\textbf{CO$_2$ emissions limit sensitivity analysis}. In a climate-neutral energy system, CDR strategies become critical only at stringent CO$_2$ emissions limits. Underground sequestration is utilised first at a 30\% emissions limit (relative to 1990 levels), closely followed by afforestation, while ERW and perennialisation are only utilised at a 20\% emissions limit. Due to solid biomass being used in higher-value processes, biochar is not selected as a viable strategy.}
    \label{supplemental:figure_cdr_usage_vs_co2_emissions_limit}
\end{figure}

Although CDRs are important in the decarbonisation effort, they are primarily required in the later stages of emissions reduction, entering the energy system at different times depending on the CO$_2$ emission limit (Figure~\ref{supplemental:figure_cdr_usage_vs_co2_emissions_limit}). In particular, underground sequestration is needed once a reduction target of approximately 30\% relative to 1990 levels is reached, while the three selected additional CDR strategies are only needed at even more stringent decarbonisation levels. This sequencing provides time to coordinate policy, establish appropriate regulations, and begin implementing the necessary infrastructure to support these CDRs across Europe. Furthermore, even when underground sequestration potential is increased to 600 MtCO$_2$/a, the selected additional CDRs continue to play a significant role in CO$_2$ removal, indicating their relative independence from underground sequestration (Figure~\ref{supplemental:figure_underground_seq_potential_co2_capture_vs_co2_sequestration_conversion}). This highlights their importance and resilience in the face of uncertainties associated with the permanent geological sequestration of CO$_2$ in Europe.

\section{Conclusion}
In this study, a highly spatially and temporally resolved PyPSA-Eur model was used to assess the role of additional CDR strategies—afforestation, perennialisation, biochar, and ERW—in decarbonising Europe by 2050. The results show that these CDR strategies are critical for achieving climate neutrality, collectively reducing total system costs by 9\%. Afforestation provides the largest contribution, accounting for a 5\% cost reduction, followed by perennialisation and ERW, each contributing approximately 2\%. All three strategies are utilised to their maximum regional available potential, reflecting their high system value under the model assumptions. In contrast, biochar is not a viable option, raising the question of whether more impactful CDR strategies should be explored and eventually implemented. The results also show that DAC is not required to achieve climate neutrality when the energy system is equipped with all the CDR strategies considered in this study and a continent-wide CO$_2$ transport network.

On the one hand, all additional CDR strategies, together with underground CO$_2$ sequestration, are subject to substantial uncertainty in both costs and removal potentials, warranting a cautious interpretation of the results. On the other hand, the conservative potentials assumed for afforestation, perennialisation, ERW, and underground sequestration are fully utilised across all scenarios, highlighting both the difficulty of offsetting industrial process emissions and the high costs associated with fully replacing fossil oil through Fischer–Tropsch production. Beyond further scaling up renewable energy and direct electrification of heat and transport, the coming decade should therefore prioritise large-scale experimental deployment of CDR strategies to reduce these uncertainties.

To address some limitations of this study, future research could model transition pathways rather than a single future time step, thereby capturing lock-in and path-dependence effects. In addition, representing further promising CDR strategies would enable a more comprehensive assessment of the role CDRs may play in Europe's decarbonisation effort.

\section{Methods}

\subsection{Model}
This study employs PyPSA-Eur \cite{Neumann2023, HORSCH2018207} (version 2025.01.0), an open-source energy system model based on PyPSA \cite{PyPSA}, to optimise capacity and dispatch across 34 European countries: the EU27 (excluding Malta and Cyprus), plus Norway, Switzerland, Great Britain, Albania, Bosnia and Herzegovina, Montenegro, North Macedonia, Serbia, and Kosovo. The model encompasses the electricity, heating, land transport, shipping, aviation, agriculture, and industry sectors, as well as industrial feedstocks, with a detailed representation of carbon capture, transport, conversion, and sequestration.

\begin{figure}[H]
    \centering
    \includegraphics[height = 13cm, angle = 90]{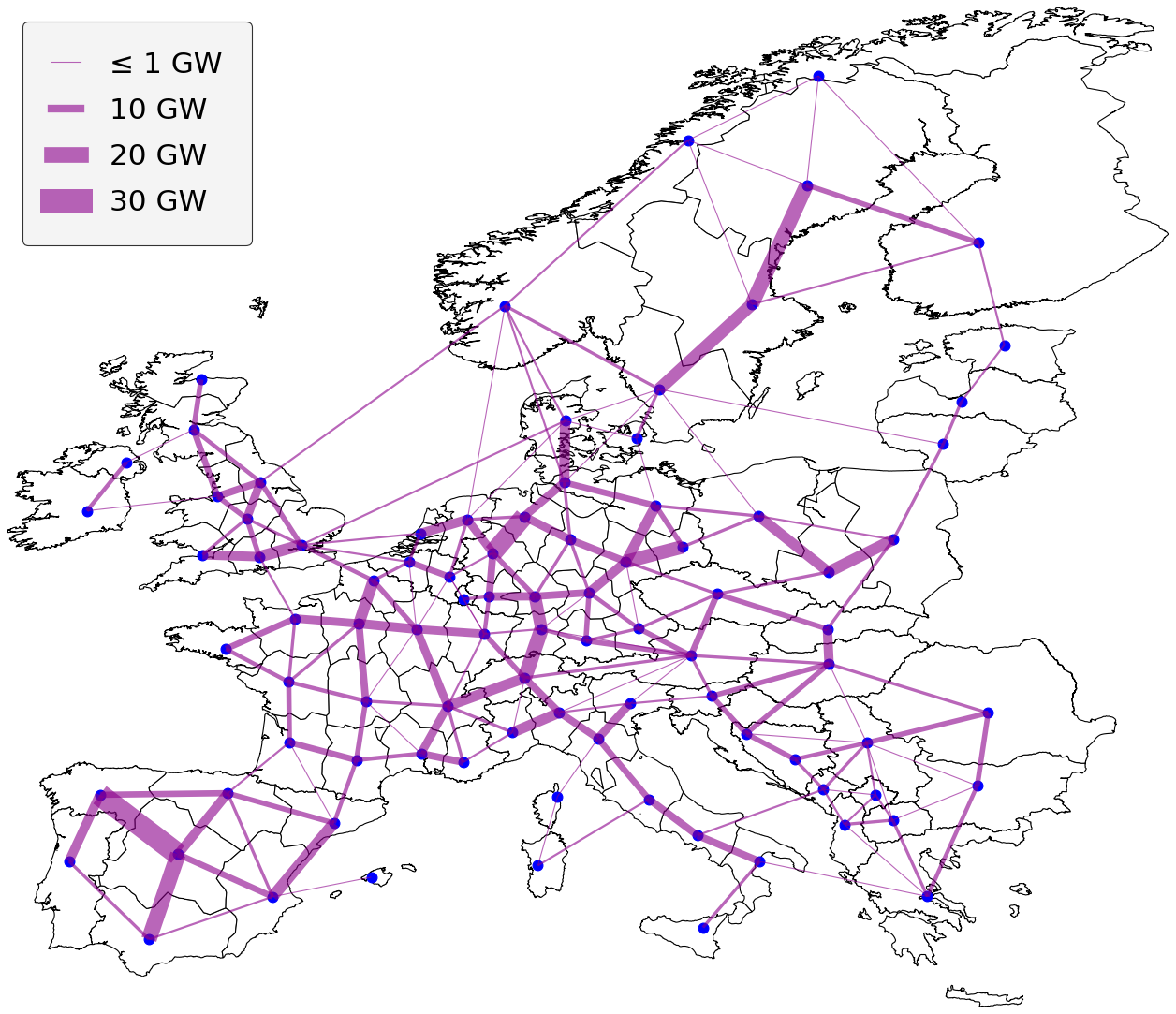}
    \caption[The European electricity transmission grid topology]{\textbf{The European electricity transmission grid topology}. The energy system is modelled with a 90-node spatial resolution, shown as blue dots. Magenta lines represent the electricity transmission grid spanning across 34 countries, including both existing power lines and new ones planned for construction as per ENTSO-E TYNDP 2020.}
    \label{supplemental:figure_electricity_transmission_grid}
\end{figure}

\begin{figure}[H]
    \centering
    \includegraphics[width = 0.84\textwidth]{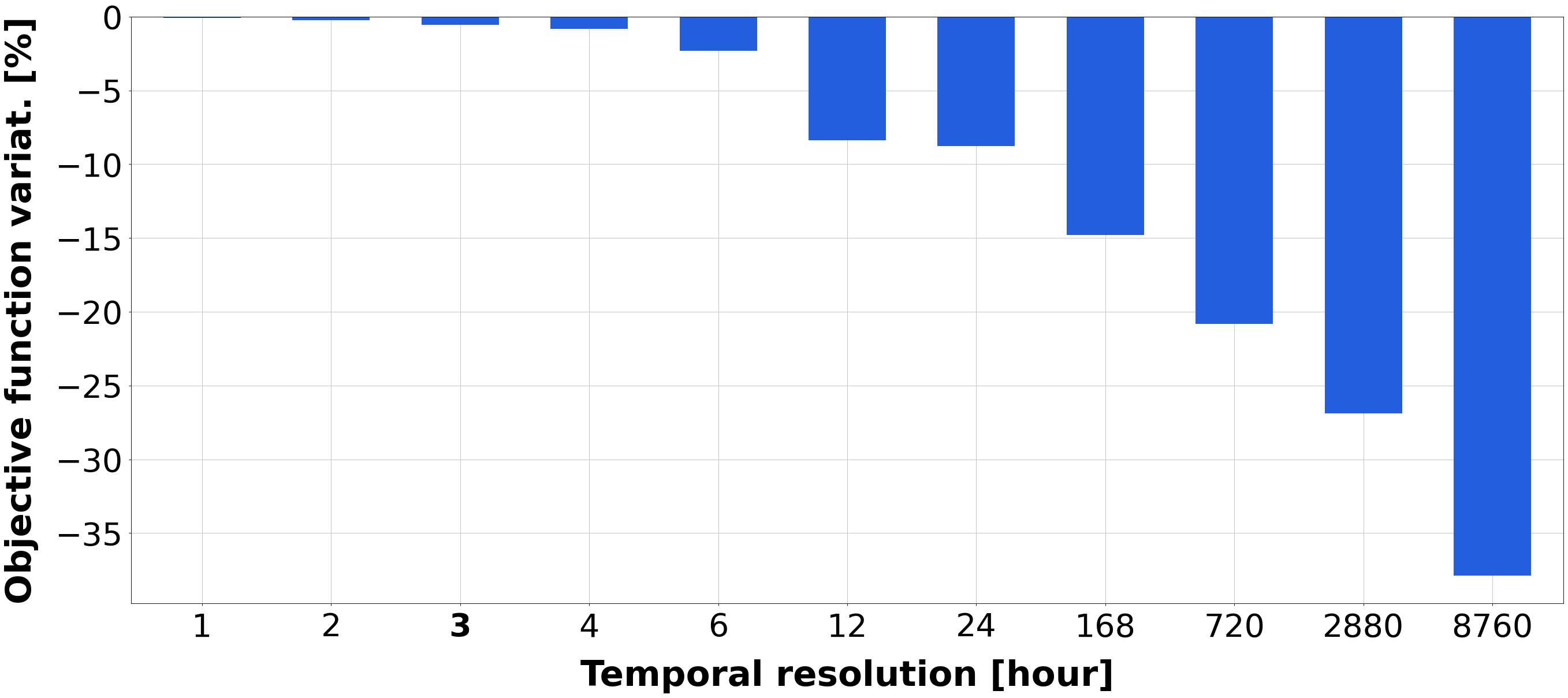}
    \caption[Model temporal resolution sensitivity analysis]{\textbf{Model temporal resolution sensitivity analysis}. In a climate-neutral energy system equipped with all CDR strategies, aggregating the model from hourly temporal resolution to a single representative year (8760 hours) reduces the objective function by 38\%. This reduction arises because lower temporal resolution smooths short-term variability in demand and renewable generation, underestimating peak capacity requirements, storage and flexibility needs, and operational constraints, resulting in an artificially lower objective function.}
    \label{supplemental:figure_model_temporal_sensitivity_analysis}
\end{figure}

PyPSA-Eur assumes long-term market equilibrium, perfect competition, and perfect foresight. It employs linear programming to optimise the cost-effective configuration of the energy system subject to various constraints. A mathematical formulation of the model is provided in Section~\ref{supplemental:material_model_mathematical_formulation}. The model minimises the annualised system cost of the European energy system under net-zero CO$_2$ emissions. It performs an overnight greenfield optimisation using a 90-node spatial resolution with 3-hourly temporal resolution over the full year (Figures~\ref{supplemental:figure_electricity_transmission_grid} and \ref{supplemental:figure_model_temporal_sensitivity_analysis}). This configuration captures European spatial heterogeneity and wind-solar variability at daily, weekly, and seasonal scales \cite{PFENNINGER20171}, addressing concerns about spatial and temporal resolution \cite{horsch_the_role_2017, FLEISCHER2020100563} while maintaining computational feasibility \cite{SCHYSKA20212606, FrysztackiRechtBrown2022_1000153979}. Renewable capacity factors are derived using atlite \cite{Hofmann2021} with ERA5 and SARAH weather data from 2013, selected for its representativeness of the average wind and solar conditions \cite{Gotske2024, COKER2020509}.

\begin{figure}[H]
    \centering
    \includegraphics[width = 0.84\textwidth]{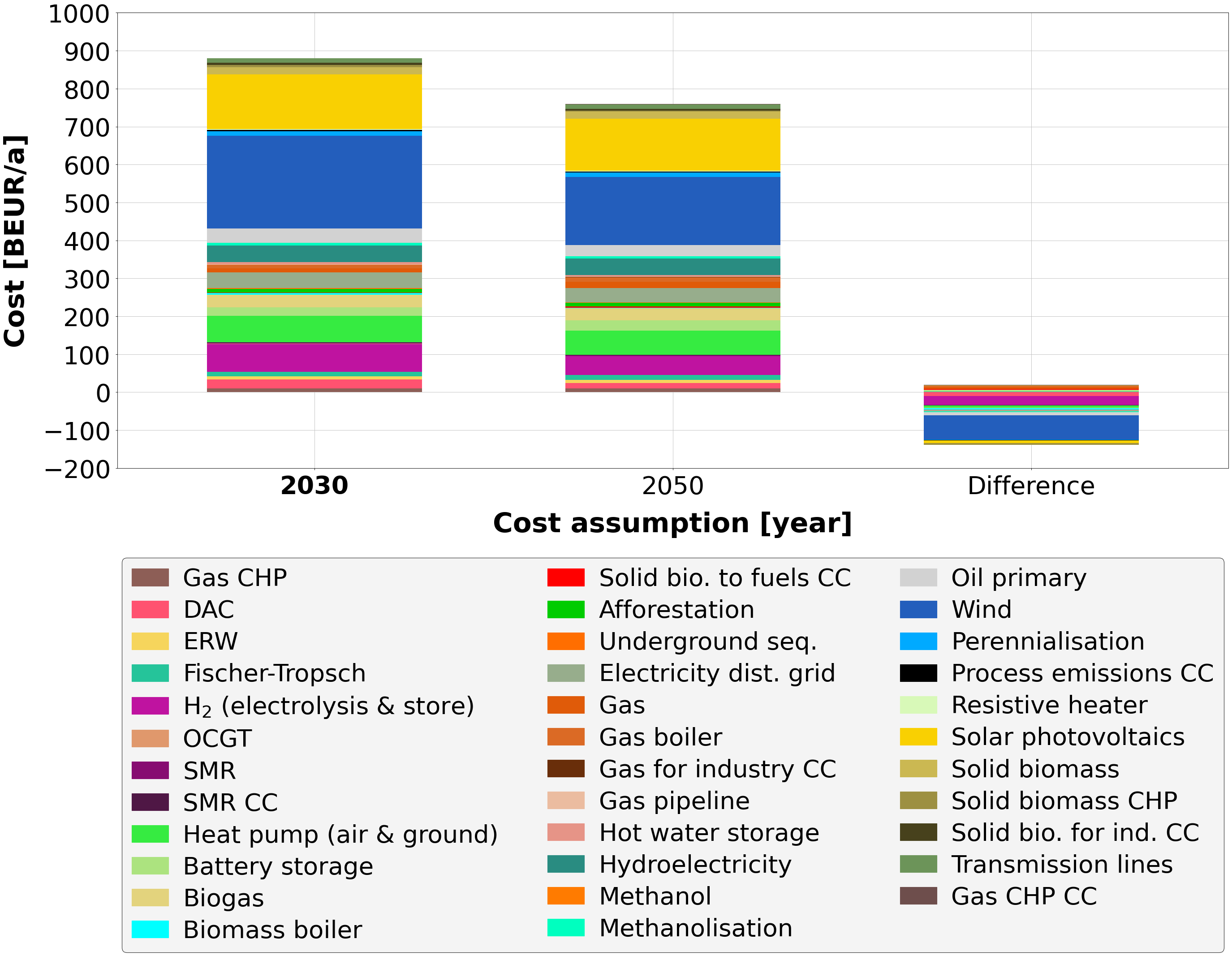}
    \caption[Cost assumptions sensitivity analysis]{\textbf{Cost assumptions sensitivity analysis}. The total cost of a climate-neutral energy system equipped with all CDR strategies is approximately 14\% lower under cost assumptions for 2050 than under those for 2030, mainly due to reduced investments in wind and electrolytic H$_2$ production. This reduction in total system costs reflects more optimistic cost projections for 2050, associated with greater technological maturity. Technologies with annual costs below 0.1 BEUR are omitted for readability.}
    \label{supplemental:figure_cost_projections_sensitivity_analysis}
\end{figure}

Technology characteristics—including costs, efficiencies, and lifetimes—are derived from the Energy System Technology Data \cite{energy_system_technology_data} (version 0.14.1), which builds primarily on data published by the Danish Energy Agency \cite{danish_energy_agency}. The model adopts projections for the year 2030, reflecting expected cost reductions for renewable and carbon-related technologies with moderate uncertainty. Cost assumptions, including a discount rate of 7\%, are applied uniformly across Europe. To assess the influence of this adoption, a sensitivity analysis of the cost projections year is presented in Figure~\ref{supplemental:figure_cost_projections_sensitivity_analysis}.

Demand for electricity (industry, agriculture, residential, services, land transport), heat (industry, agriculture, residential, services), solid biomass (industry), oil (industry, agriculture, aviation), methanol (industry, shipping), methane gas (industry), and H$_2$ (industry) is exogenously specified, with annual totals across Europe of 4362, 2989, 748, 1124, 512, 231, and 180 TWh, respectively. The model endogenously determines production technologies and fossil fuel use for electricity, heat, oil, methane gas, and H$_2$. Demand and sectoral transformation data are described in \cite{Neumann2023} and the JRC Integrated Database of the European Energy Sector (JRC-IDEES) \cite{JRC108244}.

\begin{figure}[H]
    \centering
    \includegraphics[width = 0.84\textwidth]{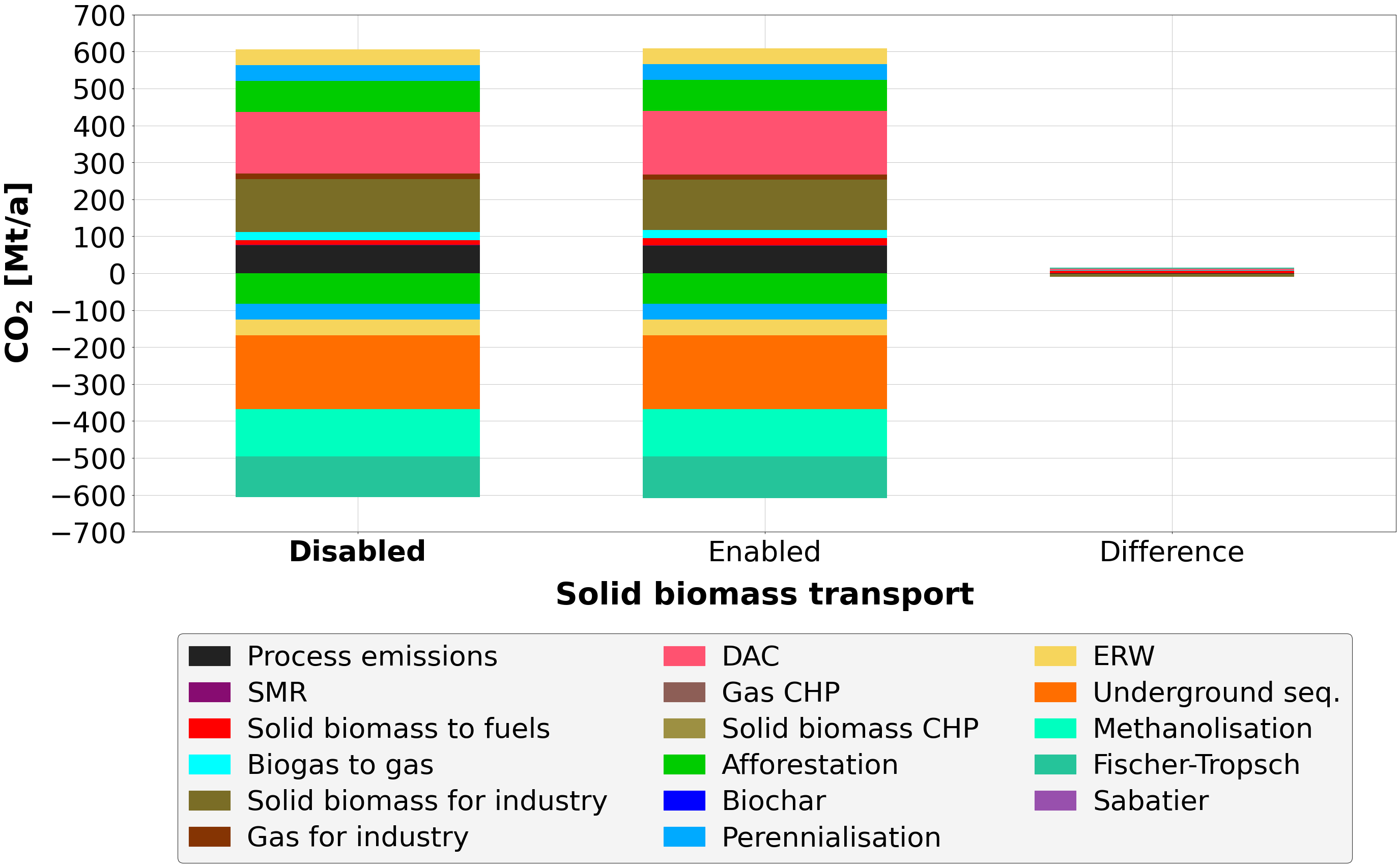}
    \caption[Solid biomass transport sensitivity analysis]{\textbf{Solid biomass transport sensitivity analysis}. In a climate-neutral energy system equipped with all CDR strategies, the levels of CO$_2$ capture, conversion, and sequestration across the various technologies and processes remain largely unchanged, regardless of the ability to transport solid biomass across Europe. Only minor differences are observed, with slightly higher CO$_2$ capture from solid biomass-to-fuels conversion and DAC, and slightly lower CO$_2$ capture from solid biomass used in industry.}
    \label{supplemental:figure_solid_biomass_transport_co2_capture_vs_co2_sequestration_conversion}
\end{figure}

\begin{figure}[H]
    \centering
    \includegraphics[width = 0.84\textwidth]{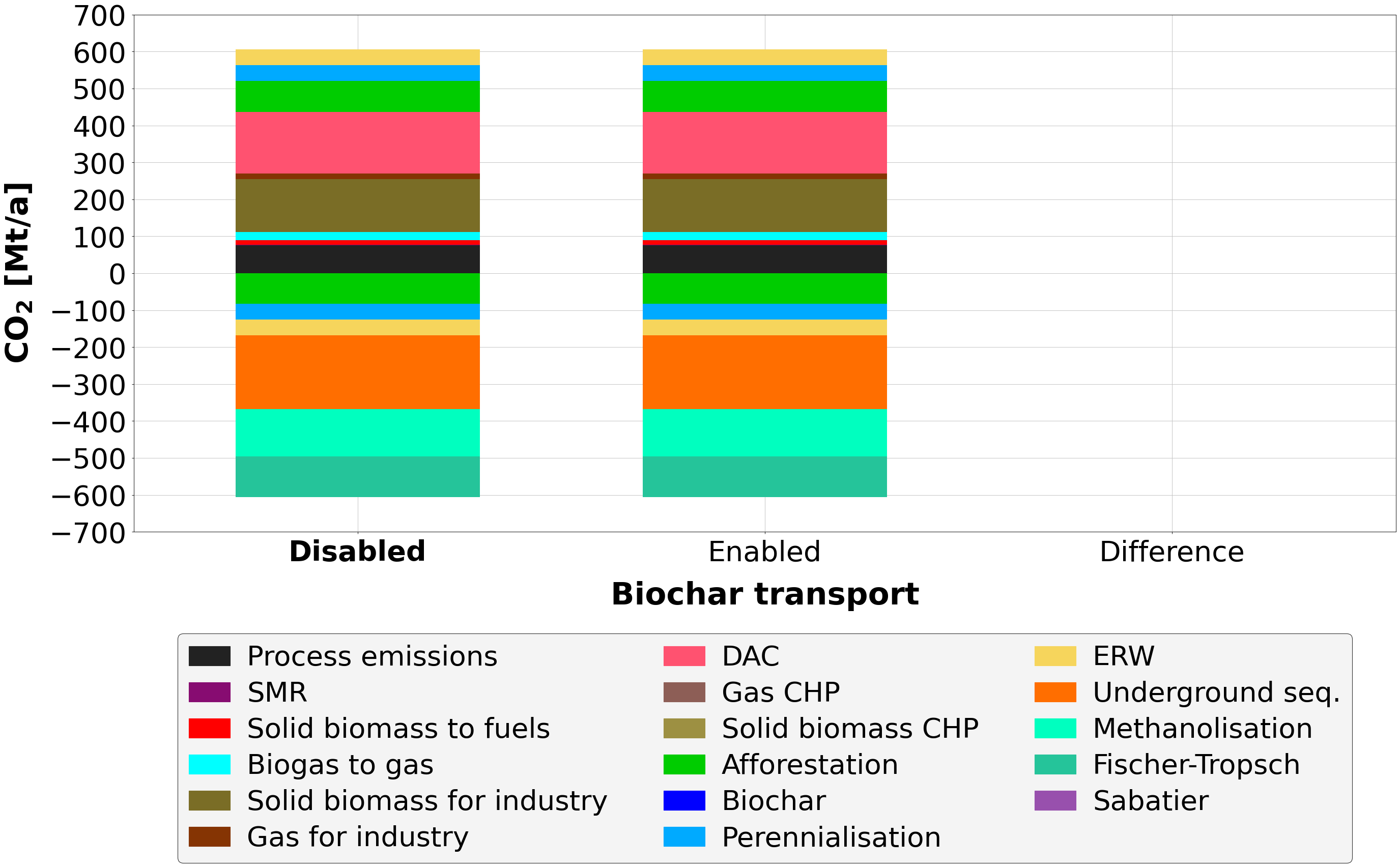}
    \caption[Biochar transport sensitivity analysis]{\textbf{Biochar transport sensitivity analysis}. In a climate-neutral energy system equipped with all CDR strategies, the levels of CO$_2$ capture, conversion, and sequestration across the various technologies and processes remain unchanged, regardless of the ability to transport biochar across Europe.}
    \label{supplemental:figure_biochar_transport_co2_capture_vs_co2_sequestration_conversion}
\end{figure}

\begin{figure}[H]
    \centering
    \includegraphics[width = 0.84\textwidth]{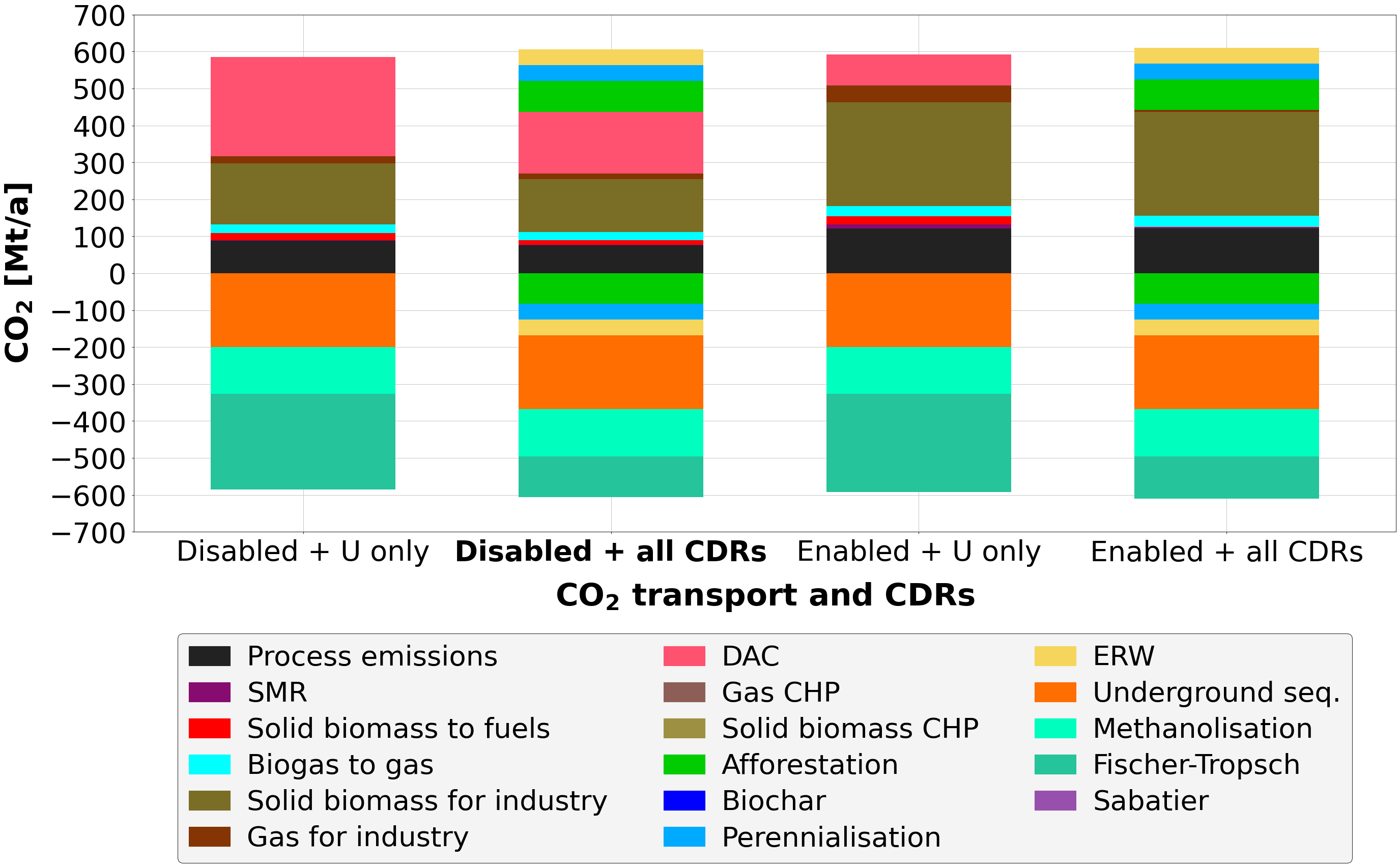}
    \caption[CO$_2$ transport and CDRs sensitivity analysis]{\textbf{CO$_2$ transport and CDRs sensitivity analysis}. A climate-neutral energy system equipped with a CO$_2$ transport network reduces the need for DAC by enabling increased CO$_2$ capture from point sources, as captured CO$_2$ can be transported to other regions for underground sequestration or synthetic fuels production. When the energy system is also equipped with the additional CDR strategies, DAC becomes entirely redundant, as the remaining atmospheric CO$_2$ is removed by these strategies instead.}
    \label{supplemental:figure_co2_transport_and_cdrs_co2_capture_vs_co2_sequestration_conversion}
\end{figure}

\begin{figure}[H]
    \centering
    \includegraphics[width = 0.84\textwidth]{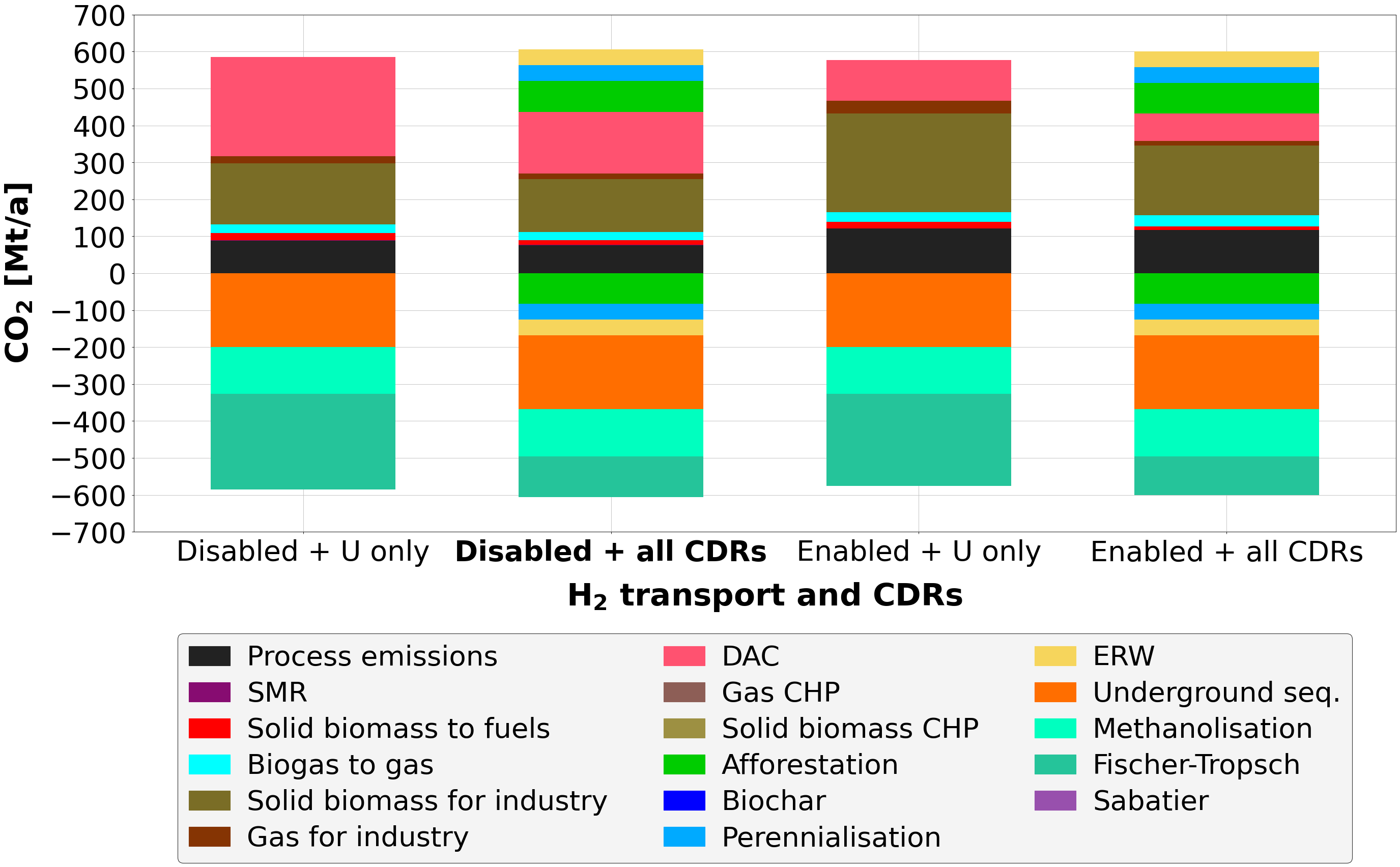}
    \caption[H$_2$ transport and CDRs sensitivity analysis]{\textbf{H$_2$ transport and CDRs sensitivity analysis}. A climate-neutral energy system equipped with an H$_2$ transport network reduces the need for DAC by enabling increased CO$_2$ capture from solid biomass used in industry and from process emission sources in regions with cost-effective electrolytic H$_2$ production. The produced H$_2$ can then be transported to regions where synthetic fuels production is more economical. When the energy system is also equipped with the additional CDR strategies, DAC is further reduced, as part of the remaining atmospheric CO$_2$ is removed by these strategies instead.}
    \label{supplemental:figure_h2_transport_and_cdrs_co2_capture_vs_co2_sequestration_conversion}
\end{figure}

Transmission capacities are fixed exogenously, encompassing existing and planned lines per ENTSO-E Ten-Year Network Development Plan 2020 \cite{entsoe_tyndp_2020} (Figure~\ref{supplemental:figure_electricity_transmission_grid}), with no capacity expansion allowed in the model. Methane gas transport occurs via pipelines with fixed capacities defined by the SciGRID\_gas project \cite{pluta_scigrid_gas_2022}, reflecting Europe's existing network. Repurposing these pipelines for H$_2$ transport is not permitted. In the baseline scenario, the transport of solid biomass, biochar, CO$_2$, and H$_2$ is not allowed. To assess the influence of this limitation, sensitivity analyses of the transport of these materials are presented in Figures~\ref{supplemental:figure_solid_biomass_transport_co2_capture_vs_co2_sequestration_conversion}, \ref{supplemental:figure_biochar_transport_co2_capture_vs_co2_sequestration_conversion}, \ref{supplemental:figure_co2_transport_and_cdrs_co2_capture_vs_co2_sequestration_conversion}, and \ref{supplemental:figure_h2_transport_and_cdrs_co2_capture_vs_co2_sequestration_conversion}, respectively. The solid biomass and biochar transport sensitivity analyses are based on the transport costs listed in Table~\ref{supplemental:table_solid_biomass_transport_cost}. Oil and methanol have unlimited cross-border transport capacity within the model; synthetic fuel imports and exports to and from Europe are not considered.

\begin{figure}[H]
    \centering
    \includegraphics[width = 0.84\textwidth]{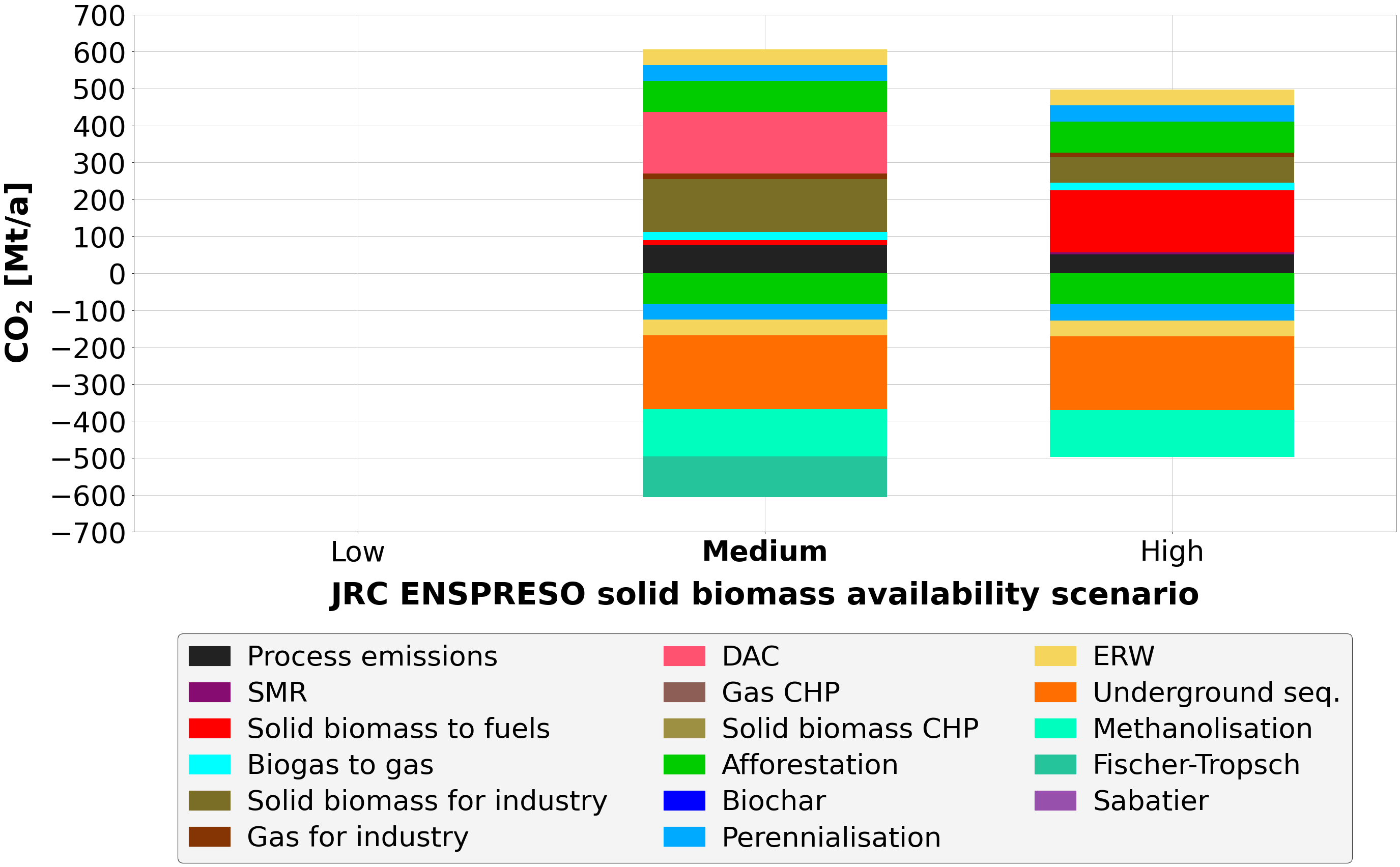}
    \caption[JRC ENSPRESO solid biomass availability scenario sensitivity analysis]{\textbf{JRC ENSPRESO solid biomass availability scenario sensitivity analysis}. With increased solid biomass availability under the JRC ENSPRESO high scenario, DAC is not selected by the model and CO$_2$ capture from solid biomass-based CHP units decreases, as it is offset by CO$_2$ capture from solid biomass used in industry and converted into fuels, relative to the medium scenario. As a consequence, Fischer–Tropsch is not used for synthetic oil production. Under the low scenario, the model is infeasible because the limited availability of solid biomass cannot satisfy exogenous demand.}
    \label{supplemental:figure_jrc_enspreso_solid_biomass_availability_scenario_co2_capture_vs_co2_sequestration_conversion}
\end{figure}

\begin{figure}[H]
    \centering
    \includegraphics[width = 0.84\textwidth]{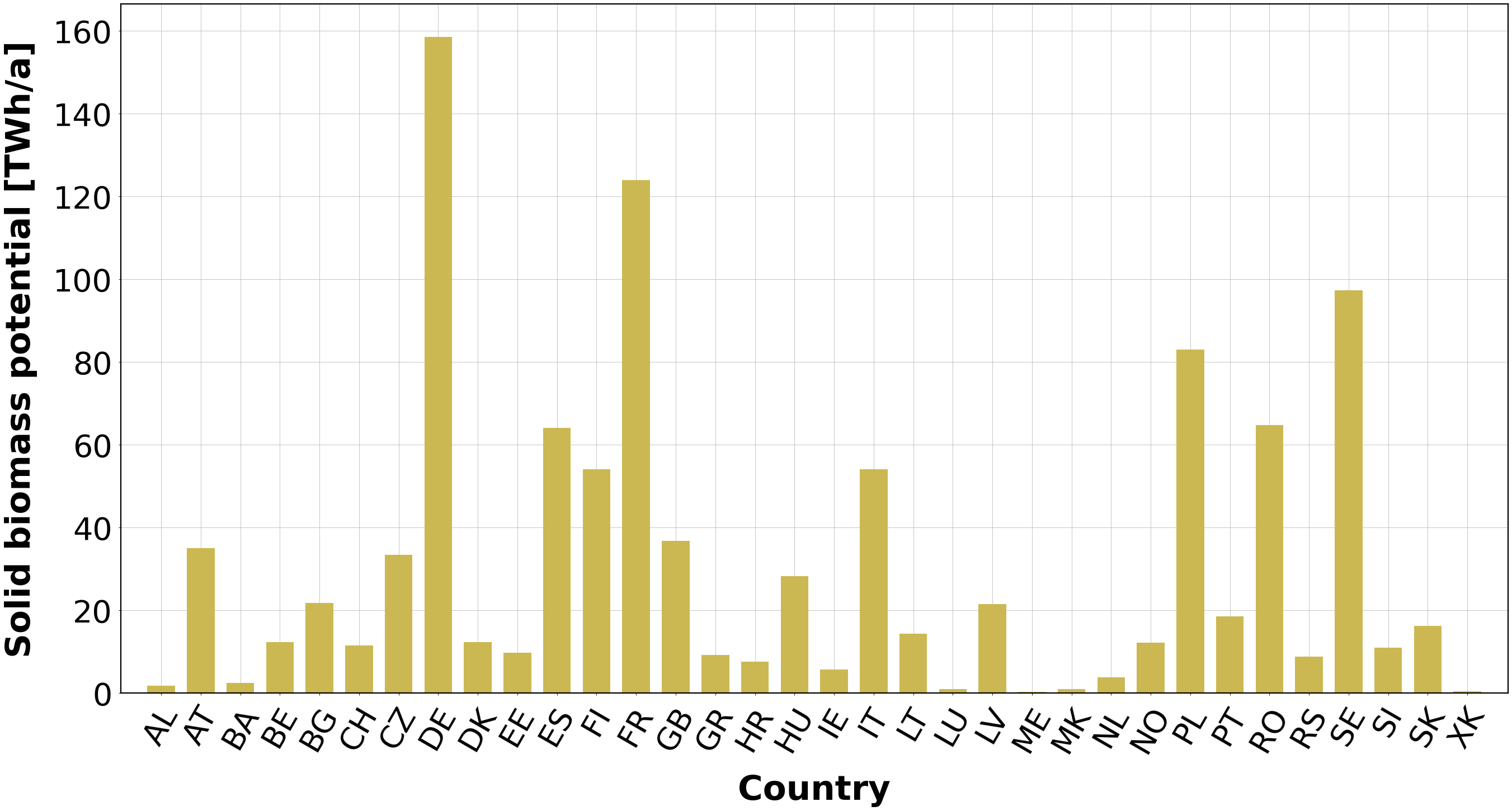}
    \caption[Solid biomass potential per country]{\textbf{Solid biomass potential per country}. Europe is assumed to have an aggregated annual solid biomass potential of 1038 TWh, based on the medium availability scenario of the JRC ENSPRESO database.}
    \label{supplemental:figure_solid_biomass_potential}
\end{figure}

The model assumes conservative biomass availability, limited to crops not competing with food production such as agricultural residues, fuelwood residues, secondary forestry residues (woodchips), sawdust, landscape residues, and biodegradable municipal waste. Crops used for first-generation biofuels production are excluded; however, their land can be reallocated to perennialisation. A solid biomass potential of 1038 TWh per year is assumed for Europe, corresponding to the medium availability scenario of the JRC ENSPRESO \cite{RUIZ2019100379} dataset for the year 2030 (Figure~\ref{supplemental:figure_solid_biomass_potential}). To assess the influence of this assumption, a sensitivity analysis of the JRC ENSPRESO solid biomass availability scenario is presented in Figure~\ref{supplemental:figure_jrc_enspreso_solid_biomass_availability_scenario_co2_capture_vs_co2_sequestration_conversion}. This biomass can be used for industrial heat generation, in CHP units and individual boilers, as well as for the production of fuels. Imports of biomass from outside Europe are excluded. Biogas potential for Europe is set at 336 TWh per year and may be upgraded to methane gas if cost-effective.

Run-of-river (ROR), hydroelectric, and pumped hydro storage (PHS) capacities are fixed exogenously based on existing European facilities.

The model includes only CO$_2$ emissions from agricultural energy use, assuming other emissions in this sector are offset by Land Use, Land Use Change, and Forestry (LULUCF).

\subsection{Carbon dioxide capture and conversion}
In the model, CO$_2$ can be captured from DAC, industrial process emissions, gas and solid biomass used in industry, CHP units combusting methane gas or solid biomass, steam methane reforming (SMR) for H$_2$ production, solid biomass upgraded to methane gas, and solid biomass converted into synthetic oil or methanol (biofuels). DAC is represented as a low-temperature heat-requiring technology that can be supplied by district heating (in which heat can be supplied by heat pumps, boilers, resistive heaters, solar thermal, CHP, and excess process heat). BECCS is modelled implicitly by capturing CO$_2$ from solid biomass used in industry and CHP units, with captured CO$_2$ either sequestered underground or used for synthetic fuels. Process emissions and gas and solid biomass demand in industry are exogenous, while CHP and DAC capacities are optimised endogenously.

Captured CO$_2$ can be converted into synthetic fuels via the Fischer–Tropsch process (synthetic oil), methanolisation (synthetic methanol), and the Sabatier reaction (synthetic methane gas). The model endogenously determines the production of synthetic oil and methane gas, as fossil options are also available. In contrast, methanol production is fixed to satisfy an exogenously specified demand. Alternatively, captured CO$_2$ can be permanently sequestered using several strategies, including geological (underground) storage, afforestation, perennialisation, biochar, and ERW, each with distinct CO$_2$ sequestration and removal potentials (Table~\ref{table_cdrs_land_and_sequestration_potentials}).

\begin{table}[!htb]
    \small
    \centering
    \caption[Land availability and CO$_2$ sequestration/removal potential of the CDR strategies]{\textbf{Land availability and CO$_2$ sequestration/removal potential of the CDR strategies}. Land availability for afforestation, biochar, and ERW is determined using the CORINE Land Cover dataset, while for perennialisation, land currently used for first-generation biofuel crops—as derived from the JRC ENSPRESO dataset—is considered fully available for replacement with perennial crops. CO$_2$ removal potential for afforestation, perennialisation, biochar, and ERW is calculated using Equations~\ref{supplemental:material_afforestation_potential}, \ref{eq:perennials_seq}, \ref{eq:biochar_potential}, and \ref{supplemental:material_erw_potential}, respectively.}
    \label{table_cdrs_land_and_sequestration_potentials}
    \begin{tabular}{p{3.4cm} p{2.8cm} p{5.3cm}}
        \toprule
        \textbf{CDR}     & \textbf{Land availability [Mha]} & \textbf{Sequestration/removal potential [MtCO$_2$/a]}\\
        \midrule
        Underground seq. & -                                & 200                                                  \\
        Afforestation    & 15                               & 83                                                   \\
        Perennialisation & 21                               & 43                                                   \\
        Biochar          & 23                               & 30                                                   \\
        ERW              & 14                               & 43                                                   \\
        \bottomrule
    \end{tabular}
\end{table}

\subsection{Carbon dioxide removal (CDR)}
\label{carbon_dioxide_removal}

\subsubsection{Underground sequestration}
\begin{figure}[H]
    \centering
    \includegraphics[width = 0.84\textwidth]{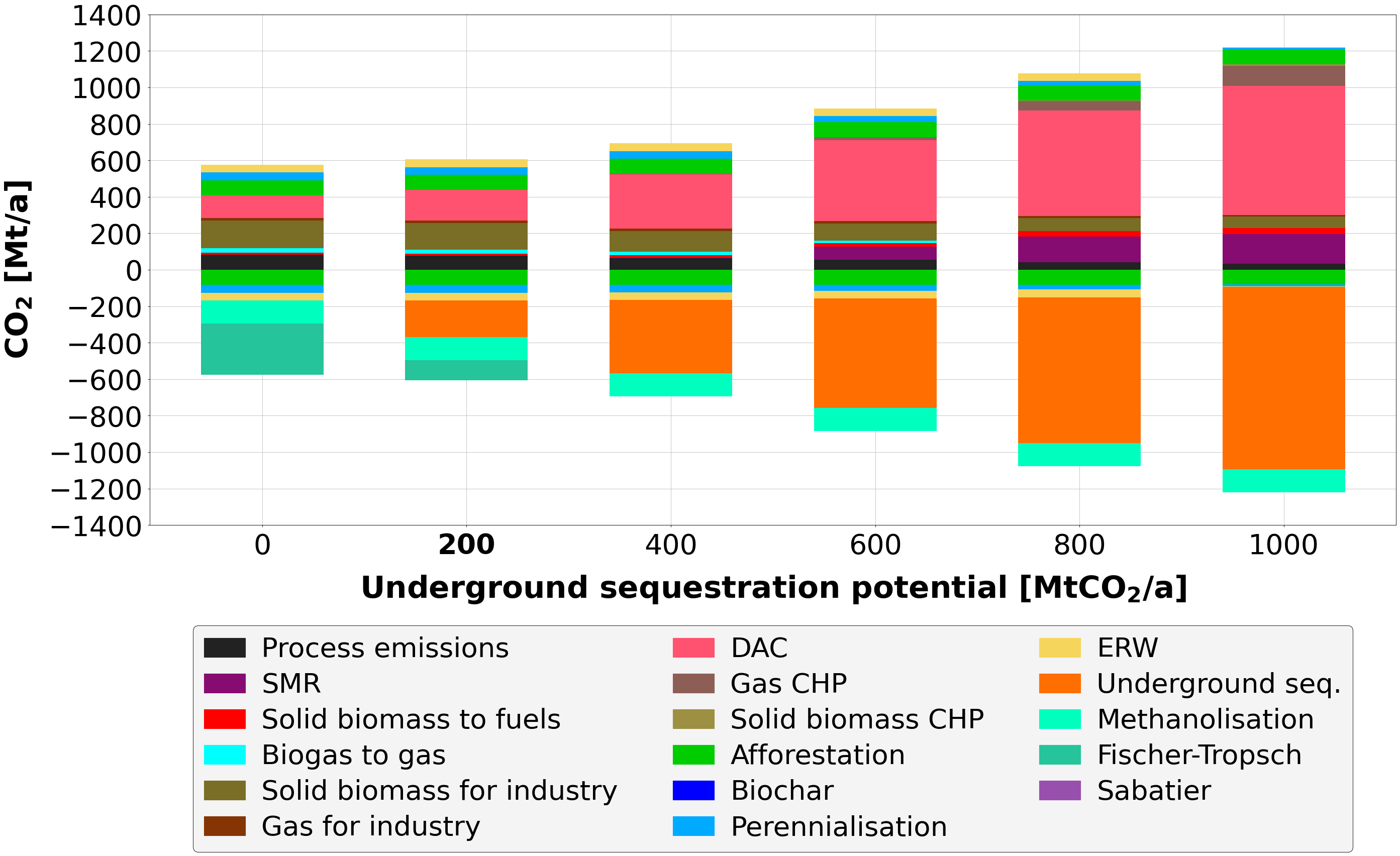}
    \caption[Underground sequestration potential sensitivity analysis]{\textbf{Underground sequestration potential sensitivity analysis}. While the role of afforestation remains largely unaffected by increases in underground sequestration potential, higher sequestration potentials progressively reduce the contribution of perennialisation and ERW, which are almost entirely pushed out from the climate-neutral energy system once underground sequestration potential reaches 1000 MtCO$_2$/a.}
    \label{supplemental:figure_underground_seq_potential_co2_capture_vs_co2_sequestration_conversion}
\end{figure}

\begin{figure}[H]
    \centering
    \includegraphics[width = 0.8\textwidth]{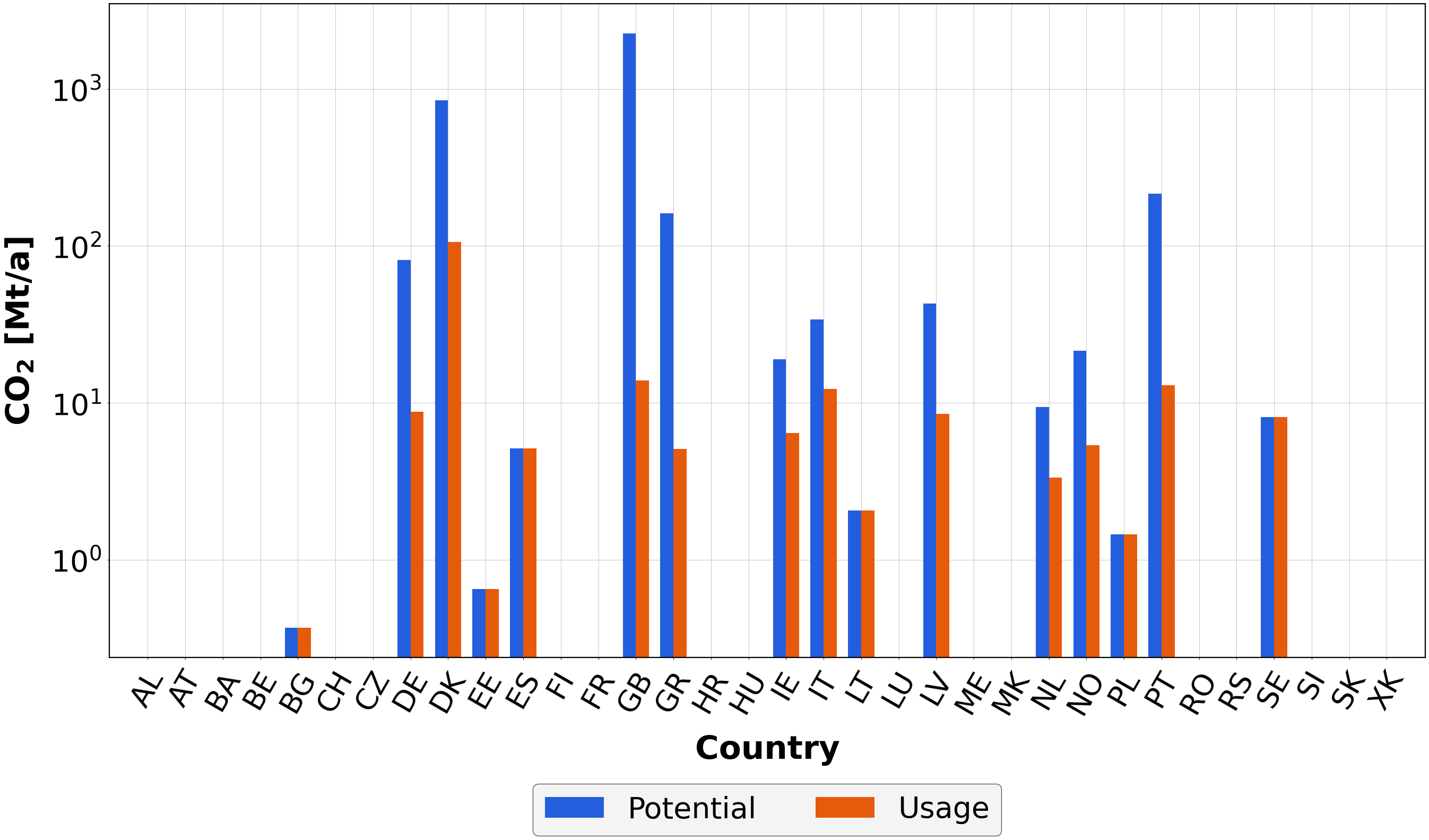}
    \caption[Underground CO$_2$ sequestration potential and usage per country]{\textbf{Underground CO$_2$ sequestration potential and usage per country}. Countries that rely heavily on DAC and point-source CO$_2$ capture typically use a larger share of their underground sequestration potential in a climate-neutral energy system equipped with all CDR strategies. While many individual countries do not fully used their own potential, the annual limit of 200 MtCO$_2$ for underground sequestration is fully reached across Europe.}
    \label{supplemental:figure_underground_co2_store_potential_and_usage}
\end{figure}

\begin{figure}[H]
    \centering
    \includegraphics[width = 0.84\textwidth]{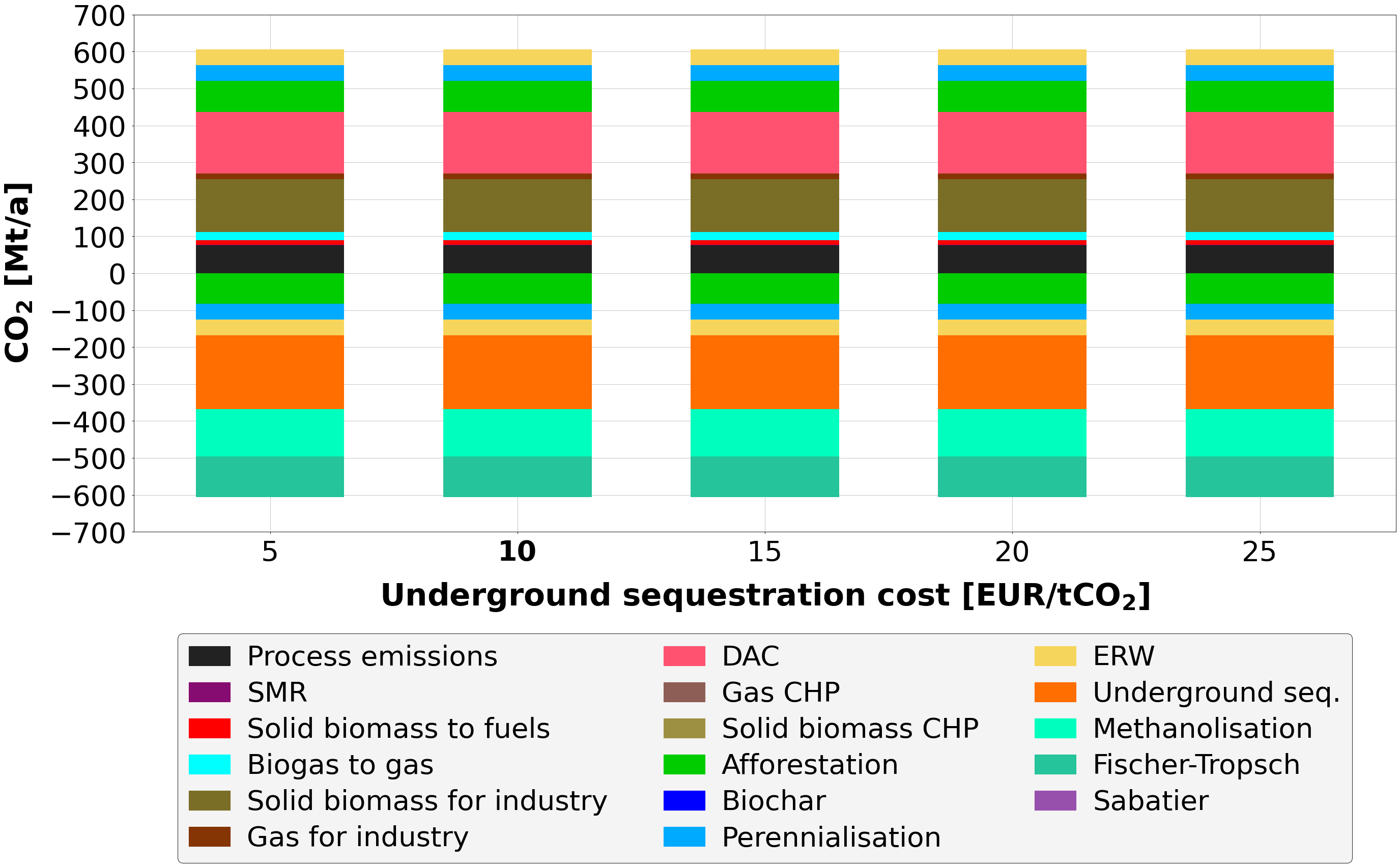}
    \caption[Underground sequestration cost sensitivity analysis]{\textbf{Underground sequestration cost sensitivity analysis}. In a climate-neutral energy system equipped with all CDR strategies, the levels of CO$_2$ capture, conversion, and sequestration across the various technologies and processes remain unchanged, regardless of the considered cost of underground sequestration.}
    \label{supplemental:figure_underground_seq_cost_co2_capture_vs_co2_sequestration_conversion}
\end{figure}

As stated previously, when combined with DAC or biogenic point-source CO$_2$ capture, underground sequestration can function as a CDR strategy by enabling net-negative emissions. Underground sequestration, also referred to as geological storage, is the process of injecting captured CO$_2$ deep underground into porous rock formations for long-term storage. Although Europe's sequestration potential is estimated at 117 Gt \cite{european_co2_sequestration_database}, the model limits this to 3.7 Gt, considering only offshore storage in deep salt caverns and depleted hydrocarbon reservoirs. Allocation across countries follows national shares (Figure~\ref{supplemental:figure_underground_co2_store_potential_and_usage} and Table~\ref{supplemental:table_underground_co2_sequestration_potential}), and the model endogenously determines utilisation. In addition, to reflect uncertainty, the maximum sequestration is limited to 200 MtCO$_2$/a throughout Europe, with a fixed cost of 10 EUR/tCO$_2$ sequestered. To assess the influence of these values, sensitivity analyses of the underground sequestration potential and cost are presented in Figures~\ref{supplemental:figure_underground_seq_potential_co2_capture_vs_co2_sequestration_conversion} and \ref{supplemental:figure_underground_seq_cost_co2_capture_vs_co2_sequestration_conversion}, respectively.

\subsubsection{Afforestation}
Afforestation refers to the establishment of forests in regions without prior significant tree cover. It is generally implemented through assisted natural regeneration, agroforestry, or active restoration, each differing by the degree of human intervention and purpose \cite{Lark2023}. In this study, PyPSA-Eur was extended to represent the afforestation process in Europe, with active restoration selected as the preferred method, as it often aligns with biodiversity-focused projects and long-term ecological management. This extension spatially resolves the afforestation CO$_2$ removal potential for each of the 90 nodes across the continent.

\begin{figure}[H]
    \centering
    \includegraphics[width = 0.84\textwidth]{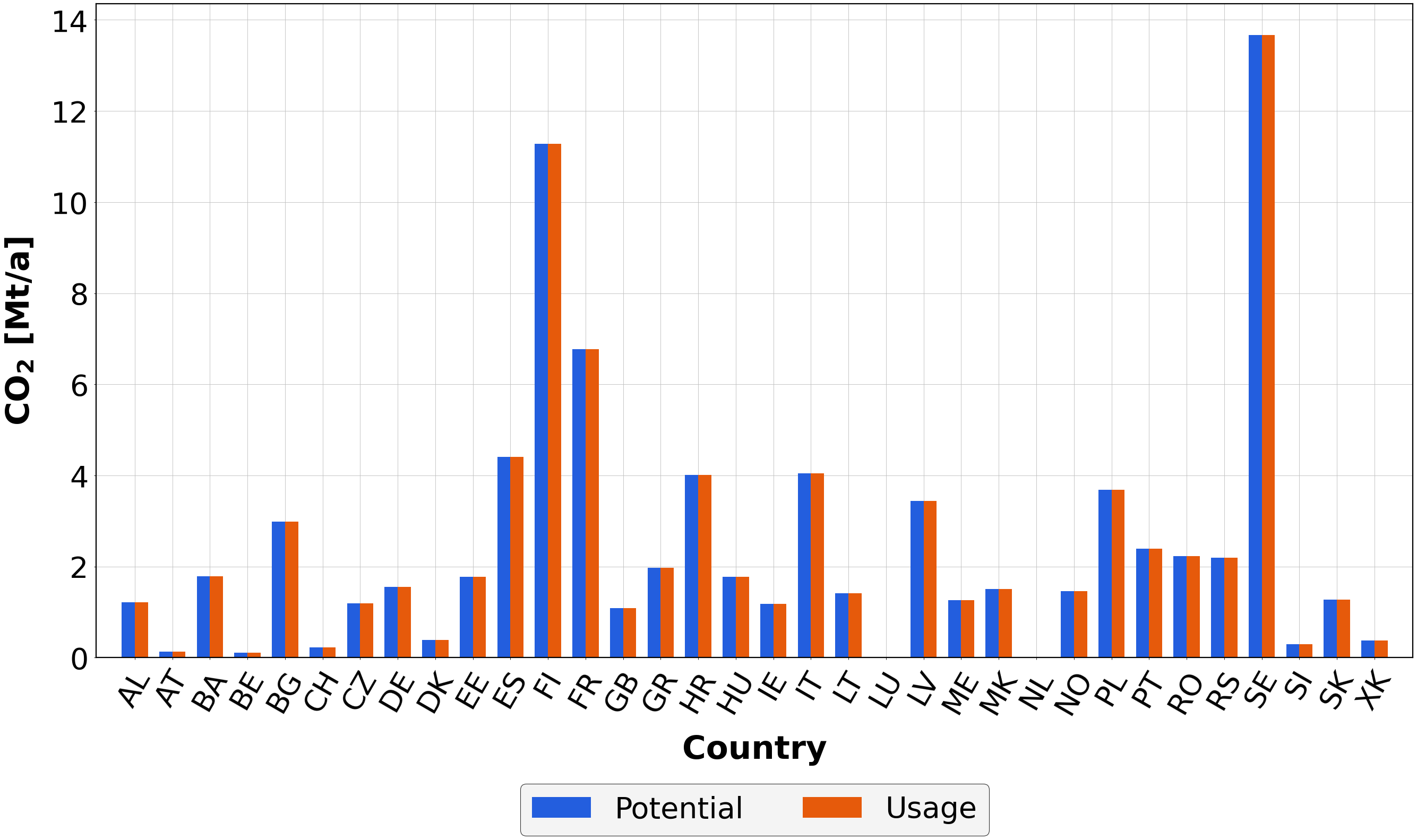}
    \caption[Afforestation CO$_2$ removal potential and usage per country]{\textbf{Afforestation CO$_2$ removal potential and usage per country}. Afforestation potential is fully used in every country in a climate-neutral energy system equipped with all CDR strategies (i.e. underground sequestration, afforestation, perennialisation, biochar, and ERW), amounting to a total of 83 MtCO$_2$/a across Europe.}
    \label{supplemental:figure_afforestation_co2_store_potential_and_usage}
\end{figure}

\begin{figure}[H]
    \centering
    \includegraphics[width = 0.84\textwidth]{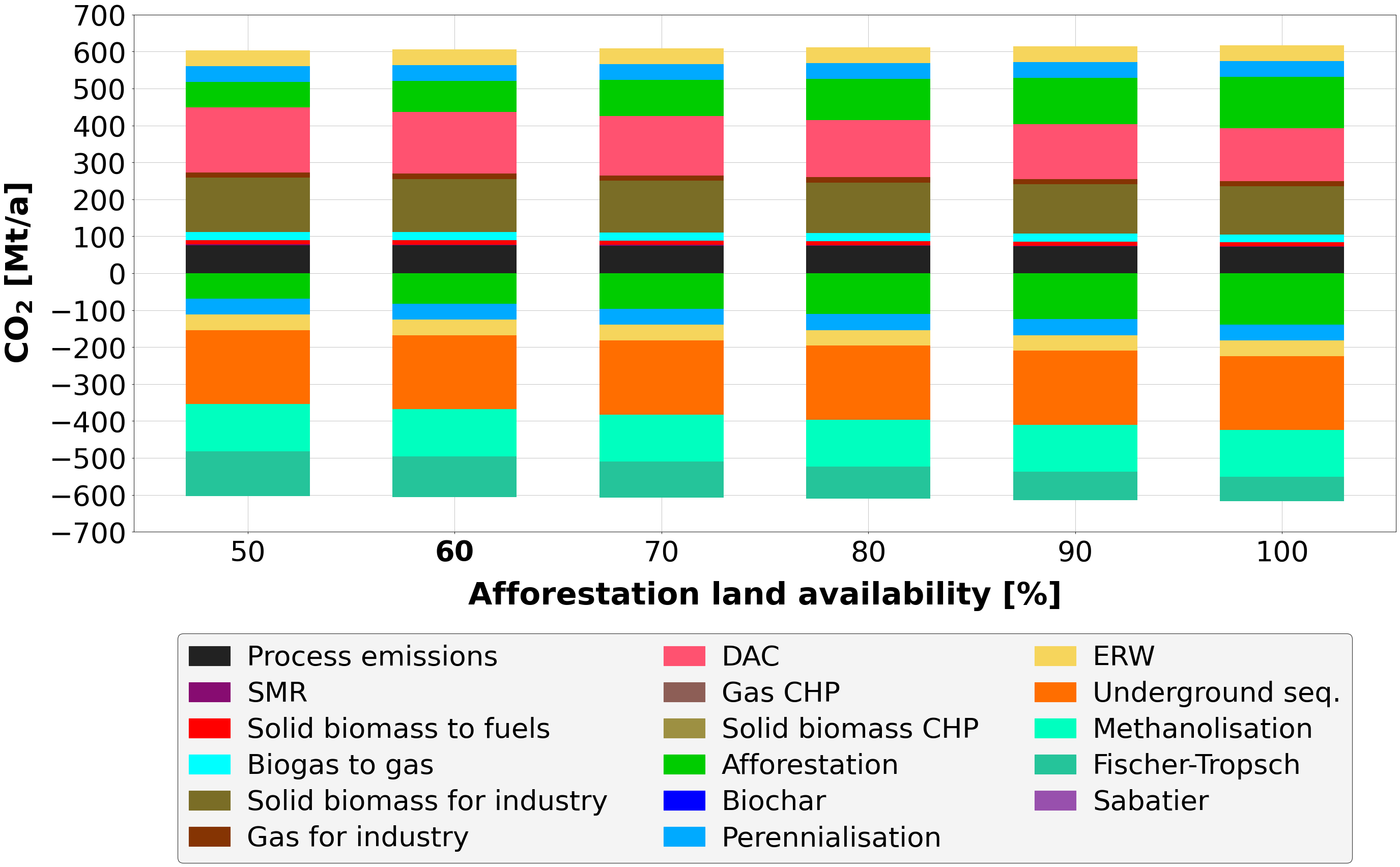}
    \caption[Afforestation land availability sensitivity analysis]{\textbf{Afforestation land availability sensitivity analysis}. Increasing the land available for afforestation leads to higher levels of CO$_2$ removal through this strategy. While this increase does not affect other CDR strategies through direct competition, it reduces reliance on DAC and lowers synthetic oil production via Fischer–Tropsch, as less captured CO$_2$ remains to be managed.}
    \label{supplemental:figure_afforestation_land_availability_co2_capture_vs_co2_sequestration_conversion}
\end{figure}

The total land available for afforestation was estimated using the 250m-resolution CORINE Land Cover dataset\cite{corine_land_cover} with the class \enquote{transitional woodland/shrub} selected due to its suitability for young plantations. To reflect potential limitations on effective CO$_2$ sequestration in young forests, such as partial land use, establishment failure, and tree mortality, only 60\% of the estimated afforestable land was considered realisable. To assess the influence of this consideration, a sensitivity analysis of the afforestable land availability is presented in Figure~\ref{supplemental:figure_afforestation_land_availability_co2_capture_vs_co2_sequestration_conversion}. Furthermore, the CO$_2$ sequestration capacity of a forest depends on its biomass density, which varies with tree species composition and geographic conditions. To address the uncertainties associated with the biomass density of young forests across Europe, the model uses country-level biomass density data from mature forests, as reported by \cite{Avitabile2024} and summarised in Table~\ref{supplemental:table_afforestation_land_potential_biomass_density_and_sequestration_potential}. Given that most of the CO$_2$ is captured during the initial growth phase and that young forests have an average age of 30 years, it is assumed that 1/30 of the biomass density grows annually. Combining these factors (Equation~\ref{supplemental:material_afforestation_potential}), a maximum CO$_2$ removal potential of afforestation of 83 MtCO$_2$/a is estimated for the whole of Europe (Figures~\ref{figure_cdr_potential} and \ref{supplemental:figure_afforestation_co2_store_potential_and_usage}). This potential is consistent with previous studies \cite{Verkerk2022, arbonics2024}, which report values ranging from 72 to 90 MtCO$_2$/a.

\begin{figure}[H]
    \centering
    \includegraphics[width = 0.84\textwidth]{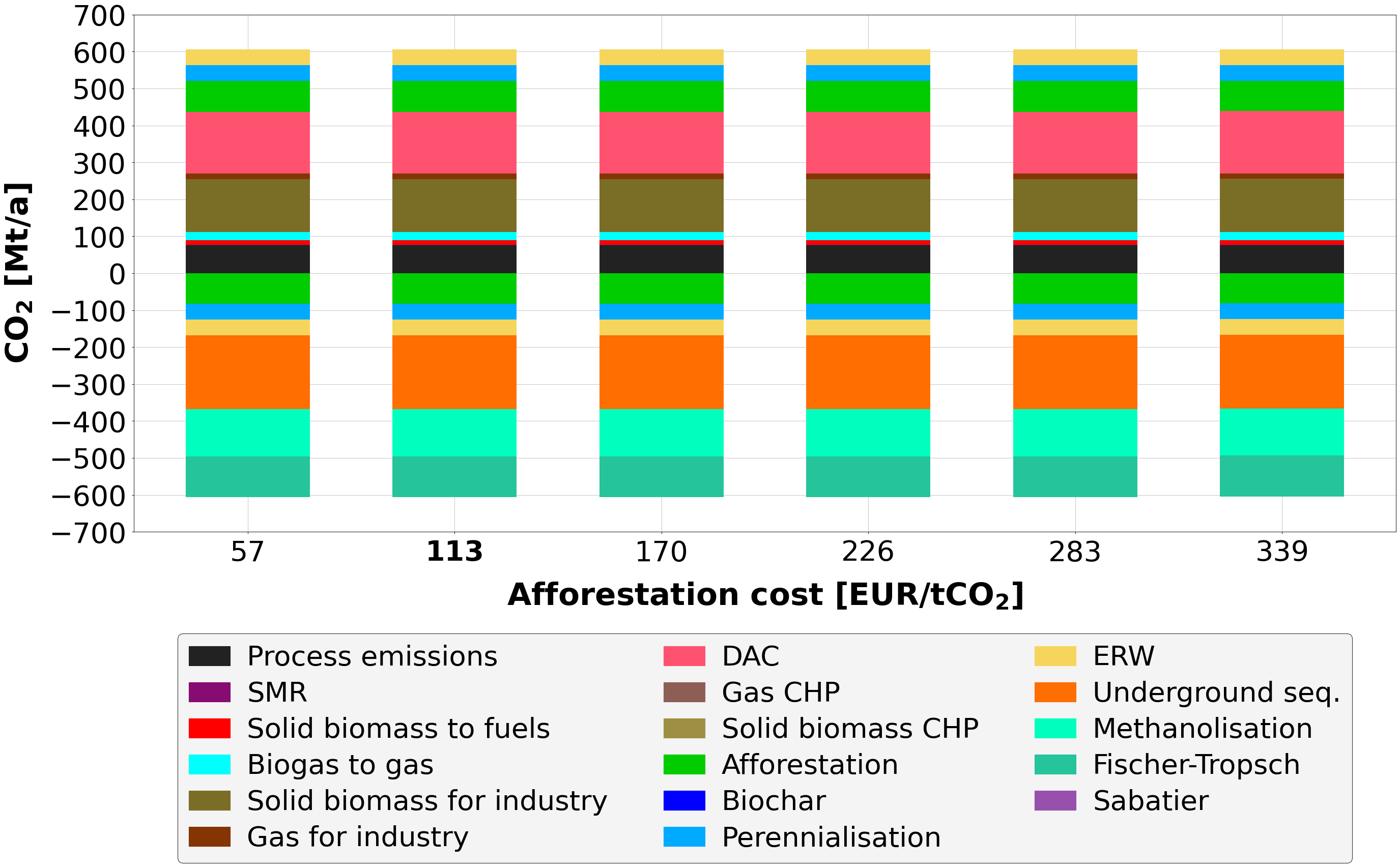}
    \caption[Afforestation cost sensitivity analysis]{\textbf{Afforestation cost sensitivity analysis}. In a climate-neutral energy system equipped with all CDR strategies, the levels of CO$_2$ capture, conversion, and sequestration across the various technologies and processes remain unchanged, regardless of the considered cost of afforestation.}
    \label{supplemental:figure_afforestation_cost_co2_capture_vs_co2_sequestration_conversion}
\end{figure}

Afforestation cost estimates from the literature span a wide range depending on implementation method, geographic context, and project objectives. These estimates have been summarised by Bodin et al. \cite{bodin_2022}, who report costs for active restoration—characterised by intensive interventions such as planting and maintenance—ranging from 95 to 23247 EUR/ha. Based on this range, an investment cost of 7749 EUR/ha was assumed for afforestation, corresponding to one third of the reported upper bound for active restoration. Annual maintenance cost was set to 388 EUR/ha, corresponding to one third of the 1165 EUR/ha reported in the same study. Assuming a land-weighted average biomass yield of 93.6 t/ha for European forests \cite{Avitabile2024} and a sequestration factor of 1.83 tonnes of CO$_2$ per tonne of biomass (derived from a dry biomass carbon content of 50\% and the stoichiometric ratio of CO$_2$ to carbon), the resulting carbon removal cost is approximately 113 EUR/tCO$_2$. To assess the influence of this result, a sensitivity analysis of the afforestation cost is presented in Figure~\ref{supplemental:figure_afforestation_cost_co2_capture_vs_co2_sequestration_conversion}.

\subsubsection{Perennialisation}
\begin{figure}[H]
    \centering
    \includegraphics[width = 0.84\textwidth]{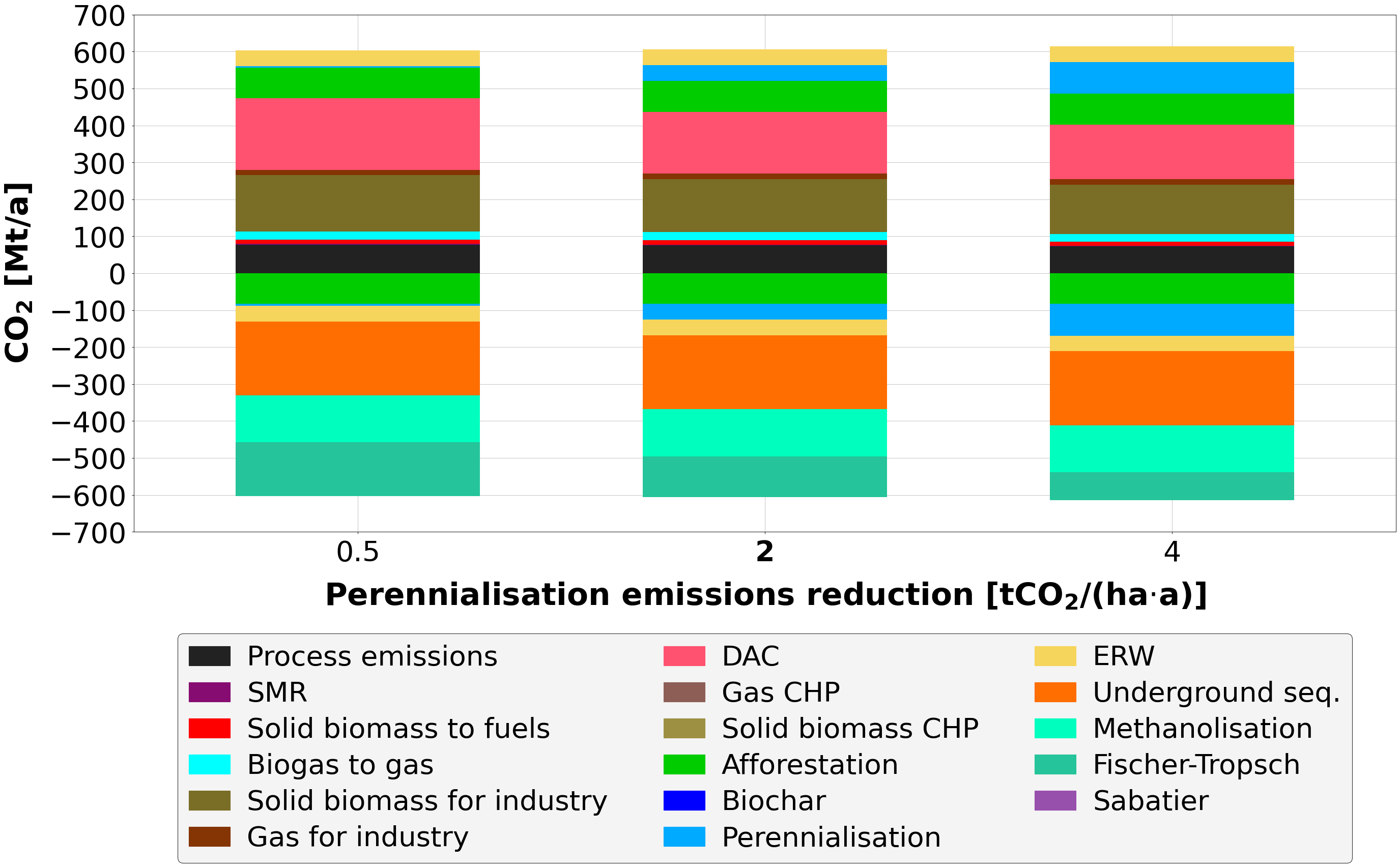}
    \caption[Perennialisation emissions reduction sensitivity analysis]{\textbf{Perennialisation emissions reduction sensitivity analysis}. Increasing the CO$_2$ emissions reduction per hectare of perennialisation leads to higher levels of CO$_2$ removal through this strategy. While this increase does not affect other CDR strategies through direct competition, it reduces reliance on DAC and lowers synthetic oil production via Fischer–Tropsch, as less captured CO$_2$ remains to be managed.}
    \label{supplemental:figure_perennials_emissions_reduction_co2_capture_vs_co2_sequestration_conversion}
\end{figure}

We investigated the introduction of perennial crops combined with green biorefining as a strategy to enhance CO$_2$ sequestration, reduce emissions, and increase the supply of bioenergy to the European energy system. Perennial crops such as clover, miscanthus, and switchgrass are an increasingly relevant solution for the production of food and feed with positive environmental impact, as they continue growing for multiple years, developing extensive root systems, which reduce fertiliser requirements and increase soil organic matter, thereby contributing to greater carbon sequestration compared to seasonal crops \cite{Jorgensen2021, GAFFEY2023108168, Glover2010, Cox2010513}. The literature estimates a combined emissions reduction ranging from 0.5 to 4 tCO$_{2e}$/(ha$\cdot$a) \cite{Jorgensen2021, Means_Crews_Souza_2022}, depending on the type of perennial crops and fertilisation levels. In this study, an average value of 2 tCO$_{2e}$/(ha$\cdot$a) was assumed for Europe as a whole, corresponding to the replacement of maize with a grass-clover mixture based on \cite{Jorgensen2021}. To assess the influence of this assumption, a sensitivity analysis of the CO$_2$ emissions reduction is presented in Figure~\ref{supplemental:figure_perennials_emissions_reduction_co2_capture_vs_co2_sequestration_conversion}.

Perennial crops have traditionally been used as fodder for ruminants. However, recent advances in green biorefinery technologies \cite{GAFFEY2023108168, Jorgensen2021, CORONA2018100, MCENIY2014335, Moller2021, SANTAMARIAFERNANDEZ2018769} enable the extraction of high-quality protein concentrate, allowing perennials to serve as feed for monogastric animals and as food ingredients for human consumption. In addition, the green biorefining process generates substantial fibre-rich side streams that can be utilised as bioenergy feedstock. Here, a simplified two-product concept is assumed, in which perennial crops are converted into protein concentrate, while the residual pulp is used for biogas production. Based on \cite{Jorgensen2021, ANDRADE2022877, CHAN2024167943, ALLANANDRADE2023124887}, we estimated that one dry tonne of perennials yields approximately 0.19 tonnes of grass protein concentrate and 0.2 MWh of biogas, while requiring 0.073 MWh of electricity for processing. This translates into a European average of  0.86 MWh of methane produced and 0.32 MWh of electricity consumed per tonne of CO$_2$ avoided.

The potential for introducing perennial crops and green biorefining in Europe depends strongly on the availability of arable land and competition with alternative land uses \cite{CORONA2018344, GAFFEY2023108168}. Here, we assumed that first-generation biofuel crops are no longer cultivated in Europe by 2050 and that perennial crops can fully replace the land previously allocated to these crops. The associated CO$_2$ sequestration potential and biogas production resulting from this land-use conversion are calculated according to Equations~\ref{eq:perennials_seq} and \ref{eq:perennials_DM}. The calculation is based on biofuel potentials from the JRC ENSPRESO \cite{RUIZ2019100379} medium availability scenario and yield data for both biofuel and perennial crops from the Eurostat database on EU crop production at standard humidity \cite{Eurostat_apro_cpshr}, aggregated at the NUTS 2 level. Under these assumptions, the total arable land available for perennial crops in Europe amounts to 21 Mha, corresponding to a total CO$_2$ removal potential of 43 MtCO$_2$/a and a biogas production of 37 TWh/a.

No costs were assumed for the conversion of cropland or changes in farming operations. The capital cost of the green biorefinery, including the additional biogas production capacity, was estimated at 6.05 MEUR/(tCO$_{2e}$/h) based on \cite{ANDRADE2022877, Jorgensen2021}. The facility is assumed to operate at 4200 full-load hours per year, reflecting the harvest period of perennials between May and October and the requirement to be processed fresh \cite{Jorgensen2021, GAFFEY2023108168}. Operational costs were estimated at 190 EUR/tCO$_2$ on average in Europe. These costs include the purchase of perennial crops from local farmers at 132 EUR/t$_{DM}$ and account for revenues from the sale of grass protein concentrate, valued at 496 EUR/t based on the market price of non-genetically modified organism (GMO) soybean \cite{Jorgensen2021}.

\subsubsection{Biochar}
Biochar is a carbon-rich material produced through the pyrolysis of agricultural and forestry residues. Thanks to its chemical and structural stability, it is resistant to microbial degradation and physical decomposition, allowing carbon to remain sequestered in soils for centuries \cite{Lehmann2006}.

Although long-term empirical data on biochar stability remain limited, existing studies indicate that approximately 70–80\% of the carbon remains stored in soils for hundreds of years \cite{Fuhrman2023, Spokas2010, Tisserant2023}. Here, a biochar sequestration rate of 70\% is assumed. Stable biochar is produced through high-temperature pyrolysis (above 500 °C), which releases volatile gases and yields a carbon-rich material with a high C/H ratio \cite{Spokas2010, Neves2011}. A biochar yield of 0.06 tonnes per MWh of solid biomass is assumed \cite{danish_energy_agency_renewable_fuels}. Under these assumptions, the combined process of solid biomass pyrolysis and subsequent soil application results in the sequestration of 0.131 tCO$_2$ per MWh of solid biomass (Equation~\ref{eq:biochar_seq}).

\begin{figure}[H]
    \centering
    \includegraphics[width = 0.84\textwidth]{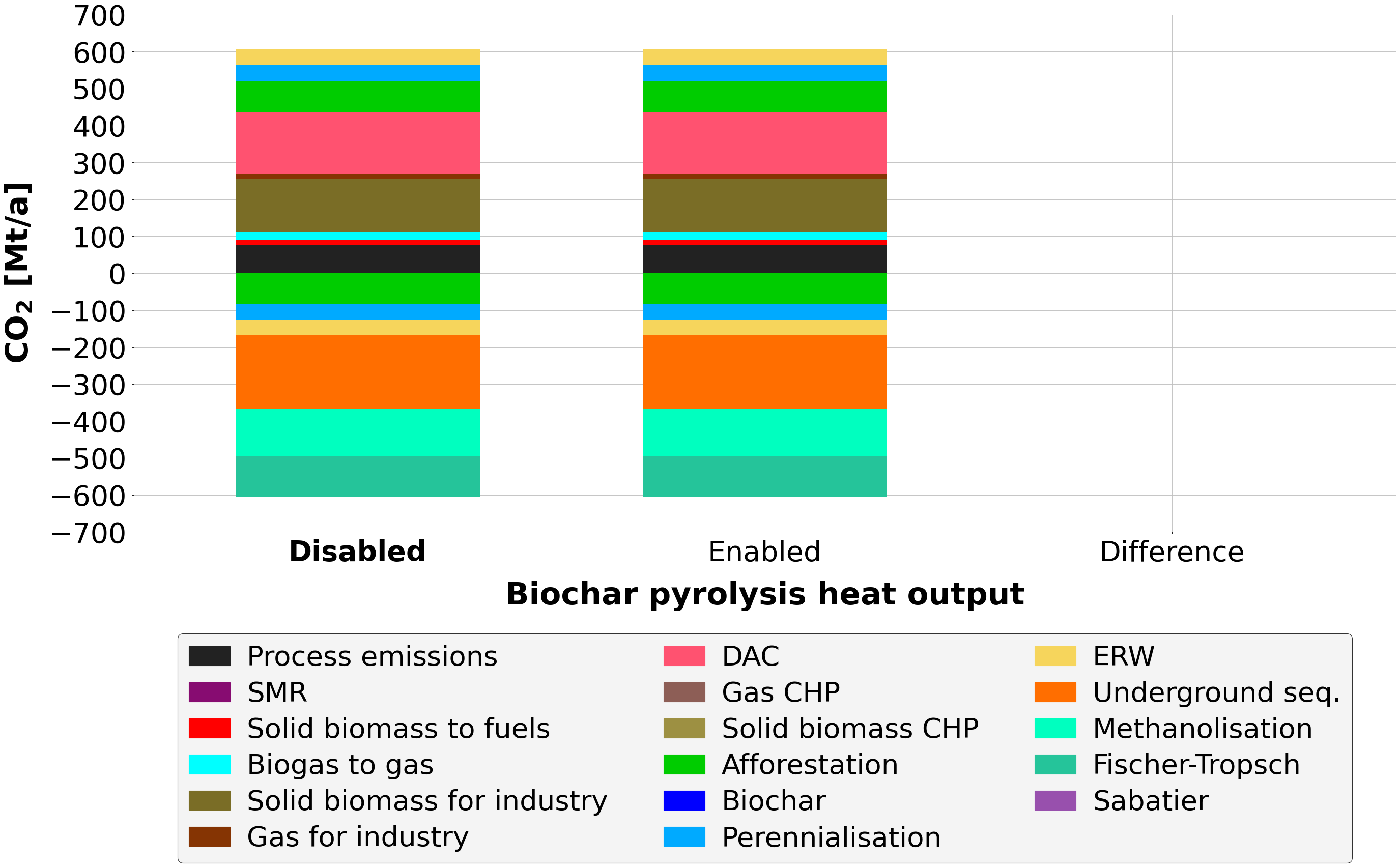}
    \caption[Biochar pyrolysis heat output sensitivity analysis]{\textbf{Biochar pyrolysis heat output sensitivity analysis}. In a climate-neutral energy system equipped with all CDR strategies, the levels of CO$_2$ capture, conversion, and sequestration across the various technologies and processes remain unchanged, regardless of whether the heat produced by biochar pyrolysis is supplied to district heating systems.}
    \label{supplemental:figure_biochar_pyrolysis_heat_output_co2_capture_vs_co2_sequestration_conversion}
\end{figure}

A valuable co-product of pyrolysis is the heat released from the combustion of pyrolysis gas, which can be supplied to district or industrial heating networks and thereby improve the economic viability of biochar production \cite{Woolf2010}. In this study, 0.49 MWh of heat per MWh of solid biomass is assumed to be available for district heating, corresponding to a heat output of 3.79 MWh$_{heat}$/tCO$_2$. In the baseline scenario, however, the heat generated from biochar pyrolysis is not supplied to district heating systems. To assess the influence of this limitation, a sensitivity analysis of the biochar pyrolysis heat supply to the heating system is presented in Figure~\ref{supplemental:figure_biochar_pyrolysis_heat_output_co2_capture_vs_co2_sequestration_conversion}. The pyrolysis process is assumed to consume 0.32 MWh$_{el}$/tCO$_2$. All technology and cost parameters for biochar pyrolysis are sourced from the Danish Energy Agency technology catalogue \cite{danish_energy_agency_renewable_fuels}, using straw as the reference feedstock.

\begin{figure}[H]
    \centering
    \includegraphics[width = 0.84\textwidth]{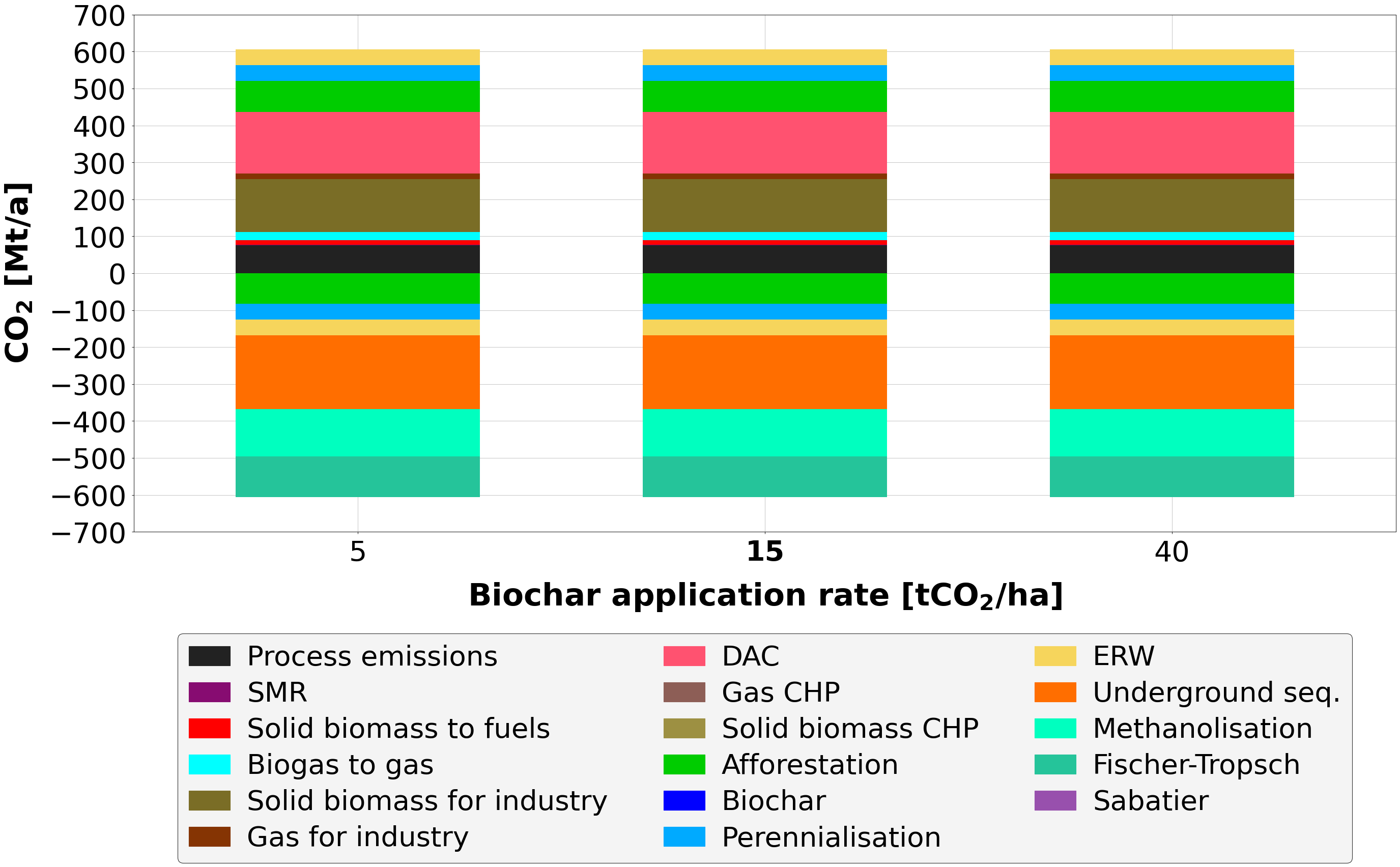}
    \caption[Biochar application rate sensitivity analysis]{\textbf{Biochar application rate sensitivity analysis}. In a climate-neutral energy system equipped with all CDR strategies, the levels of CO$_2$ capture, conversion, and sequestration across the various technologies and processes remain unchanged, regardless of the considered application rate of biochar.}
    \label{supplemental:figure_biochar_application_rate_co2_capture_vs_co2_sequestration_conversion}
\end{figure}

The CO$_2$ removal potential of biochar per NUTS 2 region in Europe was estimated using Equation~\ref{eq:biochar_potential}, assuming that biochar application is restricted to the total arable land identified in the CORINE Land Cover dataset \cite{corine_land_cover}. An application rate of biochar to soil of 15 t/ha was assumed as the base case. This parameter is highly dependent on soil type and climate, with reported literature values ranging from 5 to 40 t/ha \cite{Fuhrman2023, He2024, WIEDNER2013264, VACCARI2011231}. To assess the influence of this assumption, a sensitivity analysis of the biochar application rate is presented in Figure~\ref{supplemental:figure_biochar_application_rate_co2_capture_vs_co2_sequestration_conversion}. Land availability was further limited to 20\% of total arable land and the annual sequestration potential was calculated assuming 25 years, reflecting both a strategy to preserve carbon sinks beyond 2050 and the time required to scale up biochar production. Considering that total arable land in Europe amounts to 23 Mha, these assumptions result in a total CO$_2$ removal potential of 30 MtCO$_2$/a through biochar application (Equation~\ref{eq:biochar_potential}).

The costs of the pyrolysis process were derived from the Danish Energy Agency technology catalogue \cite{danish_energy_agency_renewable_fuels}, with a capital cost of 8.94 MEUR/tCO$_2$ and a variable operating and maintenance (VOM) cost of 48 EUR/tCO$_2$, excluding electricity and solid biomass costs. Biochar spreading is assumed to be integrated into standard tillage operations without incurring additional costs, and pyrolysis is assumed to take place in agricultural areas. Competition between the use of solid biomass for energy and for biochar production is optimised endogenously in the model.

\subsubsection{Enhanced rock weathering (ERW)}
Rock weathering is a natural process whereby atmospheric CO$_2$ dissolves in rainwater to form carbonic acid, which reacts with divalent metal ions, such as calcium and magnesium, present in rocks to produce bicarbonate. Runoff transports the dissolved bicarbonate to the ocean, where it remains stable for thousands of years \cite{Renforth2014}. High concentrations of divalent metal ions are found in silicate rocks, particularly mafic and ultramafic types such as basalt and dunite. ERW exploits the high weathering rates of these (ultra)basic rocks by crushing and spreading them over farmland to increase water contact and accelerate the weathering process.

To estimate the CO$_2$ removal potential of ERW in Europe, the assumptions of Strefler et al. \cite{Strefler2018} were followed. Basalt was selected as the rock source to avoid the potential release of toxic metals associated with dunite \cite{Rinder2021}. Crushed basalt is assumed to be spread on arable land classified in the CORINE Land Cover dataset as \enquote{non-irrigated arable land}, \enquote{permanently irrigated arable land}, and \enquote{rice fields} \cite{corine_land_cover}. Forests and unused land were excluded due to the absence of agricultural infrastructure, which would increase application costs. Because weathering rates depend strongly on temperature and water availability, only arable land located in regions with a mean annual temperature of at least 10 °C was considered, corresponding to 68 Mha across Europe. Of this area, a maximum of 20\% is assumed to be available for ERW deployment in 2050. Land used for ERW is assumed to remain available for other purposes.

\begin{figure}[H]
    \centering
    \includegraphics[width = 0.84\textwidth]{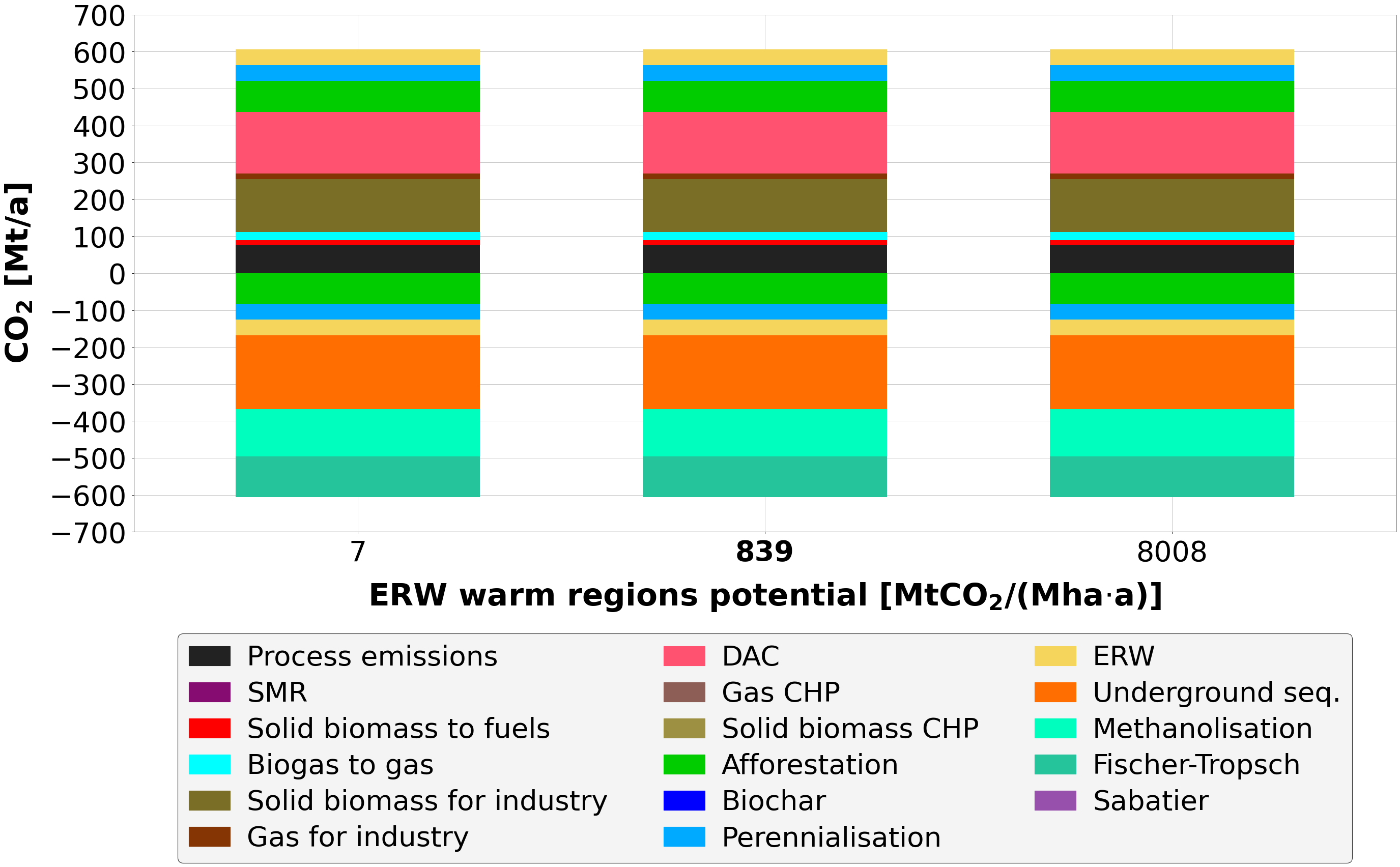}
    \caption[ERW warm regions potential sensitivity analysis]{\textbf{ERW warm regions potential sensitivity analysis}. In a climate-neutral energy system equipped with all CDR strategies, the levels of CO$_2$ capture, conversion, and sequestration across the various technologies and processes remain unchanged, regardless of the considered potential of ERW in warm regions.}
    \label{supplemental:figure_erw_warm_regions_potential_co2_capture_vs_co2_sequestration_conversion}
\end{figure}

\begin{figure}[H]
    \centering
    \includegraphics[width = 0.84\textwidth]{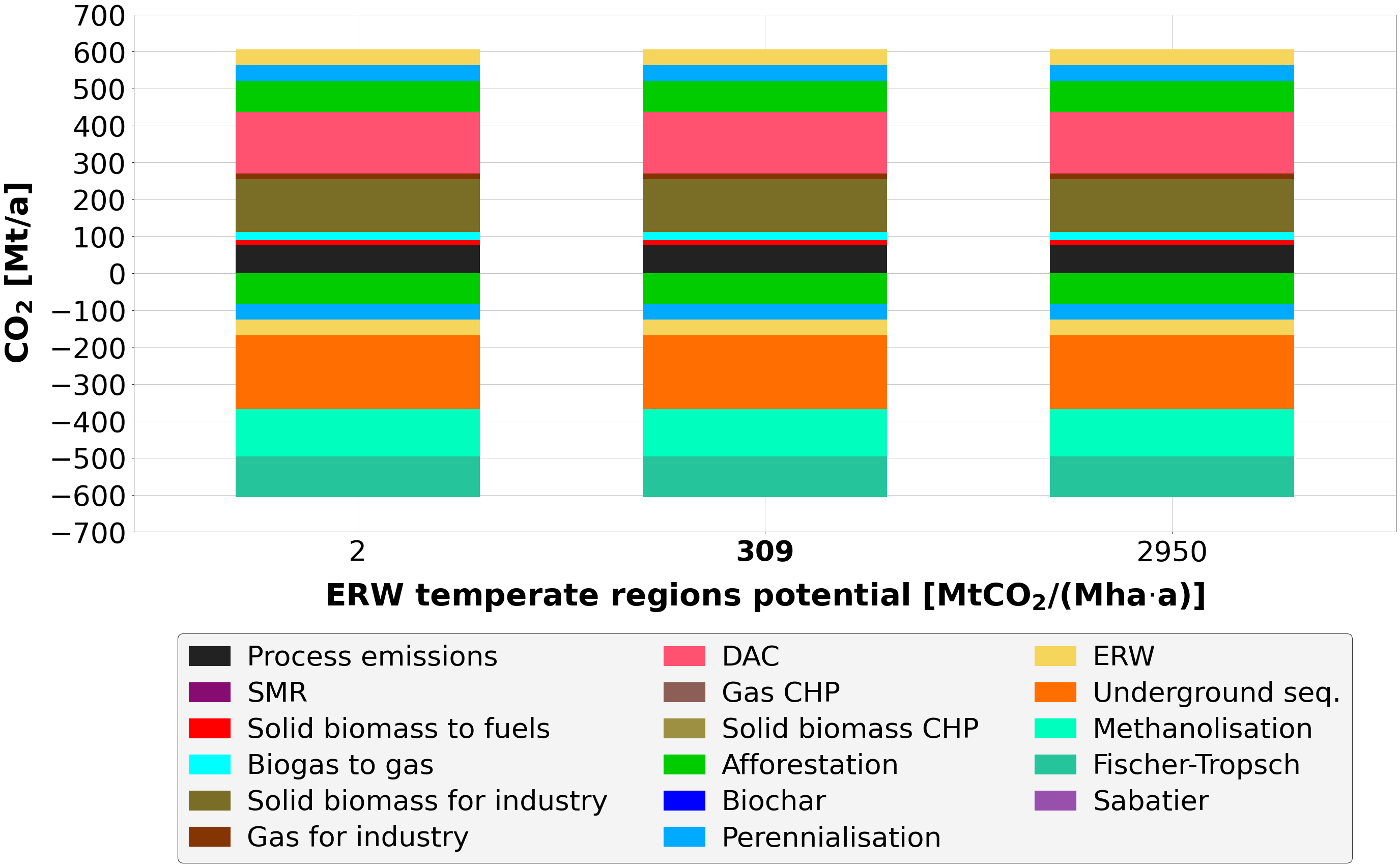}
    \caption[ERW temperate regions potential sensitivity analysis]{\textbf{ERW temperate regions potential sensitivity analysis}. In a climate-neutral energy system equipped with all CDR strategies, the levels of CO$_2$ capture, conversion, and sequestration across the various technologies and processes remain unchanged, regardless of the considered potential of ERW in temperate regions.}
    \label{supplemental:figure_erw_temperate_regions_potential_co2_capture_vs_co2_sequestration_conversion}
\end{figure}

Basalt particles with a diameter of 20 µm were assumed, providing a reasonably high weathering rate while avoiding the potential health hazards associated with finer particles \cite{Strefler2018}. Although the weathering rate is a key parameter, it remains uncertain and is the subject of ongoing research. In this study, a weathering rate based on an idealised dissolution process was adopted. As actual dissolution rates may vary by up to one order of magnitude higher or lower, their impact on the CO$_2$ removal potential was evaluated. The uncertainty in basalt weathering rates at pH 7 calculated by \cite{Strefler2018} ranges from $10^{-10.53}$ to $10^{-12.63}$ mol/(m$^2$$\cdot$s). This corresponds to a CO$_2$ removal potential range of 7-8008 MtCO$_2$/(Mha$\cdot$a) in warm regions and 2-2950 MtCO$_2$/(Mha$\cdot$a) in temperate regions. To assess the influence of these ranges, sensitivity analyses of ERW potential in warm and temperate regions are presented in Figures~\ref{supplemental:figure_erw_warm_regions_potential_co2_capture_vs_co2_sequestration_conversion} and \ref{supplemental:figure_erw_temperate_regions_potential_co2_capture_vs_co2_sequestration_conversion}, respectively. The basalt application rate was set to 150 Mt/Mha, corresponding to an 8 mm layer applied to the soil surface. Capturing one tonne of CO$_2$ requires 3.33 tonnes of basalt. Using these assumptions, the annual CO$_2$ removal potential of ERW was estimated for each region according to Equation~\ref{supplemental:material_erw_potential}. Aggregated across Europe, this yields a total removal potential of 43 MtCO$_2$/a, corresponding to a basalt demand of 141 Mt/a.

Degassing of CO$_2$ during transport of dissolved bicarbonate to the ocean is not considered, as this occurs primarily in arid regions \cite{McDermott2024}. The influence of soil pH on weathering rates is also neglected. Although basalt is globally abundant, its availability in Europe is more limited compared to other regions, and large-scale ERW deployment may therefore require alternative rock types with similar properties. Certain industrial waste products, such as cement kiln dust and crushed returned concrete, exhibit high weathering rates and low contaminant levels and could provide additional feedstock options \cite{Buckingham2024, McDermott2024}. However, their long-term availability remain highly uncertain. For simplicity, only basalt is considered in the calculations. The required annual basalt quantity is comparable in scale to the waste rock generated by European copper mining. In 2016, 0.93 Mt of copper was mined in Europe, with each tonne of ore generating 400–700 times the amount of waste rock \cite{criticalminerals}, corresponding to 372–651 Mt$_{basalt}$/a of waste rock \cite{atmineralsMetalMining}.

\begin{figure}[H]
    \centering
    \includegraphics[width = 0.84\textwidth]{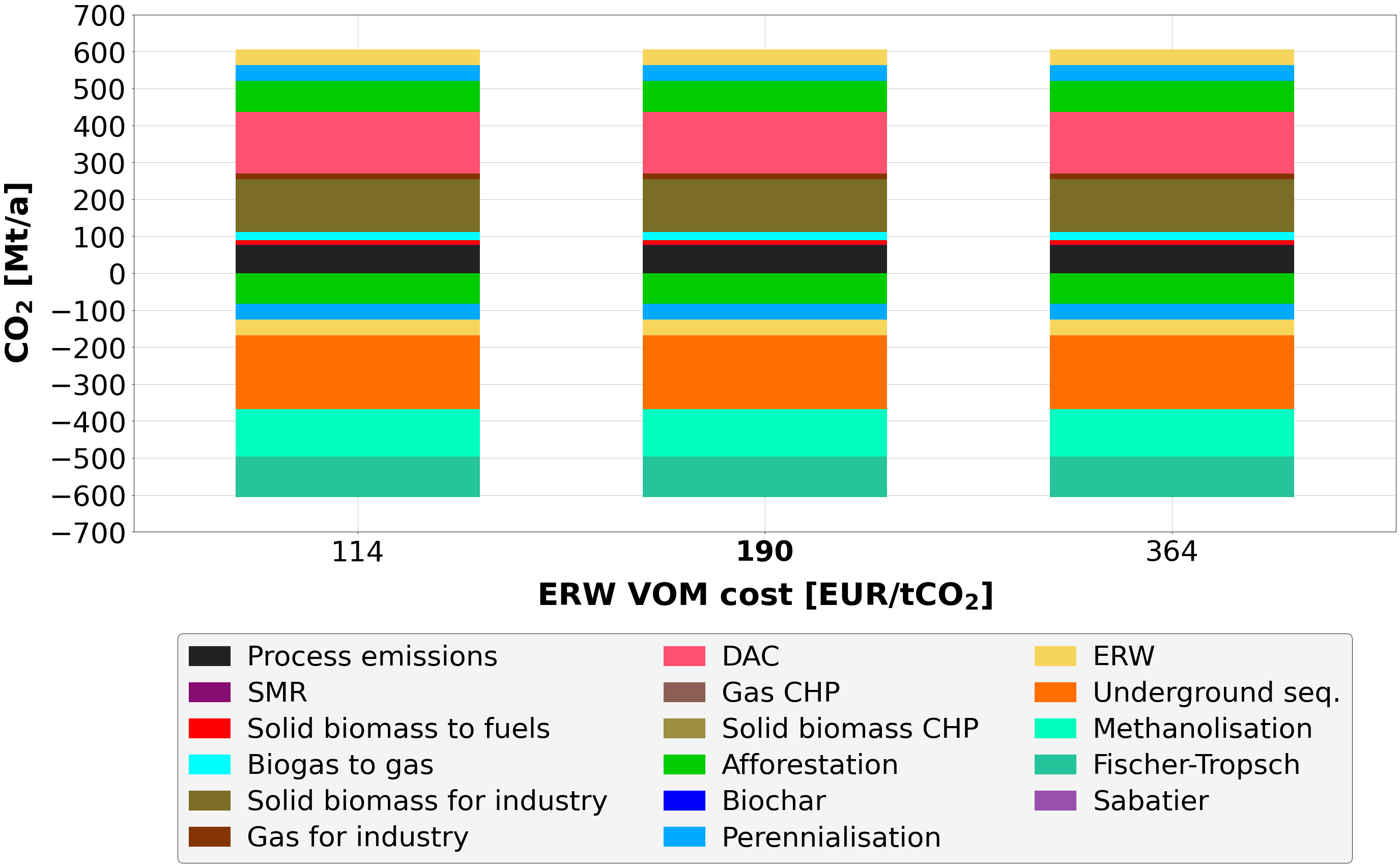}
    \caption[ERW VOM cost sensitivity analysis]{\textbf{ERW VOM cost sensitivity analysis}. In a climate-neutral energy system equipped with all CDR strategies, the levels of CO$_2$ capture, conversion, and sequestration across the various technologies and processes remain unchanged, regardless of the considered VOM cost of ERW.}
    \label{supplemental:figure_erw_vom_cost_co2_capture_vs_co2_sequestration_conversion}
\end{figure}

\begin{figure}[H]
    \centering
    \includegraphics[width = 0.84\textwidth]{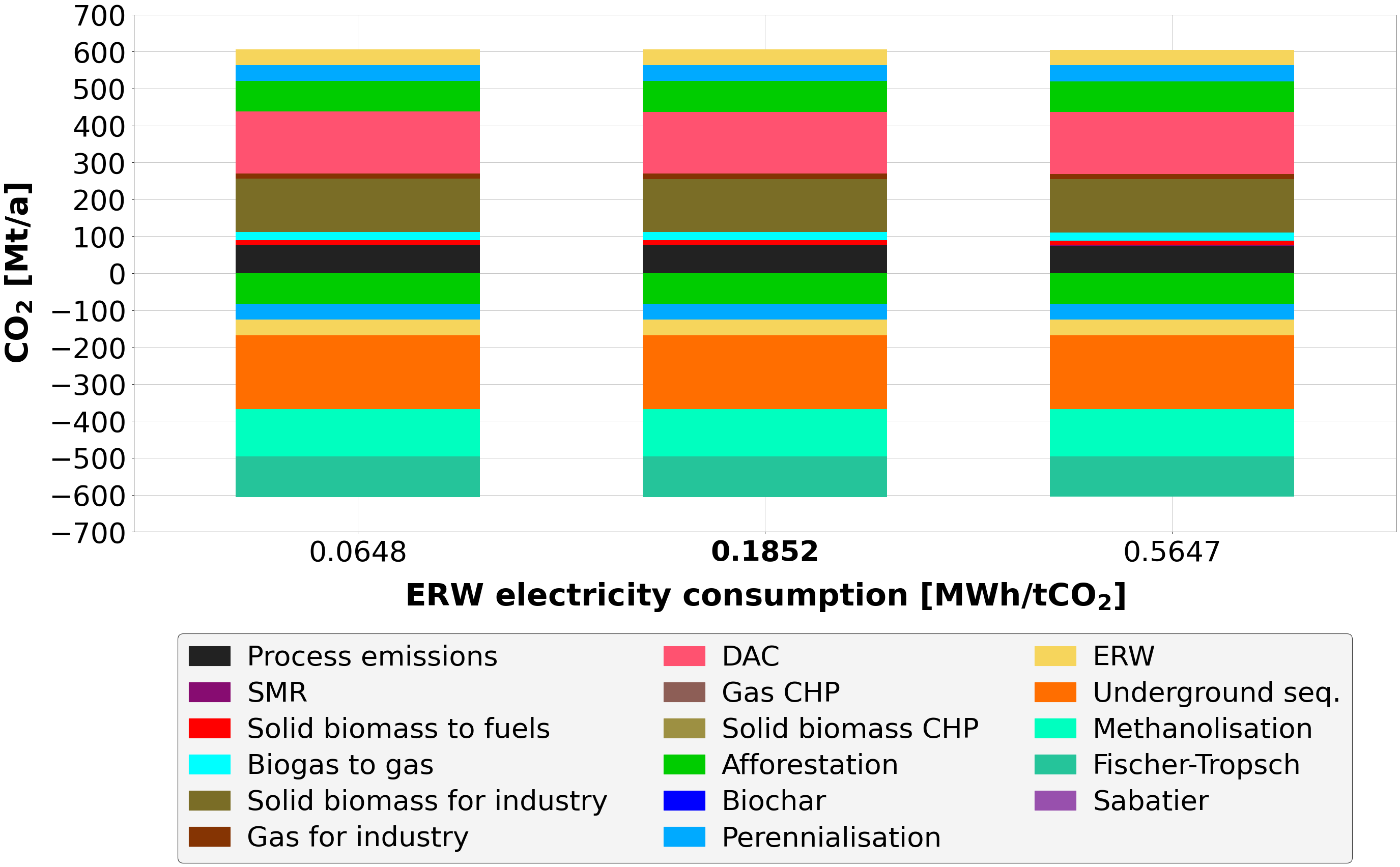}
    \caption[ERW electricity consumption sensitivity analysis]{\textbf{ERW electricity consumption sensitivity analysis}. In a climate-neutral energy system equipped with all CDR strategies, the levels of CO$_2$ capture, conversion, and sequestration across the various technologies and processes remain unchanged, regardless of the considered electricity consumption of ERW.}
    \label{supplemental:figure_erw_electricity_consumption_co2_capture_vs_co2_sequestration_conversion}
\end{figure}

Based on \cite{Strefler2018}, a VOM cost of 190 EUR/tCO$_2$ was estimated for ERW. This value is derived from open-pit mining operations, with marginal costs including operation and maintenance as well as road transport from mine to fields. Road transport assumes an average distance of 300 km, reflecting the estimate that 80\% of arable land in temperate climates lies within this distance of a suitable rock source \cite{Strefler2018}. The spreading of crushed basalt on the fields is accounted as part of the VOM cost. Electricity consumption for comminution to a particle size of 20 µm is estimated at 0.1852 MWh/tCO$_2$. To assess the influence of these estimations, sensitivity analyses of ERW VOM cost and electricity consumption are presented in Figures~\ref{supplemental:figure_erw_vom_cost_co2_capture_vs_co2_sequestration_conversion}, and \ref{supplemental:figure_erw_electricity_consumption_co2_capture_vs_co2_sequestration_conversion}, respectively.

\section{Limitations}
Here, we briefly discuss the main limitations of this study. First, the model is based on an overnight greenfield optimisation. Because the transition pathway is not explicitly modelled, potential technological lock-ins and path-dependent effects during the system transformation are not captured. While the optimisation includes the existing transmission network and hydropower capacities, it excludes current fossil-based generation assets, some of which could potentially remain in operation by 2050. Second, the study relies on exogenous cost assumptions for all technologies. These costs may differ in a future climate-neutral energy system. Technological learning is not modelled endogenously, nor are differences in the cost of capital across European countries considered. Given the substantial uncertainty surrounding future financing conditions, a uniform cost of capital is assumed across Europe to avoid masking the main drivers of optimal infrastructure deployment levels and spatial allocation by uncertain financial assumptions. Third, energy demand projections for electricity, heating, transport, and industry are based on current demand levels, assuming a substantial transformation of industrial processes to reduce CO$_2$ emissions. Electricity demand increases endogenously in response to heat electrification, electrolytic H$_2$ production, and the deployment of DAC for CO$_2$ capture. Fourth, the assumed limits on land availability for afforestation, perennialisation, and ERW are conservative, and several parameters used to estimate their removal potential are highly uncertain. These assumptions may therefore significantly influence total system costs, technology deployment levels, and the spatial distribution of technologies and processes across Europe. Fifth, several other CDR strategies that currently appear promising—such as coastal ecosystem management and direct ocean capture (DOC)—are not represented in the model. Sixth, the model assumes that Europe produces all required synthetic fuels and solid biomass domestically and does not rely on imports from other regions. Finally, the analysis focuses on the cost-optimal solution and does not explore alternative near-optimal solutions. An assessment of such solutions is provided in a separate study \cite{kalweit_2026}.

\section*{Acknowledegments}
R.F. and S.K. are fully funded by The Novo Nordisk Foundation CO$_2$ Research Center (CORC) under grant number CORC005. A.A. is fully funded by Villum Fonden under grant number 40519. We sincerely thank Uffe Jørgensen and Morten Ambye-Jensen for valuable discussions on perennialisation.

\clearpage

\end{linenumbers}

\clearpage

\bibliography{bibliography.bib}

\clearpage
\onecolumn

\beginsupplement


\section{Mathematical formulation of the model}
\label{supplemental:material_model_mathematical_formulation}
The model used in our study is based on PyPSA-Eur, which relies on a linear programme to minimise the total annualised cost of the entire European sector-coupled energy system in an optimal fashion. Formally, this minimisation is represented by the following objective function:
\begin{equation*}
    \min_{G_{n,s},E_{n,s},F_{l},g_{n,s,t}}
    \left[
        \sum_{n,s} c_{n,s} \cdot G_{n,s} +
        \sum_{n,s} \hat{c}_{n,s} \cdot E_{n,s} +
        \sum_{l} c_l \cdot F_l +
        \sum_{n,s,t} o_{n,s,t} \cdot g_{n,s,t}
    \right]
\end{equation*}
where, for technology $s$ in node $n$, $c_{n,s}$ are the fixed annualised costs for generator power capacity $G_{n,s}$, $\hat{c}_{n,s}$ are the fixed annualised costs for storage energy capacity $E_{n,s}$, $c_l$ are the fixed annualised costs for capacity $F_l$ of link $l$, and $o_{n,s,t}$ are the variable costs for generation and storage dispatch $g_{n,s,t}$ at time step $t$. The linear programme also includes multiple constraints. Amongst these, the two most relevant constraints applied to our model are succinctly described next.
\newline

At every time step of the model's temporal resolution, the demand for electricity must be met by the supply, which can be expressed as follows:
\begin{equation}
    \label{supplemental:material_electricity_constraint}
    \sum_{s} g_{n,s,t} +
    \sum_{l} \alpha_{n,l,t} \cdot f_{l,t} = d_{n,t}
    \hspace{0.15cm} \leftrightarrow \hspace{0.15cm} \lambda_{n,t}
    \hspace{0.5cm} \forall n,t
\end{equation}
where the sum of generation and storage dispatch $g_{n,s,t}$ of technology $s$, added to the sum of power flow $f_{l,t}$ in link $l$ with a direction and efficiency $\alpha_{n,l,t}$, equals demand $d_{n,t}$ in every node $n$ at every time step $t$. The dual value $\lambda_{n,t}$ of this constraint represents the electricity shadow price for node $n$ at time step $t$.
\newline

In addition, the model is subject to a continental-level (Europe) CO$_2$ emissions constraint, requiring all model nodes to collectively comply with a predefined limit. This constraint can be expressed as follows:
\begin{equation}
    \label{supplemental:material_co2_constraint}
    \sum_{n,s,t} \varepsilon_{s} \frac{g_{n,s,t}}{\eta_{s}} \leq LIMIT_{CO_2}
    \hspace{0.15cm} \leftrightarrow \hspace{0.15cm} \mu_{CO_2}
\end{equation}
where the sum of CO$_2$ emissions $\varepsilon_{s}$ in tonnes per each MWh$_\text{th}$ produced by technology $s$ with efficiency $\eta_{s}$ in all nodes $n$ of the model must be equal to or lower than CO$_2$ limit $LIMIT_{CO_2}$. The dual value $\mu_{CO_2}$ of this constraint represents the CO$_2$ shadow price for the entire Europe.

\clearpage

\section{Carbon dioxide removal (CDR)}
\label{supplemental:carbon_dioxide_removal_cdr}

\subsection{Afforestation}
In the model, the annual potential for CO$_2$ removal via afforestation in each node $n$, denoted $P_{CO_2afforestation,n}$, is calculated as follows:

\begin{equation}
    \label{supplemental:material_afforestation_potential}
    \begin{aligned}
        P_{CO_2afforestation,n} \left[\frac{tCO_2}a\right] = \; & A_n \left[ha\right] \cdot \frac{D_{n}}{30} \left[\frac{t_{biomass}}{ha}\right] \\
        & \cdot C_{biomass} \left[\frac{t_C}{t_{biomass}}\right] \cdot \frac{44}{12} \left[\frac{tCO_{2}}{t_C}\right] \cdot 0.6
    \end{aligned}
\end{equation}

where $A_n$ represents the total land area available for afforestation within the node, and $D_n$ denotes the biomass density of mature forests in the node. The biomass density is divided by the typical age of young forests, which is assumed to be 30 years in Europe, to estimate annual biomass accumulation. $C_{biomass}$ represents the carbon content in solid biomass on a dry basis (50\%), while the coefficient 44/12 converts carbon mass to its CO$_2$ equivalent using the stoichiometric ratio. Finally, a conservative land-use coefficient of 0.6 is applied to reflect constraints such as partial land use and tree mortality. Aggregating the removal CO$_2$ potential via afforestation across all nodes results in a total estimated of 83.1 MtCO$_2$/a in Europe.

For simplicity, the CO$_2$ removal rate is modelled as constant throughout the year, although in reality CO$_2$ uptake associated with forest growth exhibits seasonal variability, with substantially higher removal rates during summer months.

\subsection{Perennialisation}
In the model, the annual potential for CO$_2$ removal via perennialisation in each node $n$, denoted $P_{CO_2pgbr,n}$, is calculated as follows:

\begin{equation}
    \label{eq:perennials_seq}
    \begin{aligned}
        P_{CO_2pgbr,n} \left[\frac{tCO_2}a\right] = \; & \sum_i \left(\frac{P_{i,n}\left[MWh\right]}{Y_{i,n}\left[MWh/ha\right]}\right) \cdot r_{pgbr} \left[\frac{tCO_{2e}}{ha \cdot a}\right]
    \end{aligned}
\end{equation}

where the land area allocated to biofuel crops is calculated as the ratio of energy potential for first-generation biofuels $P_{i,n}$ to the corresponding crop yields $Y_{i,n}$. The energy potential is obtained from the JRC ENSPRESO \cite{RUIZ2019100379} medium scenario for three categories $i$ representing sugar beet (bioethanol production), rapeseed and other oil crops, and starchy crops. The yields $Y_i$ are calculated from the Eurostat database on EU crop production at standard humidity \cite{Eurostat_apro_cpshr}, aggregated at the NUTS 2 level, by mapping the JRC ENSPRESO categories to the corresponding Eurostat classes. When multiple Eurostat classes correspond to a single JRC ENSPRESO category, yields are calculated as production-weighted averages based on annual output. The mean value over the period 2017–2020 was used as the representative yield for each biofuel crop in the model. $r_{pgbr}$ represents the combined emission reduction and carbon sequestration associated with converting biofuel crops to perennial crops (2 tCO$_{2e}$/(ha$\cdot$a)) \cite{Jorgensen2021}. Aggregating the removal CO$_2$ potential via perennialisation across all nodes results in a total estimated of 42.9 MtCO$_2$/a in Europe.
\newline

In addition, the annual biogas production potential from the green refining of perennial crops in each node $n$, denoted $P_{CH_4pgbr,n}$, is calculated as follows:

\begin{equation}
    \label{eq:perennials_DM}
    \begin{aligned}
        P_{CH_4pgbr,n} \left[\frac{MWh_{CH_4}}a\right] = \; & \sum_i \left(\frac{P_{i,n}\left[MWh\right]}{Y_{i,n}\left[MWh/ha\right]}\right) \cdot Y_p \left[\frac{t_{DM}}{ha \cdot a}\right]  \\
        & \cdot Y_{CH_4} \left[\frac{MWh_{CH_4}}{t_{DM}}\right]
    \end{aligned}
\end{equation}

where $Y_p$ represents the yields of perennial crops within the node, obtained from the Eurostat database. The following Eurostat classes were considered: G0000 (\enquote{Plants harvested green}), G1000 (\enquote{Temporary grasses and grazings}), G2000 (\enquote{Leguminous plants harvested green}), G2100 (\enquote{Lucerne}), and G2900 (\enquote{Other leguminous plants harvested green, including clover}). For the substitution of biofuels crops, the yield of the most prolific crop was applied, and a broad range of classes was considered to account for the substantial variation in productivity across European climate conditions. $Y_{CH_4}$ represents the amount of biogas produced per dry tonne of perennial crops (0.2 MWh$_{CH_4}$/t$_{DM}$ \cite{Jorgensen2021, ANDRADE2022877,CHAN2024167943, ALLANANDRADE2023124887}). Aggregating the biogas production potential via green refining of perennial crops across all nodes results in a total estimated of 36.8 TWh/a in Europe.

In the model, CO$_2$ removal and biogas production are temporally coupled for simplicity, such that annual CO$_2$ removal occurs only during the summer months when perennial crops are harvested and biogas is produced. This modelling choice preserves the temporal correlation between biogas production and energy demand.

\subsection{Biochar}
In the model, the annual potential for CO$_2$ removal via biochar in each node $n$, denoted $P_{CO_2biochar,n}$, is calculated as follows:

\begin{equation}
    \label{eq:biochar_potential}
    \begin{aligned}
        P_{CO_2biochar,n} \left[\frac{tCO_2}a\right] = \; & A_n \left[ha\right] \cdot C_{char} \left[\frac{t_C} {t_{char}}\right] \cdot a_{char} \left[\frac{t_{char}}{ha}\right] \\
        & \cdot r_{char} \left[\frac{t_{Cseq}}{t_{Cchar}}\right] \cdot f \left[a^{-1}\right] \cdot \frac{44}{12} \left[\frac{tCO_2}{t_C}\right] \cdot 0.2
    \end{aligned}
\end{equation}

where $A_n$ represents the total land area available for biochar within the node, corresponding to arable land classified in the CORINE Land Cover dataset \cite{corine_land_cover} as \enquote{non-irrigated arable land}, \enquote{permanently irrigated arable land}, and \enquote{rice fields}. $C_{char}$ represents the carbon content in biochar and is calculated according to Equation~\ref{eq:biochar_C_bal}. $a_{char}$ represents the biochar application rate to the soil (15 t$_{char}$/ha), and $r_{char}$ the biochar sequestration rate (70\%). $f$ is the rate used to convert the total biochar potential to an annual value (1/25 years), while the coefficient 44/12 converts carbon mass to its CO$_2$ equivalent using the stoichiometric ratio. A conservative land-use coefficient of 0.2 is applied to reflect that only a fraction of the available arable land is used for biochar application. Aggregating the removal CO$_2$ potential via biochar across all nodes results in a total estimated of 30.2 MtCO$_2$/a in Europe.
\newline

The properties of solid biomass in our model are based on the aggregation of several categories of forest and agricultural residues in the JRC ENSPRESO \cite{RUIZ2019100379} dataset, namely MINBIOAGRW1, MINBIOFSR1, MINBIOWOO1, MINBIOWOO1a, MINBIOFSR1a, and MINBIOMUN1. To maintain coherence between the technology input parameters from the Danish Energy Agency \cite{danish_energy_agency_renewable_fuels} and the definition of solid biomass in the model, the carbon content of the resulting biochar was calculated using the carbon balance of the pyrolysis process, as expressed in the following equation:

\begin{equation}
    \label{eq:biochar_C_bal}
    \begin{aligned}
        C_{char} \left[\frac{t_C}{t_{char}}\right] = \; & C_{biomass} \left[\frac{t_{C b}}{t_b}\right] \cdot \left(1 - moist \left[\frac{t_{H_2O}}{t_b}\right]\right) \cdot \frac{1}{LHV_b} \left[\frac{t_b}{MWh_b}\right] \\
        & \cdot \frac{1}{Y_{char}} \left[\frac{MWh_b}{t_{char}}\right] \cdot Y_{C,pyro} \left[\frac{t_{C char}}{t_{C b}}\right]
    \end{aligned}
\end{equation}

where $C_{biomass}$ represents the carbon content in solid biomass on a dry basis (50\%), assuming an average moisture content of 18\%. $LHV$ is the assumed lower heating value (18 GJ/t$_b$). The Danish Energy Agency provides data about the ratio of carbon in the solid biomass retained in the biochar, Y$_{C,pyro}$ (estimated at 50\%), as well as the biochar output expressed on an energy basis. As the agency assumes an input moisture content of 13\% to the pyrolysis reactor, the additional heat required for biomass drying (0.83 MWh$_{heat}$/t$_{H_2O}$ \cite{ALAMIA201592}) was subtracted from the district heating production. This results in 0.85 t$_C$/t$_{char}$, which is consistent with values reported in the literature for high-temperature pyrolysis \cite{Neves2011}.
\newline

The process of sequestering carbon from solid biomass through biochar is quantified as follows:

\begin{equation}
    \label{eq:biochar_seq}
    \begin{aligned}
        S_{CO_2char} \left[\frac{tCO_2}{MWh_{biomass}}\right] = \; & Y_{char} \left[\frac{t_{char}}{MWh_{biomass}}\right] \cdot C_{char} \left[\frac{t_C}{t_{char}}\right] \\
        & \cdot r_{char} \left[\frac{t_{C seq}}{t_{C char}}\right] \cdot \frac{44}{12} \left[\frac{tCO_2}{t_C}\right]
    \end{aligned}
\end{equation}

where $Y_{char}$ represents the yield of biochar from the pyrolysis process (0.06 t$_{char}$/MWh$_{biomass}$), $C_{char}$ the carbon content of the biochar (0.85 t$_C$/t$_{char}$), and $r_{char}$ the biochar sequestration rate (70\%). The coefficient 44/12 converts carbon mass to its CO$_2$ equivalent using the stoichiometric ratio. This results in the sequestration of 0.131 tCO$_2$ per MWh of solid biomass.
\newline

In the model, CO$_2$ removal via biochar is assumed to occur simultaneously with the pyrolysis process, as the subsequent spreading of biochar onto fields is not explicitly represented. This assumption is adopted to preserve a strict temporal coupling between heat production from pyrolysis and heat demand in the rest of the energy system.

\subsection{Enhanced rock weathering (ERW)}
In the model, the annual potential for CO$_2$ removal via ERW in each node $n$, denoted $P_{CO_2ERW,n}$, is calculated as follows:

\begin{equation}
    \label{supplemental:material_erw_potential}
    \begin{aligned}
        P_{CO_2ERW,n} \left[\frac{tCO_2}a\right] = \; & \left(0.95 \cdot A_{warm,n} + 0.35 \cdot A_{temp,n}\right) \left[ha\right] \cdot a_{basalt} \left[\frac{t_{basalt}} {ha}\right] \\
        & \cdot r_{basalt} \left[\frac{tCO_{2}}{t_{basalt}}\right] \cdot SSA \left[\frac{m^2}{g_{basalt}}\right] \cdot WR_{grain} \left[\frac{g_{basalt}}{m^2\cdot s}\right] \cdot t \left[\frac{s}{a}\right] \cdot 0.2
    \end{aligned}
\end{equation}

where $A_{warm,n}$ is the total land area in warm regions and $A_{temp,n}$ is the total land area in temperate regions available for ERW within the node. This corresponds to 0.3 Mha in warm regions and 67.6 Mha in temperate regions across Europe as a whole. Each area is adjusted using a specific coefficient (0.95 and 0.35, respectively) to account for the fact that the other weathering parameters were calculated for a region with an average temperature of 25 °C. $a_{basalt}$ represents the amount of basalt that can be spread per hectare (150 t$_{basalt}$/ha), whereas $r_{basalt}$ represents the amount of CO$_2$ that can be captured per tonne of crushed basalt (0.3 tCO$_2$/t$_{basalt}$). $SSA$ represents the specific surface area of the basalt grain with a diameter of 20 µm (69.18$\cdot$20$^{-1.24}$ m$^2$/g$_{basalt}$), $WR_{grain}$ its weathering rate at pH 7 (125$\cdot$10$^{-10.53}$ g$_{basalt}$/(m$^2$$\cdot$s)), and $t$ the number of seconds in a year (3.155$\cdot$10$^7$ s/a). Finally, a conservative land-use coefficient of 0.2 is applied to reflect that only a fraction of the available land is used for spreading crushed basalt. Aggregating the removal CO$_2$ potential via ERW across all nodes results in a total estimated of 42.5 MtCO$_2$/a in Europe, corresponding to a basalt demand of 141 Mt/a.

In the model, the temporal activation of ERW is linked to the electricity consumption associated with basalt processing, and therefore does not explicitly represent the actual temporal dynamics of the CO$_2$ removal process.

\clearpage

\section{Figures}\label{supplemental:figures}

\vspace{20pt}

\vspace{20pt}

\vspace{20pt}

\vspace{20pt}

\vspace{20pt}

\vspace{20pt}

\vspace{20pt}

\vspace{20pt}

\vspace{20pt}

\vspace{20pt}

\vspace{20pt}

\vspace{20pt}

\vspace{20pt}

\vspace{20pt}

\vspace{20pt}

\vspace{20pt}

\vspace{20pt}

\vspace{20pt}

\vspace{20pt}

\vspace{20pt}

\vspace{20pt}

%

\vspace{20pt}

\vspace{20pt}

\vspace{20pt}

\vspace{20pt}

\vspace{20pt}

\vspace{20pt}

\vspace{20pt}

\vspace{20pt}

\vspace{20pt}

\vspace{20pt}

\vspace{20pt}

\vspace{20pt}

\vspace{20pt}

\vspace{20pt}

\clearpage

\section{Tables}
\label{supplemental:tables}

\noindent

\begin{table}[H]
    \small
    \centering
    \caption[Total cost per country of an energy system equipped with underground CO$_2$ sequestration only]{\textbf{Total cost per country of an energy system equipped with underground CO$_2$ sequestration only}. The variation refers to the increase or decrease (in percentage) of this cost compared with the cost of a climate-neutral energy system equipped with all CDR strategies (i.e. underground sequestration, afforestation, perennialisation, biochar, and ERW).}
    \label{supplemental:table_total_system_cost_per_country}
    \begin{tabular}{p{4cm} p{2.5cm} p{2.2cm}}
        \toprule
        \textbf{Country}       & \textbf{Cost [BEUR/a]} & \textbf{Variation [\%]} \\
        \midrule
        Albania                & 1.58                   & +6.3                    \\
        Austria                & 21.51                  & -7.2                    \\
        Bosnia and Herzegovina & 2.59                   & -8.5                    \\
        Belgium                & 17.74                  & -1.2                    \\
        Bulgaria               & 6.29                   & +6                      \\
        Switzerland            & 16.33                  & -0.6                    \\
        Czech Republic         & 13.64                  & +2.9                    \\
        Germany                & 127.78                 & -8.7                    \\
        Denmark                & 78.27                  & -33                     \\
        Estonia                & 3.37                   & -18.1                   \\
        Spain                  & 64.54                  & -13.9                   \\
        Finland                & 30.59                  & -14.1                   \\
        France                 & 109.7                  & -3.5                    \\
        Great Britain          & 142.98                 & -29.5                   \\
        Greece                 & 12.11                  & -11.1                   \\
        Croatia                & 5.44                   & +5.3                    \\
        Hungary                & 14.06                  & -9                      \\
        Ireland                & 7.59                   & -29.1                   \\
        Italy                  & 82.01                  & -6.8                    \\
        Lithuania              & 2.94                   & +18                     \\
        Luxembourg             & 1.34                   & -4.5                    \\
        Latvia                 & 3.53                   & +22.1                   \\
        Montenegro             & 1.45                   & +6.2                    \\
        North Macedonia        & 1.3                    & +3.8                    \\
        The Netherlands        & 44.14                  & -10.7                   \\
        Norway                 & 17.42                  & -14.1                   \\
        Poland                 & 57.04                  & -10.5                   \\
        Portugal               & 8.56                   & -2.7                    \\
        Romania                & 14.15                  & +2                      \\
        Serbia                 & 3.79                   & +3.2                    \\
        Sweden                 & 30.31                  & -9.3                    \\
        Slovenia               & 1.24                   & +4                      \\
        Slovakia               & 5.04                   & +3.8                    \\
        Kosovo                 & 0.45                   & +8.9                    \\
        \bottomrule
    \end{tabular}
\end{table}


\begin{table}[H]
    \small
    \centering
    \caption[Total cost per technology of an energy system equipped with underground CO$_2$ sequestration only]{\textbf{Total cost per technology of an energy system equipped with underground CO$_2$ sequestration only}. The variation refers to the increase or decrease (in percentage) of this cost compared with the cost of a climate-neutral energy system equipped with all CDR strategies. Technologies with annual costs below 0.1 BEUR are omitted for readability.}
    \label{supplemental:table_cost_per_technology}
    \begin{tabular}{p{5cm} p{2.5cm} p{2.2cm}}
        \toprule
        \textbf{Technology}            & \textbf{Cost [BEUR/a]} & \textbf{Variation [\%]} \\
        \midrule
        Wind                           & 284.74                 & -14.3                   \\
        Solar photovoltaics            & 171.81                 & -16.3                   \\
        H$_2$ (electrolysis and store) & 117.84                 & -38.1                   \\
        Heat pump (air and ground)     & 74.18                  & -5                      \\
        Hydroelectricity               & 43.81                  & +0                      \\
        Electricity distribution grid  & 40.99                  & -1.4                    \\
        DAC                            & 37.81                  & -37.8                   \\
        Biogas                         & 32.46                  & +3                      \\
        Battery storage                & 27.85                  & -20.8                   \\
        Fischer-Tropsch                & 27.61                  & -56.9                   \\
        Solid biomass                  & 18                     & +1.1                    \\
        Transmission lines             & 11.75                  & +0                      \\
        Gas                            & 10.63                  & +10.4                   \\
        Gas CHP                        & 9.08                   & +14.1                   \\
        Methanolisation                & 7.45                   & -5.6                    \\
        Gas boiler                     & 7.28                   & +21                     \\
        Solid biomass for industry CC  & 6.9                    & -11.9                   \\
        Hot water storage              & 5.83                   & -11.3                   \\
        Biomass boiler                 & 5.01                   & -0.8                    \\
        Solid biomass CHP              & 4.88                   & +22.7                   \\
        Oil primary                    & 4.87                   & +674.7                  \\
        Process emissions CC           & 3.66                   & -13.4                   \\
        SMR                            & 3.63                   & -20.4                   \\
        Resistive heater               & 2.86                   & -3.1                    \\
        Underground sequestration      & 2.58                   & -11.6                   \\
        Solid biomass to fuels CC      & 1.99                   & -39.7                   \\
        Gas pipeline                   & 1.06                   & +0                      \\
        Gas for industry CC            & 0.81                   & -21                     \\
        OCGT                           & 0.59                   & +13.6                   \\
        SMR CC                         & 0.54                   & -18.5                   \\
        Methanol                       & 0.31                   & -16.1                   \\
        \bottomrule
    \end{tabular}
\end{table}


\begin{table}[H]
    \small
    \centering
    \caption[Underground CO$_2$ sequestration potential per country]{\textbf{Underground CO$_2$ sequestration potential per country}. The potentials are based on the European CO$_2$ storage database (CO2StoP) and are available at \url{https://energy.ec.europa.eu/publications/assessment-co2-storage-potential-europe-co2stop_en}. The model only considers the offshore underground potential of deep salt caverns and depleted hydrocarbon reservoirs.}
    \label{supplemental:table_underground_co2_sequestration_potential}
    \begin{tabular}{p{4.3cm} p{5.7cm}}
        \toprule
        \textbf{Country}       & \textbf{Sequestration potential [MtCO$_2$]} \\
        \midrule
        Albania                & 0                                           \\
        Austria                & 0                                           \\
        Bosnia and Herzegovina & 0                                           \\
        Belgium                & 0                                           \\
        Bulgaria               & 0.4                                         \\
        Switzerland            & 0                                           \\
        Czech Republic         & 0                                           \\
        Germany                & 81.1                                        \\
        Denmark                & 845.2                                       \\
        Estonia                & 0.7                                         \\
        Spain                  & 5.1                                         \\
        Finland                & 0                                           \\
        France                 & 0                                           \\
        Great Britain          & 2255.4                                      \\
        Greece                 & 161.3                                       \\
        Croatia                & 0                                           \\
        Hungary                & 0                                           \\
        Ireland                & 18.9                                        \\
        Italy                  & 33.9                                        \\
        Lithuania              & 2.1                                         \\
        Luxembourg             & 0                                           \\
        Latvia                 & 42.9                                        \\
        Montenegro             & 0                                           \\
        North Macedonia        & 0                                           \\
        The Netherlands        & 9.4                                         \\
        Norway                 & 21.4                                        \\
        Poland                 & 1.4                                         \\
        Portugal               & 214.5                                       \\
        Romania                & 0                                           \\
        Serbia                 & 0                                           \\
        Sweden                 & 8.1                                         \\
        Slovenia               & 0                                           \\
        Slovakia               & 0                                           \\
        Kosovo                 & 0                                           \\
        Europe                 & 3701.8                                      \\
        \bottomrule
    \end{tabular}
\end{table}


\begin{table}[H]
    \small
    \centering
    \caption[Afforestation land availability, biomass density, and CO$_2$ removal potential per country]{\textbf{Afforestation land availability, biomass density, and CO$_2$ removal potential per country}. Land availability for afforestation is determined using the CORINE Land Cover dataset, and its biomass density is specified in \cite{Avitabile2024}. CO$_2$ removal potential for afforestation is calculated using Equation~\ref{supplemental:material_afforestation_potential}.}
    \label{supplemental:table_afforestation_land_potential_biomass_density_and_sequestration_potential}
    \begin{tabular}{p{3.8cm} p{2.8cm} p{2.7cm} p{3cm}}
        \toprule
        \textbf{Country}       & \textbf{Land availability [Mha]} & \textbf{Biomass density [t/ha]} & \textbf{Removal potential [MtCO$_2$/a]} \\
        \midrule
        Albania                & 0.21                             & 94.3                            & 1.2                                     \\
        Austria                & 0.01                             & 186.7                           & 0.1                                     \\
        Bosnia and Herzegovina & 0.24                             & 121.6                           & 1.8                                     \\
        Belgium                & 0.01                             & 178.7                           & 0.1                                     \\
        Bulgaria               & 0.43                             & 113.9                           & 3                                       \\
        Switzerland            & 0.02                             & 196.8                           & 0.2                                     \\
        Czech Republic         & 0.09                             & 209.9                           & 1.2                                     \\
        Germany                & 0.12                             & 204.2                           & 1.5                                     \\
        Denmark                & 0.05                             & 120.2                           & 0.4                                     \\
        Estonia                & 0.27                             & 108.9                           & 1.8                                     \\
        Spain                  & 1.16                             & 62.1                            & 4.4                                     \\
        Finland                & 2.77                             & 66.8                            & 11.3                                    \\
        France                 & 0.75                             & 147.3                           & 6.8                                     \\
        Great Britain          & 0.15                             & 115.4                           & 1.1                                     \\
        Greece                 & 0.62                             & 51.8                            & 2                                       \\
        Croatia                & 0.35                             & 189.8                           & 4                                       \\
        Hungary                & 0.2                              & 148.6                           & 1.8                                     \\
        Ireland                & 0.17                             & 117.3                           & 1.2                                     \\
        Italy                  & 0.61                             & 108                             & 4                                       \\
        Lithuania              & 0.16                             & 144.7                           & 1.4                                     \\
        Luxembourg             & 0                                & 169.8                           & 0                                       \\
        Latvia                 & 0.43                             & 131.6                           & 3.4                                     \\
        Montenegro             & 0.18                             & 117.2                           & 1.3                                     \\
        North Macedonia        & 0.26                             & 95.7                            & 1.5                                     \\
        The Netherlands        & 0                                & 192.8                           & 0                                       \\
        Norway                 & 0.36                             & 66.3                            & 1.5                                     \\
        Poland                 & 0.33                             & 181.9                           & 3.7                                     \\
        Portugal               & 0.91                             & 42.9                            & 2.4                                     \\
        Romania                & 0.18                             & 197.5                           & 2.2                                     \\
        Serbia                 & 0.31                             & 117.5                           & 2.2                                     \\
        Sweden                 & 3                                & 74.8                            & 13.7                                    \\
        Slovenia               & 0.02                             & 213.8                           & 0.3                                     \\
        Slovakia               & 0.1                              & 215.5                           & 1.3                                     \\
        Kosovo                 & 0.05                             & 117                             & 0.4                                     \\
        Europe                 & 14.54                            & 117                             & 83.1                                    \\
        \bottomrule
    \end{tabular}
\end{table}


\begin{table}[H]
    \small
    \centering
    \caption[Perennialisation land availability and CO$_2$ removal potential per country]{\textbf{Perennialisation land availability and CO$_2$ removal potential per country}. Land currently used for first-generation biofuel crops—as derived from the JRC ENSPRESO dataset—is considered fully available for replacement with perennial crops. CO$_2$ removal potential for perennialisation is calculated using Equation~\ref{eq:perennials_seq}.}
    \label{supplemental:table_perennials_land_and_sequestration_potentials}
    \begin{tabular}{p{3.8cm} p{3.8cm} p{4.9cm}}
        \toprule
        \textbf{Country}       & \textbf{Land availability [Mha]} & \textbf{Removal potential [MtCO$_2$/a]} \\
        \midrule
        Albania                & 0                                & 0                                       \\
        Austria                & 0.23                             & 0.5                                     \\
        Bosnia and Herzegovina & 0                                & 0                                       \\
        Belgium                & 0.28                             & 0.6                                     \\
        Bulgaria               & 1.03                             & 2.1                                     \\
        Switzerland            & 0                                & 0                                       \\
        Czech Republic         & 0.5                              & 1                                       \\
        Germany                & 1.69                             & 3.4                                     \\
        Denmark                & 0.09                             & 0.2                                     \\
        Estonia                & 0.25                             & 0.5                                     \\
        Spain                  & 2.29                             & 4.6                                     \\
        Finland                & 0.07                             & 0.1                                     \\
        France                 & 4.6                              & 9.2                                     \\
        Great Britain          & 1.79                             & 3.6                                     \\
        Greece                 & 0.51                             & 1                                       \\
        Croatia                & 1.14                             & 2.3                                     \\
        Hungary                & 1.01                             & 2                                       \\
        Ireland                & 0.02                             & 0                                       \\
        Italy                  & 1.64                             & 3.3                                     \\
        Lithuania              & 0.72                             & 1.4                                     \\
        Luxembourg             & 0.01                             & 0                                       \\
        Latvia                 & 0.17                             & 0.3                                     \\
        Montenegro             & 0                                & 0                                       \\
        North Macedonia        & 0                                & 0                                       \\
        The Netherlands        & 0.12                             & 0.2                                     \\
        Norway                 & 0                                & 0                                       \\
        Poland                 & 1.47                             & 2.9                                     \\
        Portugal               & 0.02                             & 0                                       \\
        Romania                & 1.39                             & 2.8                                     \\
        Serbia                 & 0                                & 0                                       \\
        Sweden                 & 0.08                             & 0.2                                     \\
        Slovenia               & 0.01                             & 0                                       \\
        Slovakia               & 0.28                             & 0.6                                     \\
        Kosovo                 & 0                                & 0                                       \\
        Europe                 & 21.43                            & 42.9                                    \\
        \bottomrule
    \end{tabular}
\end{table}


\begin{table}[H]
    \small
    \centering
    \caption[Biochar land availability and CO$_2$ removal potential per country]{\textbf{Biochar land availability and CO$_2$ removal potential per country}. Land availability for biochar is determined using the CORINE Land Cover dataset, and its CO$_2$ removal potential is calculated using Equation~\ref{eq:biochar_potential}.}
    \label{supplemental:table_biochar_land_and_sequestration_potentials}
    \begin{tabular}{p{3.8cm} p{3.8cm} p{4.9cm}}
        \toprule
        \textbf{Country}       & \textbf{Land availability [Mha]} & \textbf{Removal potential [MtCO$_2$/a]} \\
        \midrule
        Albania                & 0.03                             & 0                                       \\
        Austria                & 0.26                             & 0.3                                     \\
        Bosnia and Herzegovina & 0.04                             & 0                                       \\
        Belgium                & 0.14                             & 0.2                                     \\
        Bulgaria               & 0.78                             & 1                                       \\
        Switzerland            & 0.15                             & 0.2                                     \\
        Czech Republic         & 0.6                              & 0.8                                     \\
        Germany                & 2.72                             & 3.6                                     \\
        Denmark                & 0.55                             & 0.7                                     \\
        Estonia                & 0.14                             & 0.2                                     \\
        Spain                  & 2.54                             & 3.3                                     \\
        Finland                & 0.34                             & 0.4                                     \\
        France                 & 3.11                             & 4.1                                     \\
        Great Britain          & 1.34                             & 1.8                                     \\
        Greece                 & 0.42                             & 0.5                                     \\
        Croatia                & 0.08                             & 0.1                                     \\
        Hungary                & 0.98                             & 1.3                                     \\
        Ireland                & 0.07                             & 0.1                                     \\
        Italy                  & 1.69                             & 2.2                                     \\
        Lithuania              & 0.43                             & 0.6                                     \\
        Luxembourg             & 0.01                             & 0                                       \\
        Latvia                 & 0.21                             & 0.3                                     \\
        Montenegro             & 0                                & 0                                       \\
        North Macedonia        & 0.06                             & 0.1                                     \\
        The Netherlands        & 0.15                             & 0.2                                     \\
        Norway                 & 0.11                             & 0.1                                     \\
        Poland                 & 2.75                             & 3.6                                     \\
        Portugal               & 0.23                             & 0.3                                     \\
        Romania                & 1.76                             & 2.3                                     \\
        Serbia                 & 0.44                             & 0.6                                     \\
        Sweden                 & 0.6                              & 0.8                                     \\
        Slovenia               & 0.02                             & 0                                       \\
        Slovakia               & 0.33                             & 0.4                                     \\
        Kosovo                 & 0.02                             & 0                                       \\
        Europe                 & 23.09                            & 30.2                                    \\
        \bottomrule
    \end{tabular}
\end{table}


\begin{table}[H]
    \small
    \centering
    \caption[ERW land availability and CO$_2$ removal potential per country]{\textbf{ERW land availability and CO$_2$ removal potential per country}. Land availability for ERW is determined using the CORINE Land Cover dataset, and its CO$_2$ removal potential is calculated using Equation~\ref{supplemental:material_erw_potential}.}
    \label{supplemental:table_ERW_land_and_sequestration_potentials}
    \begin{tabular}{p{3.8cm} p{3.8cm} p{4.9cm}}
        \toprule
        \textbf{Country}       & \textbf{Land availability [Mha]} & \textbf{Removal potential [MtCO$_2$/a]} \\
        \midrule
        Albania                & 0.02                             & 0.1                                     \\
        Austria                & 0.23                             & 0.7                                     \\
        Bosnia and Herzegovina & 0.04                             & 0.1                                     \\
        Belgium                & 0.13                             & 0.4                                     \\
        Bulgaria               & 0.15                             & 0.5                                     \\
        Switzerland            & 0.09                             & 0.3                                     \\
        Czech Republic         & 0.52                             & 1.6                                     \\
        Germany                & 2.33                             & 7.2                                     \\
        Denmark                & 0.01                             & 0                                       \\
        Estonia                & 0                                & 0                                       \\
        Spain                  & 0.29                             & 1                                       \\
        Finland                & 0                                & 0                                       \\
        France                 & 3.09                             & 9.5                                     \\
        Great Britain          & 1.16                             & 3.6                                     \\
        Greece                 & 0.05                             & 0.3                                     \\
        Croatia                & 0.08                             & 0.2                                     \\
        Hungary                & 0.31                             & 1                                       \\
        Ireland                & 0.07                             & 0.2                                     \\
        Italy                  & 1.23                             & 3.9                                     \\
        Lithuania              & 0.05                             & 0.2                                     \\
        Luxembourg             & 0                                & 0                                       \\
        Latvia                 & 0                                & 0                                       \\
        Montenegro             & 0                                & 0                                       \\
        North Macedonia        & 0.01                             & 0                                       \\
        The Netherlands        & 0.14                             & 0.4                                     \\
        Norway                 & 0                                & 0                                       \\
        Poland                 & 2.63                             & 8.1                                     \\
        Portugal               & 0.06                             & 0.2                                     \\
        Romania                & 0.46                             & 1.4                                     \\
        Serbia                 & 0.2                              & 0.6                                     \\
        Sweden                 & 0.02                             & 0.1                                     \\
        Slovenia               & 0.02                             & 0.1                                     \\
        Slovakia               & 0.23                             & 0.7                                     \\
        Kosovo                 & 0                                & 0                                       \\
        Europe                 & 13.66                            & 42.5                                    \\
        \bottomrule
    \end{tabular}
\end{table}


\begin{table}[H]
    \small
    \centering
    \caption[Solid biomass transport cost per country]{\textbf{Solid biomass transport cost per country}. The costs are based on the JRC European TIMES Energy System Model (JRC-EU-TIMES) and are available at \url{https://data.jrc.ec.europa.eu/collection/id-00287}. Due to the material similarity between solid biomass and biochar, these costs are also applied to biochar transport across Europe.}
    \label{supplemental:table_solid_biomass_transport_cost}
    \begin{tabular}{p{4.3cm} p{4cm}}
        \toprule
        \textbf{Country}       & \textbf{Cost [EUR/(MWh$\cdot$km)]} \\
        \midrule
        Albania                & 0.0552                             \\
        Austria                & 0.1365                             \\
        Bosnia and Herzegovina & 0.0594                             \\
        Belgium                & 0.1406                             \\
        Bulgaria               & 0.0635                             \\
        Switzerland            & 0.1771                             \\
        Czech Republic         & 0.0875                             \\
        Germany                & 0.1333                             \\
        Denmark                & 0.1906                             \\
        Estonia                & 0.0865                             \\
        Spain                  & 0.1198                             \\
        Finland                & 0.1490                             \\
        France                 & 0.1427                             \\
        Great Britain          & 0.1438                             \\
        Greece                 & 0.1115                             \\
        Croatia                & 0.0802                             \\
        Hungary                & 0.0729                             \\
        Ireland                & 0.1333                             \\
        Italy                  & 0.1323                             \\
        Lithuania              & 0.0740                             \\
        Luxembourg             & 0.1458                             \\
        Latvia                 & 0.0813                             \\
        Montenegro             & 0.0604                             \\
        North Macedonia        & 0.0510                             \\
        The Netherlands        & 0.1458                             \\
        Norway                 & 0.1615                             \\
        Poland                 & 0.0781                             \\
        Portugal               & 0.0927                             \\
        Romania                & 0.0625                             \\
        Serbia                 & 0.0573                             \\
        Sweden                 & 0.1615                             \\
        Slovenia               & 0.0979                             \\
        Slovakia               & 0.0833                             \\
        Kosovo                 & 0.0583                             \\
        \bottomrule
    \end{tabular}
\end{table}

\end{document}